VIETNAM NATIONAL UNIVERSITY HANOI
UNIVERSITY OF ENGINEERING AND TECHNOLOGY

Le Viet Ha

# ENHANCING WEBSHELL DETECTION WITH DEEP LEARNING-POWERED METHODS

PHD DISSERTATION IN INFORMATION SYSTEMS

**SUPERVISORS**

Nguyen Ngoc Hoa    Phung Van On

Ha Noi - 2024

# DECLARATION OF AUTHORSHIP

I, Le Viet Ha, declare that this dissertation titled, "ENHANCING WEBSHELL DETECTION WITH DEEP LEARNING-POWERED METHODS" and the work presented in it are my own. I confirm that:

- This work was done mainly while in candidature for the degree of Ph.D at VNU University of Engineering and Technology.

- This dissertation has not previously been submitted for any degree.

- The results in my dissertation are my independent work, except where works in the collaboration have been included. Other appropriate acknowledgments are given within this dissertation by explicit references.

Signed:

Date:

# ACKNOWLEDGEMENTS

This dissertation would not have been possible without the support, guidance, and encouragement of many individuals.

First and foremost, I would like to express my deepest gratitude to my supervisors, Associate Professor Nguyen Ngoc Hoa and Doctor Phung Van On, whose expertise, patience, and unwavering support have been instrumental in the completion of this research. Your insightful feedback and continuous motivation have pushed me to refine my work and think critically, for which I am profoundly grateful.

I am deeply appreciative of the support from my colleagues and friends, whose encouragement and camaraderie have provided me with the energy and resilience to persevere through the challenges of this journey.

Lastly, but most importantly, I owe a great debt of gratitude to my family, whose love and understanding have been my constant source of strength. This accomplishment would not have been possible without you.

Thank you all for your contributions to this work and to my life.

# ABSTRACT

The increasing prevalence of webshell attacks poses a significant threat to web application security, necessitating the development of robust detection mechanisms. The dissertation clearly identifies two research directions: scanning web application source code and in-depth analysis of HTTP traffic to detect webshells. First, the dissertation proposes an advanced DL-Powered Source-Code Scanning Framework, called ASAF, that integrates signature-based techniques with deep learning algorithms to enhance the detection of both known and unknown webshells. We design the framework to facilitate the creation of customized detection models for various programming languages. For the interpreted language, the study chose PHP; for the compiled language, the dissertation chose ASP.NET to build a complete ASAF-based model for experimentation and comparison with other research results to prove its effectiveness.

Second, the dissertation introduces a deep neural network that utilizes real-time HTTP traffic analysis of web applications to detect webshells. The study proposes an algorithm to improve the loss function applied in the deep learning model to solve the problem of data imbalance. To demonstrate its effectiveness, we experimented with and compared the model to other studies on the same CSE-CIC-IDS2018 dataset. We have also integrated the model with the NetIDPS system to improve its capacity to identify new webshells. From there, proactively prevent these attacks by automatically adding attack source IPs to the blacklist and creating rules to block URIs querying webshells on the web server.

This research contribution has been demonstrated through 01 national patent, 2 SCI-E journals, 1 E-SCI journal, 1 national journal, 2 WoS conference papers and 1 pending patent, as well as being practically applied in the national research project, code number KC01.19/16-20, granted by Ministry of Science and Technology of Vietnam.

# TABLE OF CONTENTS

# LIST OF FIGURES

# LIST OF TABLES

# ABBREVIATIONS

| APT | Advanced Persistent Threat |
|---|---|
| ANN | Artificial Neural Network |
| AES | Advanced Encryption Standard |
| CNN | Convolutional Neural Network |
| DNN | Deep Neural Network |
| DT | Decision Tree |
| DL | Deep Learning |
| HTTP | HyperText Transfer Protocol |
| IDS | Intrusion Detection System |
| IPS | Intrusion Prevention System |
| GBDT | Gradient Boosted Decision Trees |
| LSTM | Long Short-Term Memory |
| ML | Machine Learning |
| MLP | Multilayer Perceptron |
| NB | Naive Bayes |
| OpCode | Operation Code |
| RNN | Recurrent Neural Network |
| RSA | Rivest–Shamir–Adleman |
| SVM | Support Vector Machine |
| SSL | Secure Sockets Layer |
| TLS | Transport Layer Security |
| TF-IDF | Term Frequency - Inverse Document Frequency |
| RF | Random Forest |
| WAF | Web Application Firewall |

# INTRODUCTION

## Research Motivations

***Webshell Attack*** Nowadays, digital transformation is considered an important and inevitable trend for many countries around the world. In Vietnam, digital transformation has become a topic of interest in recent years and is most clearly demonstrated through the National Digital Transformation Program that has been issued. The advancement of web development [22, 11] technology has made web applications more and more popular, gradually replacing traditional native applications because they do not depend on the operating system. Most applications serving e-government and digital transformation in Vietnam today are built on web platforms, typically the National Public Service Portal system [1]. Along with this, the issues of information security for the web system have become increasingly important. Malicious code injection (webshell) attacks [33, 95, 68] are the most common and also the most hazardous sort of web application attack [28]. According to the recent Microsoft 365 Defender data [2], the use of webshell attacks not only continued but also accelerated every day. Webshell attacks [103] pose a severe threat to organisations due to the extensive damage and vulnerabilities they introduce after compromising web-facing servers.

As pieces of malicious code written in common web development programming languages (e.g., ASP, PHP, and JSP) that are installed on web servers, webshells allow attackers to remotely execute arbitrary system commands, exfiltrate sensitive files, install additional payloads, and pivot laterally into internal networks. Attackers can also use webshells to maintain stealthy persistence in order to prolong exploitation after the initial breach. Many advanced webshells feature extensive capabilities via graphical user interfaces, including brute-forcing credentials, uploading malware, and interacting with databases. Once a webshell is uploaded, attackers have an unrestricted foothold within the victim's infrastructure. Webshells are especially dangerous due to their ability to bypass conventional network perimeter defences by using allowed protocols like HTTP or HTTPS [96]. Their flexible and compact nature also allows webshells to evade detection through obfuscation and polymorphism [3, 65]. Overall, webshells represent a serious threat due to their role as a pivot point, enabling an unimpeded gateway for attackers.

---

[1] https://dichvucong.gov.vn/p/home/dvc-trang-chu.html

[2] Web shell attacks continue to rise, https://www.microsoft.com/en-us/security/blog/2021/02/11/web-shell-attacks-continue-to-rise

Advances in detection techniques have struggled to keep pace as attackers continually release new, heavily obfuscated webshell tools to evade defenses. Manual inspection is time-consuming, given that a single webshell update could require hours of expert reverse engineering. Detecting obfuscated webshells poses significant challenges for security research. Attackers are continuously adapting exploitation techniques to evade detection, deploying webshells encoded by means such as base64 or hex encoding, and using custom encryption schemes. According to analysis from Cloudflare, over two-thirds of webshells exhibit some form of obfuscation. Advanced polymorphic webshells such as "Chameleon" can rapidly mutate appearances across attacks while maintaining core malicious functions. The ease of automating webshell obfuscation and morphing has outpaced improvements in detection approaches tailored to discerning underlying patterns amid intentionally distorted malcode. Defenders also face challenges in obtaining robust datasets spanning various obfuscation schemas needed to train machine learning models.

**Webshell Detection**

Two primary approaches exist across the spectrum of webshell detection: Source Code Analysis and Network-based Analysis.

Source code analysis takes yet another approach by directly analysing web application source code for webshell using analysis tools. Code analysis works by inspecting repositories for suspicious functions, commands, file inclusions, or other constructs indicative of a webshell payload. This enables identifying inactive webshells injected into the code before production deployment. Analysing source code rather than running software provides the ability to catch webshells compiled directly into applications. However, code analysis faces challenges in detecting highly obfuscated or customised webshells designed to mask their malicious intent. Without runtime context, benign code can also generate false positives.

Network-based analysis webshell detection [98] operates by analysing web traffic as it enters or exits the network perimeter. This is commonly implemented through Web Application Firewalls (WAFs) [10, 36] or Intrusion Detection and Prevention Systems (IDPSs) [67, 8, 7, 15] examining packets and connections. Detection works by identifying anomalous patterns in network traffic that are different from how a legitimate application should work. For example, the WAF could detect unusual HTTP request parameter names, uncommon user-agent strings, excessively long request content lengths, or other characteristics that signal a webshell payload. Network security devices build detection rules based on common webshell patterns or by observing benign traffic to flag outliers. Rules must be constantly updated to catch new and modified webshells in the endless evolution of attacker tradecraft. A major advantage of network-based webshell detection is the ability to catch both inbound attacks as well as outbound commands and control. Monitoring perimeter traffic enables catching webshell upload attempts and immediately blocking malicious IPs. The network view also enables examining outbound connections for communications indicative of an active webshell, such as unexpected shell prompts or terminal commands. Network detection faces limitations in identifying highly obfuscated webshells designed to mimic legitimate traffic through extensive encoding and morphism. Skilled attackers can often craft webshells that evade signature-based detection systems.

*AI-Powered for Webshell Detection*

Today, the advent of Machine Learning (ML) and Deep Learning (DL) techniques has revolutionised the field of cybersecurity, particularly in the domain of webshell detection.

Code analysis techniques, leveraging ML/DL algorithms, involve the extraction and analysis of features from webshell code, encompassing syntax, semantic structures, and behavioural patterns in identifying malicious code. The authors in [30] utilise vectorized opcode sequences extracted from PHP webshells to evaluate an NB-opcode model, achieving 97.4% accuracy. The authors in [20] propose a combination of random forest classifier and GBDT classifier to detect PHP webshells, achieving accuracy and false positive rates of 99.169% and 0.682%, respectively. The authors in [4] propose an ensemble detection model consisting of Logistic Regression, Support Vector Machine, Multi-layer Perceptron, and Random Forest to detect PHP webshell. The experiment demonstrates that their model could improve the accuracy rate up to 99.14%. The authors in [69] propose a matrix decomposition algorithm using statistical features extracted from known webshells. This matrix is used to weigh the features applied to the supervised machine learning algorithm to detect multilingual webshells, but their accuracy was not high. It can be seen that most of the research focuses on the PHP language, because PHP is the most popular server-side programming language today. Very few studies have the ability to detect other types of webshells, such as ASP.NET, JSP, Perl, or Python, or if they do, the accuracy is not high and cannot be applied in practice.

ML/DL also facilitates webshell detection by enabling deep inspection of network traffic patterns. These models can analyse network packet attributes like source and destination IP addresses, payload data, communication protocols, and timing information to extract sophisticated features that indicate webshell activity. Furthermore, their ability to analyse large volumes of normal background network traffic to identify abnormal activity makes them suitable for real-time detection in high-traffic environments. The authors in [99] combine the characteristics of convolutional neural networks and long short-term memory networks to detect the existence of Webshell in network traffic. To do so, they propose a character-level method to transform the Webshell content feature to archive 98.51% F1-score. The authors in [96] propose a runtime webshell detection system analysing HTTP requests. It uses a Support Vector Machine (SVM) classifier that was trained on preprocessed and vectorized HTTP requests. The preprocessing includes decoding the GET and POST parameters, and the SVM sorts the requests into three groups: suspicious, attack, and benign. The authors in [71] proposed a Word2Vec-based method to vectorize the HTTP requests. These vectors will be used as input data for the CNN model to classify malicious HTTP requests from normal types. The model shows a better result compared to other traditional machine learning models such as NB, DT, and SVM. There are a lot of studies that detect webshells by analysing network traffic, but more than half of the studies focus on intrusion detection and the classification of types of cyber attacks and are unlikely to detect new types of webshells.

*The increasing sophistication and variety of webshells, particularly those designed to evade traditional detection methods, underscores the urgent need for advanced techniques to improve their detection. Many studies demonstrate the effectiveness of using ML and DL algorithms to improve the ability to detect new variants of webshells. However, these studies still have certain limitations, and there is much room for improvement. This is the main motivation for us to conduct this dissertation.*

## Research Challenges

The research context analysed above shows that webshell detection algorithms that can be applied in practice still face a number of challenges, as follows:

1. **The diversity of webshell languages** underscores the dynamic landscape of cybersecurity threats and defensive measures. These languages encompass a spectrum ranging from widely used scripting languages like PHP, Python, and Perl to more specialised ones such as ASP, JSP, and Ruby. Each language offers unique features and capabilities; therefore, how to represent the source files so that the functionalities of the webshell can be fully expressed is the issue. There are various lines of research being pursued right now to find webshells in source code files utilising AI techniques applied to image recognition or natural language processing. However, the downside of most of these methods is that they convert almost the original content of the source files into a matrix, which will no longer be valid for the webshell that equipped code obfuscation, code encryption, or evasion techniques. Understanding this diversity is paramount for cybersecurity professionals to develop robust defence strategies and safeguard against webshellbased attacks.

2. **Advanced webshells** often exhibit complex functionalities and evasion techniques, leveraging obfuscation, encryption, and polymorphism to conceal their presence and evade detection by traditional security measures [77, 9, 26, 60]. They are obfuscated to obscure the original source code while maintaining functionality. The primary purpose is to evade detection by security mechanisms and make it harder to identify and remove the malicious code. Webshells use polymorphic techniques, where the code changes its structure each time it is executed. Traditional signature-based methods relying on fixed patterns or codes are less effective against polymorphic webshells. A fileless webshell is a sort of webshell that does not write any files to the disc of the target machine like typical webshells. They exploit legitimate system tools and utilities to execute arbitrary commands without the need for persistent files. As a result, detection requires more advanced techniques, such as behaviour-based monitoring, anomaly detection, and runtime analysis of memory activities.

3. **Quality of the datasets** is one of the key factors in the development of webshell defence techniques. However, since webshell is sensitive data, there will not be many official, reliable data sources willing to share it. Furthermore, new webshell variants and techniques are constantly emerging, and hackers never share these new webshells for free until they become obsolete. There are currently some open webshell datasets such as Tenc [3] and WebSHark 1.0 [42] datasets. Another difficulty is that, in the case of false positives and negatives, we cannot guarantee 100% of the collected data to be classified correctly.

4. **The effectiveness of the detection method** is demonstrated by three criteria: accuracy, detection time and resource usage. These three criteria are always closely linked, but in opposite directions. In webshell detection problems, the big challenge is to build a solution while attaining both criteria of accuracy in the detection of advanced webshells, detection speed fast enough to minimise damage to the system, and optimising resource use. Signature-based webshell detection solutions can give very fast results but are only effective against known webshells. As for unknown webshells, it is necessary to use deep analysis techniques that often consume a lot of resources and time.

---

[3] tennc, Tennc, https://github.com/tennc/WebShell, 2021

5. **Practical application** with information security systems is an important factor if the solution is to be applied to practical use. For example, network-based analysis for webshell detection must be able to integrate with IDPS to automatically update rules so that the system blocks IP hackers as soon as anomaly signs are detected.

## Objectives of Dissertation

The main objective of the dissertation is to propose webshell detection methods that employ the deep learning models in order to improve the performance in term of accuracy and effective. To achieve the main objective of the dissertation, four specific objectives are as follows:

- Objective 1: Overview of webshell, the most advanced techniques used by hackers to hide or evade their webshell attack. Research webshell detection techniques and analyse the advantages and disadvantages of each method. Evaluate the results of the latest research on the problem of detecting webshell attacks.

- Objective 2: Proposing an DL-Powered Source Code Analysis Framework, namely ASAF, that combines signature-based techniques with deep learning algorithms. This hybrid approach enables the rapid and accurate detection of both known and unknown webshell types. The proposed framework provides a guideline for developing specific models tailored to various programming languages.

- Objective 3: Based on the proposed framework above, develop two comprehensive systems tailored to detect webshell attacks using PHP (interpreted language) and ASP.NET (compiled language). The deep learning models integrated into the systems must be optimised for their specific webshell detection problems to ensure effective detection with minimal computational resources. The detection results of the systems must be compared with those of other studies to prove their effectiveness.

- Objective 4: Proposing a deep learning model for webshell attacks that perform in-depth analysis of HTTP queries directed at web application systems, effectively identifying queries that indicate both known and unknown webshell attacks. The model is capable of seamlessly integrating into NetIDPS, demonstrating its practical applicability for automatic blocking of suspicious webshell attack source addresses in real-time.

## Research Scope

To achieve the objectives of this dissertation, we focus on the following key areas:

1. Researching methods and techniques for analyzing web application source code to detect webshells.

2. Researching methods and techniques to deeply analyze HTTP network traffics exchange with web application servers to detect queries to webshell.

## Methodologies

The research methodology for the dissertation is conducted in a systematic manner, as described in the following:

- **Theoretical Methodology:** We take a survey, synthesize, and evaluate previous research relevant to the dissertation in order to analyze the achieved result and the remaining problems that need further research in the direction of the dissertation. Documents and information are mainly collected from articles in prestigious scientific journals on the ISI/Scopus list and proceedings of specialized scientific conferences from reputable online libraries. IEEE Xplorer [4], ACM Digital Library [5], SpringerLink [6], ScienceDirect [7], Wiley Online Library [8].From there, select the remaining problems in webshell detection models and use machine learning to research and propose more effective detection models.

- **Experimental Methodology:**

To evaluate the effectiveness of webshell detection models using web application source scanning, we used the self-built datasets consisting of 11,362 PHP files (7,275 benign files and 4,087 Webshell files) and 5411 ASP.NET files (3,347 benign files and 2,064 Webshells). These are sets of data we collected from reputable, selected, and verified sources, including github source-sharing libraries and domestic and international webshell research groups.

---

[4] https://ieeexplore.ieee.org
[5] https://dl.acm.org
[6] https://link.springer.com
[7] https://www.sciencedirect.com/
[8] https://onlinelibrary.wiley.com

To evaluate the effectiveness of a webshell detection model using deep analysis of network traffic, we used the CSE-CIC-IDS2018 data set from the Canadian Institute for Cybersecurity to objectively compare the efficiency of the proposed model with other related studies and the self-built dataset through our test bed to assess the ability of the model to detect advanced webshell attacks.

We test the models proposed in the dissertation using the above data set and compare their effectiveness with the results of other studies related to webshell attack detection.

## Research Contributions

With clearly defined research objectives and research scope, the dissertation has the following contributions:

1. Proposing an DL-powered source code scanning framework for webshell detection that combines signature-based techniques with deep learning algorithms. This framework provides guidance for developing specific models for accurate and efficient webshell detection in a variety of programming languages. For each type of interpreted and compiled programming language, we chose PHP and ASP.NET as the most popular languages of each type to build a webshell detection model based on ASAF. We conducted experiments and compared the above model with other studies to prove the effectiveness of ASAF.

2. Propose a deep learning model to thoroughly analyze the HTTP traffic to the web application server in order to quickly detect webshell queries. To solve the problem of data imbalance for training sets, we also propose an algorithm to improve the quality of training sets employed in the deep learning model. To demonstrate its effectiveness, we experimented and compared the model to other studies. The deep learning model can work with the intrusion detection and prevention system to add attack source IPs to a blacklist and proactively block URI queries to webshell on the web server before they happen.

## Dissertation Structure

This dissertation has three main chapters, each with the following main contents, in addition to the introduction and conclusion.

- Chapter 1 provides an overview of webshells, methods for detecting webshell attacks, the use of ML/DL in webshell detection, a review of scientific literature, and criteria for evaluating the effectiveness of ML/DL models.

- Chapter 2 proposes an DL-powered source code scanning framework that combines signature-based techniques with deep learning algorithms. This hybrid approach enables the rapid and accurate detection of both known and unknown webshell types. The proposed framework provides a guideline for developing specific models tailored to various programming languages. Based on the proposed framework above, we will develop two comprehensive systems tailored to detect webshell attacks using PHP and ASP.NET.

- Chapter 3 proposes deep learning model for webshell attacks that perform indepth analysis of HTTP queries directed at web application systems, effectively identifying queries that indicate both known and unknown webshell attacks. We experiment with the model on two datasets and compare the results with those of other studies to showcase its effectiveness. We have also integrated the model into a NetIDPS system to automatically block suspicious source addresses in real-time, ensuring its practical applicability.

# Chapter 1

# THEORETICAL BACKGROUND AND PRELIMINARIES

In this chapter, we will present some basic knowledge about webshell, webshell evasion techniques, and webshell detection approaches, as well as analyze and evaluate the results achieved by related works.

The dissertation then identifies scientific gaps in this context, determines the research direction of the dissertation, and serves as a basis for highlighting new contributions in the following chapters.

## 1.1 Fundamental Concepts

### 1.1.1 Webshell Overview

A typical web application architecture consists of three fundamental components:

- Web Browser: This is the client-side component that serves as the primary interface for user interaction. It receives user input, manages presentation logic, and controls user interactions with the application. Additionally, it may validate user inputs for data integrity and security purposes when necessary.

- Web Server: also known as the server-side component, is in charge of handling the application's business logic. It routes user requests to the appropriate components and oversees the application's overall operation. It has the capability to handle requests from a diverse range of clients.

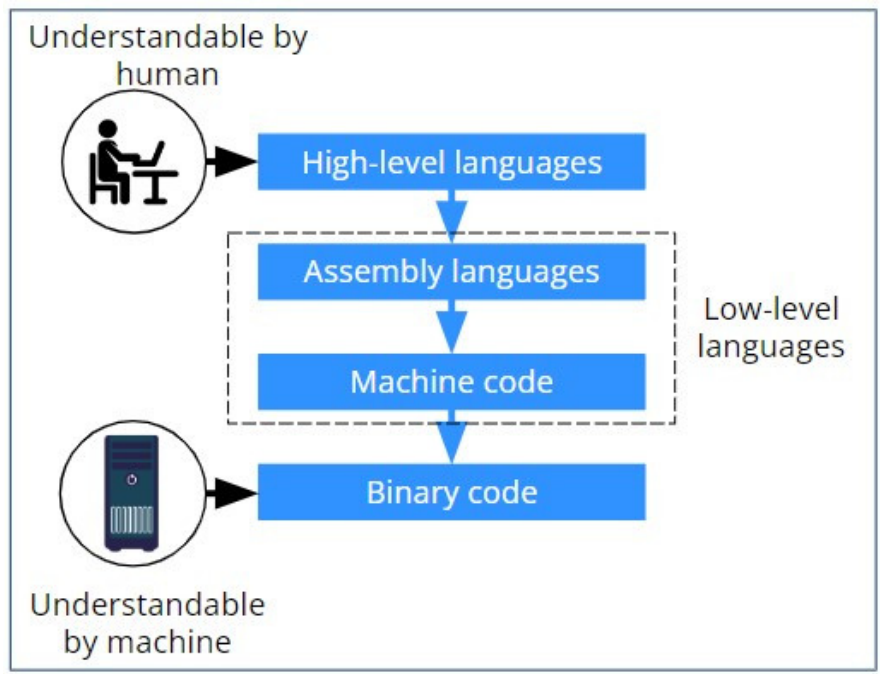

Figure 1.1: The conversion process from programming languages to machine code

- Database Server: is responsible for providing the application with the necessary data. It manages data-related tasks such as storage, retrieval, and manipulation. In multi-tiered architectures, database servers may also handle business logic tasks with the assistance of stored procedures.

The web server utilizes the HyperText Transfer Protocol (HTTP) along with other protocols to receive user requests through a web browser. The web server then processes these requests, applying business logic to fulfill them. The web server then delivers the requested content to the end-user, facilitating their interaction with the web application. As a result, the server side will need a programming language to handle requests sent from the client side (browser). A programming language consists of a set of commands, syntax, and semantics used to control a computer and perform a specific job. Several systematic steps transform a high-level programming language into machine language, transforming human-readable code into executable instructions that a computer's hardware can comprehend and execute. Initially, lexical analysis breaks down the high-level language source code into individual tokens such as keywords, identifiers, operators, and literals. After lexical analysis, the program undergoes syntax analysis to conform to the programming language's syntactic rules, creating a hierarchical structure known as the abstract syntax tree. We then conduct semantic analysis to confirm the meaning and context of statements, identifying any errors or inconsistencies in the program's logic. Once the semantic analysis successfully completes, it generates an intermediate representation of the program, not only providing a platform-independent representation of the program's logic but also facilitating optimization and code generation. Before the final step of code generation, you can apply optional optimization steps to enhance the efficiency and performance of the program. During code generation, the program is translated into machine code or bytecode, which converts high-level language constructs into equivalent instructions in the target machine architecture. The computer's processor then executes this generated machine code, enabling the desired computations or actions specified by the original high-level program. The comprehensive conversion process, is shown in the Fig. 1.1, that entails the coordination of various components, such as compilers, interpreters, and runtime environments, to ensure accurate and efficient translation from high-level programming languages to machine language. Before execution, a compiler compiles the entire program into machine code. An interpreter interprets the program during execution, line by line or statement by statement. In addition to the web server software (such as Apache, IIS, or Nginx) responsible for establishing a connection and transferring data using the HTTP protocol between the browser and the web server, a web server consists of numerous other components. The Fig. 1.2 represents Apache Web Server architecture. The PHP interpreter is responsible for converting source code into operation code (OpCode), helping the server perform logical processing and return results to the web client.

- **Interpreter**

An interpreter is a computer program that helps the CPU directly execute instructions written in a programming or scripting language without requiring prior compilation into a machine language program as shown in the Fig. 1.3. If the web application is programmed in an interpreted language, the result file of the translation process will not be visible, and the website will include source code files in programming language form.

- **Compiler**

A compiler is a computer program that translates a series of statements of a programming language, called source code, into an equivalent program but in the form of a computer language as shown in the Fig. 1.4. The CPU, or virtual machine, will execute the object code or machine code generated by the compiler.

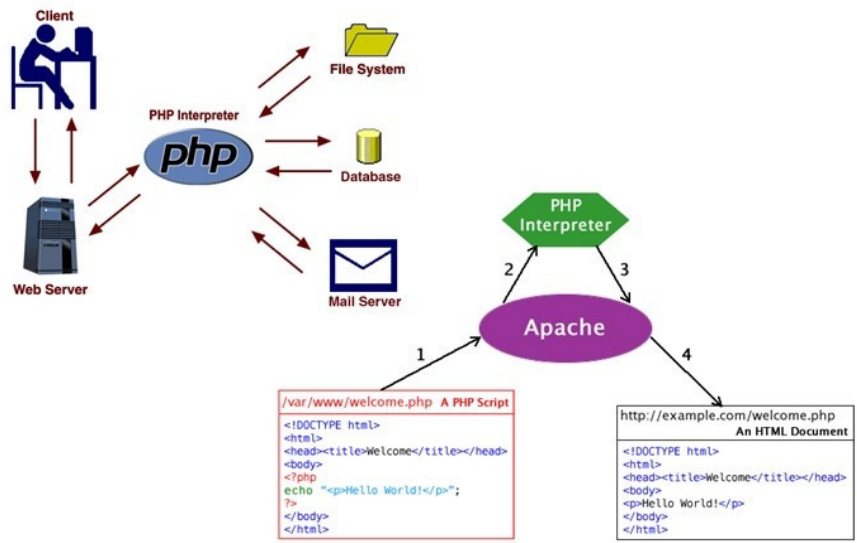

Figure 1.2: Example of Apache web server architecture

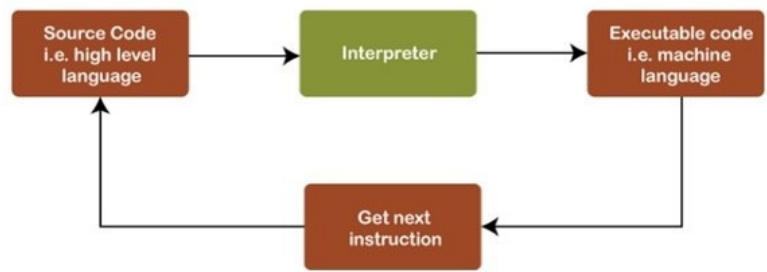

Figure 1.3: Interpreter process

A decompiler is a type of compiler that converts from a low-level programming language to a high-level language, while a layered compiler converts from one high-level language to another high-level language or to an intermediate language for further processing.

A website programmed in a compiled language will see the results of the translation process. The website will include source code files in programming language form, intermediate code files, or machine code files.

#### 1.1.1.1 Webshell Definition

A webshell is a small piece of malicious code injected on web servers by attackers to grant remote access and code execution that is written in popular web development programming languages (e.g., ASP, PHP, and JSP) [33, 95, 68]. Webshell is an entry point that gives hackers the ability to execute commands on servers to steal data or to utilize the server as a base for other attacks like lateral movement, the deployment of additional payloads, or hands-on keyboard activity, all while remaining undetected inside a target. However, a webshell itself cannot attack or exploit a remote vulnerability, so it is always the second stage of an attack, named post-exploitation.

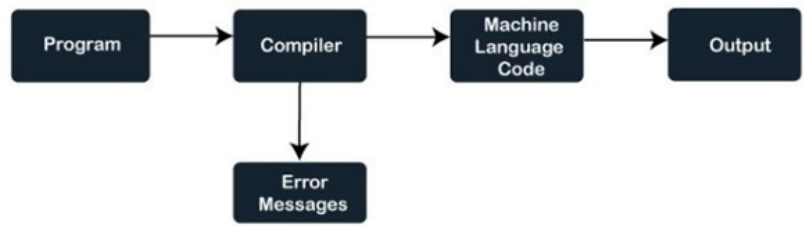

Figure 1.4: Compiler process

As a typical example, Hafnium, a Chinese cybercriminal organization, carried out the recent significant webshell attack that made news in March 2021. A malware program called China Chopper [1], was used in the attack as the webshell, and it was injected through a critical vulnerability in Microsoft Exchange Servers. The China Chopper comes in a little box but packs a big punch. In just 4 kilobytes of space, the easy-to-use webshell graphical user interface allows even beginners to handle files and databases, obfuscate code, and more. Even after patching the server vulnerability, the China Chopper webshell's backdoor remained in the compromised system, rendering it exceptionally harmful.

The ChinaChopper webshell attack is shown in Fig. 1.5, where a hacker exploits an existing vulnerability in the webserver to install a malicious code file into the webserver directory and then commands it to run the file by making a request through the web browser. Executing infected files causes a persistent breach in the web server, allowing a hacker to launch any cyberattack.

To find servers to target, common hackers search the internet, frequently utilizing scanning tools like Shodan.io. They have been known to quickly exploit recently discovered vulnerabilities, as well as previously patched ones that still exist on many servers. As for APT attacks [101], they are highly sophisticated and targeted cyberattacks conducted by well-funded and organized threat actors, often with nation-state backing or significant resources. APT attacks are characterized by their long-term and persistent nature, as the attackers focus on maintaining access to a compromised system or network over an extended period of time to achieve specific objectives. Due to their complexity, APT attacks require a comprehensive cybersecurity strategy involving proactive monitoring, threat intelligence, intrusion detection systems, regular security assessments, employee training, and incident response plans. To minimize potential damage, organizations must be prepared to detect and respond to APT attacks quickly and effectively.

---

[1]China Chopper Webshell Report, https://www.mandiant.com/sites/default/files/2021-09/rpt-china-chopper.pdf

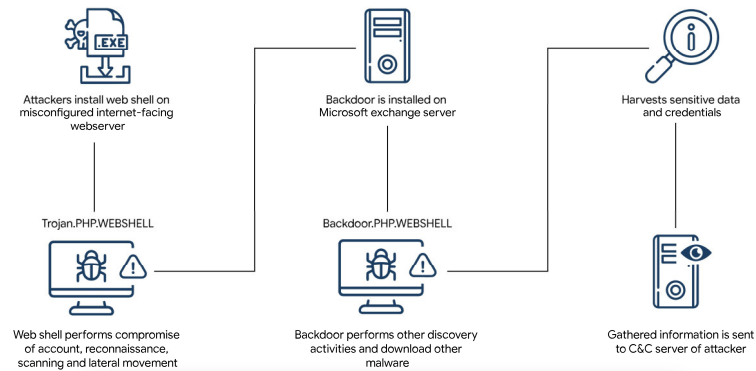

Figure 1.5: China Chopper webshell attack stages

Basically, a webshell attack is divided into four stages, as shown in the Fig. 1.6: Finding and Exploiting Vulnerabilities, Persistent Remote Access, Privilege Escalation, Pivoting, and Launching Attacks.

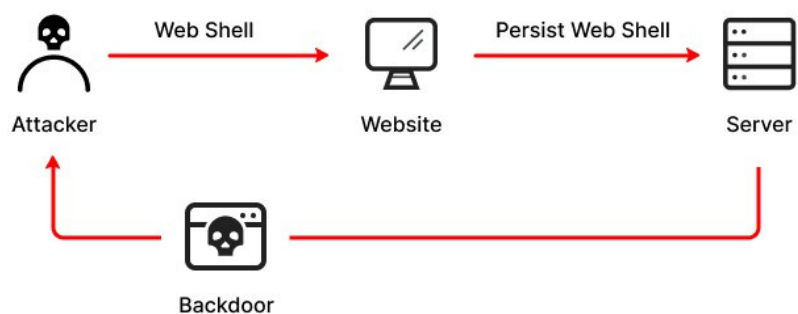

Figure 1.6: Four stages of webshell attack

1. **Finding and Exploiting Vulnerabilities:** Hackers gather information about their target by using various tools and techniques. This includes identifying open ports, services running on those ports, and potential weaknesses in those services in the target's infrastructure. Once hackers identify potential weaknesses, they assess them to determine if they are actual vulnerabilities. If a vulnerability is confirmed, the hacker proceeds to exploit it.

2. **Persistent Remote Access:** Hackers inject webshells into web servers by exploiting vulnerabilities like file upload flaws, remote and local file inclusion, SQL injection, and command injection. They also take advantage of misconfigured permissions, outdated software, and insecure APIs to place malicious scripts on the server Webshell scripts provide a backdoor allowing attackers to remotely access an exposed server. Persistent attackers do not have to exploit a new vulnerability for each malicious activity. Some attackers even fix the vulnerability they exploit to prevent others from doing the same and avoid detection. Some webshells use techniques such as password authentication to ensure that only specific attackers can access them. Typically, webshells conceal themselves with code, preventing search engines from blacklisting the website hosting the shell.

3. **Privilege Escalation:** Webshells normally run with user permissions, which can be limited. Attackers can escalate privileges through webshells by exploiting system vulnerabilities to acquire root privileges. Root account access allows attackers to perform almost any action—they can install software, change permissions, add or remove users, read emails, steal passwords, etc.

4. **Pivoting and Launching Attacks:** Attackers can use webshells to pivot to additional targets, both in and out of the network. Sniffing network traffic to identify live hosts, firewalls, or routers (enumeration) can be a time-consuming process for attackers, taking weeks to complete. An attacker who successfully persists on a network will move patiently, possibly even using a compromised system to attack other targets. This allows the attacker to remain anonymous, and pivoting through several systems can make it virtually impossible to trace attacks to the source. An attacker can use webshells to connect servers to a botnet, a network of systems under their control. The affected servers execute commands sent by attackers through a command-and-control server connected to the webshell. This is a common technique for DDoS attacks that require extensive bandwidth. Attackers are not directly targeting the system where they installed the webshell, but rather exploiting its resources to attack more valuable targets.

Many web application programming languages implement functions such as exec(), eval(), system(), and os(), or process strings as syntax with special characters (such as "`", or backtick, in the case of PHP) that can be used to execute system commands. In cyberattacks, threat groups abuse this functionality by smuggling these default functions and commands via webshells, allowing for remote tasking and code execution. The scope of code execution are arbitrary, limited only by the capabilities of the underlying victim server operating system shell. The following are some common post-installation reconnaissance commands that attackers initially use:

- whoami

- netstat

- ip route or route print

- ls –latr or dir

- uname –a or systeminfo

- ifconfig or ipconfig

This set of commands allows the attackers to get their bearings within the victim system and understand what kind of privileges are available from the perspective of the compromised server. Additionally, attackers gain the ability to discover what applications and data reside on the local file system and perform additional reconnaissance to determine their next action in relation to escalating access or moving laterally to another host.

To enable webshell functionality, attackers may choose to upload new files to the compromised web servers, or they may add webshell functionality and code to an already-existing resource on the server. The attacker opts for this action to prevent any potential suspicion in the event of monitoring file creation events.

It gets even more complicated when an attacker finds a web application parameter that is already being used as input in one of these risky default functions (like a web form or an interactive application). This lets the attacker use webshell without having to upload a backdoor to the victim server. The downside of this approach is that it allows remote tasking input and output to flow across the network without any obfuscation, potentially leading to detection by monitoring services. However, we would briefly use this capability to transition remote access to a more covert method.

Webshell behavior is highly dependent on the configuration of the compromised web service. Rather than opening a new network service, like a traditional bind implant (which would be relatively simple to detect and alert on), webshells most often use the preexisting HTTP (S) service already hosted on the victim system to facilitate backdoor access. For instance, if the web service operates on HTTP 80/TCP, the webshell will also be reachable through the same protocol. However, if the web service is hosted on HTTPS 443/TCP, the webshell will also use 443/TCP and inherit any existing SSL/TLS configuration, including using the legitimate victim web application SSL/TLS certificate and all associated metadata for connections flowing to the webshell. This is one of the reasons why webshells, compared to other types of implants, have the potential to remain undetected for longer periods. Simply put, they become lost in the daily HTTP noise.

Threat actors commonly chain together obfuscation techniques to conceal the true functionality of the webshell and avoid detection. These techniques are often used in combination and include, but are not limited to:

- String rotations
- Array segmentation
- Hex encoding
- Base64 encoding
- Compression
- Whitespace removal

Many webshells found in the wild also encrypt remote command input and output with hard-coded pre-shared keys. While code obfuscation or encryption isn't a new concept in cyber attacks, it introduces an additional layer of challenge when it comes to detecting and investigating webshell implants.

#### 1.1.1.2  Webshell Classification

Webshells can have many different classifications based on characteristics, scripting languages, capabilities, etc [51, 61, 66]. The following is the most common classification based on programming language:

1. **PHP webshells:** These are written in PHP, a most widely used scripting language for web development, according to statistics from W3Techs [9]. When accessed through a web browser, PHP webshells [32] embedded within PHP files execute on the server.

2. **ASP/ASPX webshells:** Written in Active Server Pages (ASP) or ASP.NET (the second most popular server-side programming language after PHP), These webshells target Windows servers running Microsoft technologies.

3. **Others Server-side Programming Language webshells:** In addition to PHP and ASP.NET, there are some webshells written in other server-side programming languages, such as Java Server Pages (JSP) webshells [16], which are deployed on servers running Java technologies; Perl webshells, which can be executed on servers that support Perl scripting; Python and Ruby webshells, which are less common than other types but still pose a threat; and so on.

4. **Shell Script webshells:** These webshells are based on various shell scripting languages, like Bash or PowerShell, and are used to execute system commands on the server.

Besides, another classification method is based on the way the webshell communicates with the hacker's control computer as shown in the Fig. 1.7.

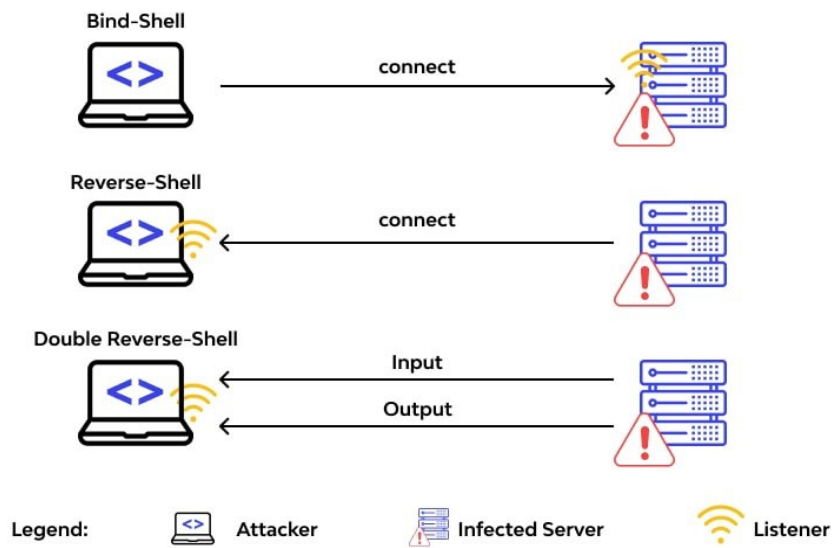

Figure 1.7: Webshell classification based on communication

---

[9] *https : //w3techs.com/technologies/overview/programming_language*

1. **Bind Shell:** is a type of shell that listens to a specific port and waits for incoming connections. The bind shell establishes a connection and provides a shell interface to the remote machine, enabling the user to execute commands on the target system. Legitimate purposes, like remote administration, commonly utilize this type of shell. The user can connect to the target machine from a remote location and perform tasks, just as if they were sitting at the console. System administrators often use bind shells to remotely manage servers, networked devices, and other systems. By connecting to the bind shell, the administrator can access the target system's shell and perform tasks such as monitoring system performance, updating software, and managing configurations. You can set up bind shells using a variety of methods, including network protocols like Telnet or SSH. To set up a bind shell, the administrator must specify the port to listen on and configure the firewall to allow incoming connections. After setting up the bind shell, the administrator can connect to the target system remotely and carry out tasks as if they were physically present at the console. While bind shells can be useful for remote administration, they also present a security risk. An attacker could gain unauthorized access to the target system if you fail to properly secure the binding shell. To prevent this, it is important to secure the bind shell with strong authentication methods, such as passwords or public key authentication, and to limit access to the bind shell to trusted users.

2. **Reverse Shell:** is a type of shell that creates a connection between a remote machine and a target machine. Once the attacker establishes the connection, they gain control over the target system, enabling them to execute commands, run scripts, and carry out various tasks. This allows the attacker to take control of the target system without the need for physical access. Typically, attackers combine reverse shells with other attack types like malware infections or vulnerabilities in web applications. The attacker may use a reverse shell to gain access to sensitive information, such as passwords and confidential data, or perform actions that compromise the security of the target system, such as installing malware or modifying system configurations. To prevent reverse shell attacks, it is important to implement secure network practices, such as firewalls, intrusion detection systems, and anti-malware solutions. Additionally, it is important to keep software and systems up-to-date with the latest security patches and to educate users about the dangers of reverse shell attacks. With the proper precautions in place, it is possible to reduce the risk of reverse shell attacks and protect sensitive information and systems from unauthorized access.

3. **Double Reverse Shell:** is a reverse shell, which separates the standard input and output channels. So two connections to the same port are created to the attacker's computer. Multiple layers of connection can help evade detection by security systems. Even if the first reverse shell is detected and terminated, the second one might still provide a backdoor into the system. It can help in bypassing network restrictions and firewall rules more effectively

Each type of webshell has its own features and requires specific detection and mitigation strategies. Some webshells consist of *multiple files* or components, making them harder to detect. They might include various scripting languages and files for different purposes. Some hackers use *fileless webshells* to carry out attacks. These are a type of webshell that operates without leaving any traceable files on the compromised system's disk. Unlike traditional webshells, which involve uploading a script or code to a server and saving it as a file, fileless webshells execute directly in memory or use existing system tools and processes to achieve their objectives. Organizations should be aware of the various types of webshells and take steps to protect their web servers against these threats.

#### 1.1.1.3 Webshell Evasion

Hackers employ various techniques to evade detection and enable webshells to bypass security defenses [12, 78, 100, 6, 59, 92, 102]. These evasion tactics manipulate the characteristics of the webshell code, communication channels, and execution environment to avoid detection by security systems.

One common tactic is obfuscation [43, 21] of the webshell payload source code through methods like encryption, encoding, and polymorphism. Encryption hides the true form of the code until runtime decryption. To transform the byte stream, encoding uses schemes like base64. Polymorphic webshells dynamically mutate on each request while preserving core functionality. These obfuscation tactics allow webshells to avoid signature matching against known patterns.

Another evasion strategy manipulates communication to disguise webshell traffic as legitimate. Tactics include realistic impersonation of browser user agents, mimicking common web parameters, blending timing and request distributions, and other forms of traffic shaping. HTTPS encryption also hides malicious payloads. Webshells may also exploit trust in authenticated paths to bypass firewall rules. Slow traffic rates avoid flooding detection. Stealthy launch points, such as compromised servers, client pivoting, and anonymous VPNs, mask the attack origin.

On the server side, attackers exploit environment attributes to hide webshells. Naming files with extensions identical to those of legitimate scripts evades behavioral detection. Targeting interpreted languages like PHP with loose runtime coupling enables masking malicious logic within application code. Writing webshells in native languages enhances stealth.

Advanced webshells employ a variety of techniques to evade debugging or runtime analysis attempts aimed at security inspection. To avoid inspection, Webshells remain passive, disable core functions, or self-delete when they detect analysis. These antianalysis tactics complicate efforts to study and detect sophisticated webshells.

In summary, webshell authors leverage numerous techniques across the cyberkill chain to evade defenses. Robust multi-layer detection combining network, host, authentication, and code analysis is required to counter evasion threats. Models adaptive to zero days, such as behavior profiling and deep learning, provide needed resilience against obfuscation. Webshell detection must evolve as rapidly as attackers' evasion tactics.

The Fig. 1.8 is example of Behinder obfuscated webshell template. Behinder webshell accepts attacker input from HTTP POST requests. The Behinder client shapes the attacker's input into a valid class using the target web server's syntax, in this case PHP.

To recover attacker instructions from network traffic, the hardcoded pre-shared key from the webshell script must be recovered. In this case, the default AES key supplied by the source code is *"e45e329feb5d925b"* (first 16 characters of the MD5 hash of the "rebeyond" string). Before using the AES encryption key, the contents undergo base64 encoding, necessitating the decoding of the string.

Deobfuscating the string as shown in the Fig. 1.10, reveals the arbitrary instructions passed to the server as a PHP class. Operator instructions for the webshell are encoded inside the *$cmd* parameter:

Before evaluation, the cmd parameter's value undergoes base64 decoding. In our example, the command *"Y2QgL3Zhci93d3cvaHRtbC87d2hvYW1p"* decodes to *cd /var/www/html/;whoami*:

Figure 1.8: Behinder webshell sample

While obfuscation techniques can mask the contents of a script, in cases where TLS is not being used, the query responses from the server will be displayed in plain text via the web logs and PCAPs. To remain stealthy under these conditions, attackers opt to also encrypt their webshell responses using the same hardcoded pre-shared key. Successfully deobfuscating the script explains what it is capable of. However, by obtaining the pre-shared key, one can further comprehend the input and output generated from a compromised asset. Analysts can leverage this information when they generate packet capture or HTTP application content logs of the event.

We can broadly classify evasion techniques into four main categories:

1. **Obfuscation** webshells refer to the practice of deliberately making a webshell's code more complex, convoluted, and difficult to read or understand. Obfuscation's primary goal is to hinder the detection and analysis of malicious code by security tools and human analysts. Hackers often obfuscate webshell code by adding random characters, whitespace, and "dead" code segments, renaming variables and functions to non-descriptive names, or encoding it using various methods (e.g., base64 encoding). Furthermore, they can modify the flow of control in the webshell code by adding conditional branches, loops, or jumps that are not necessary for the program's actual functionality. An encryption algorithm, such as AES, RSA, DES, or another method, also encrypts the malicious code within the webshell. The chosen encryption algorithm and key encrypt the webshell's payload, including its core functionality and commands. This transforms the original code into ciphertext, which appears as random data. The webshell embeds a decryption function. When the webshell executes, this function incorporates the necessary logic and key(s) to decrypt the encrypted payload. Upon access or execution, the webshell dynamically decrypts its payload in memory. This implies that the webshell never stores the encrypted code in its decrypted form on disk, posing a challenge for detection through file-based analysis.

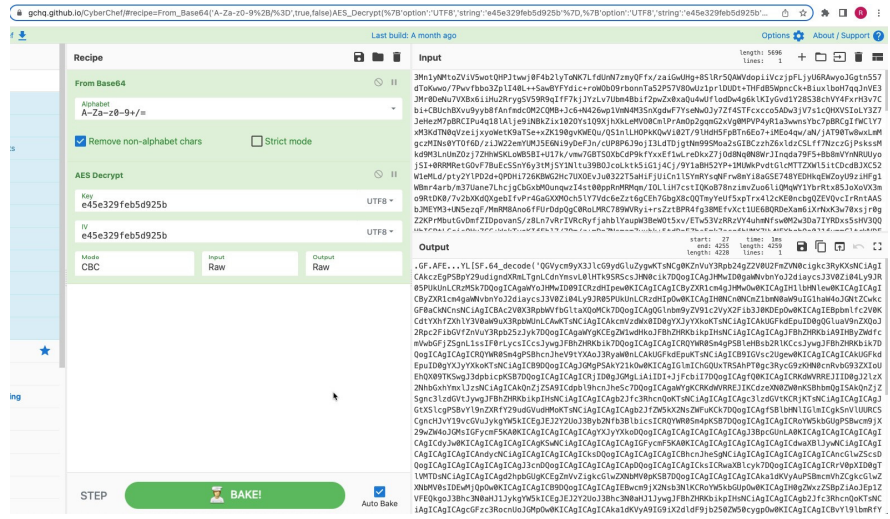

Figure 1.9: Decoding and decrypting the obfuscated string

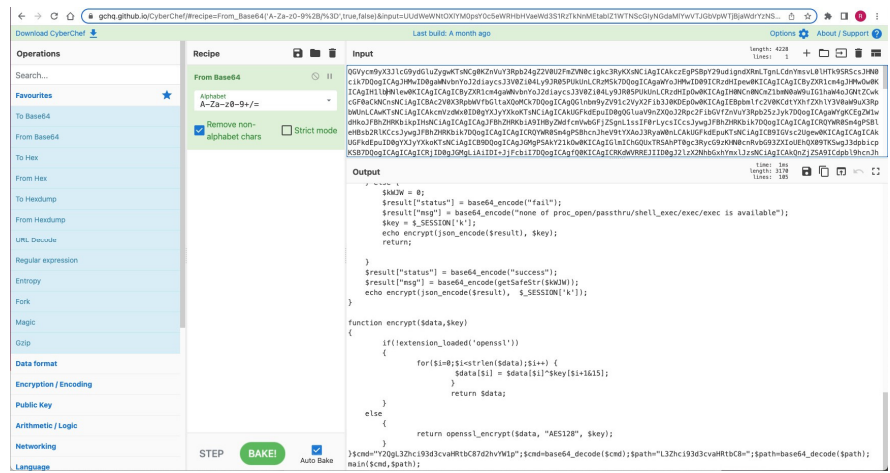

Figure 1.10: Contents of the deobfuscated function

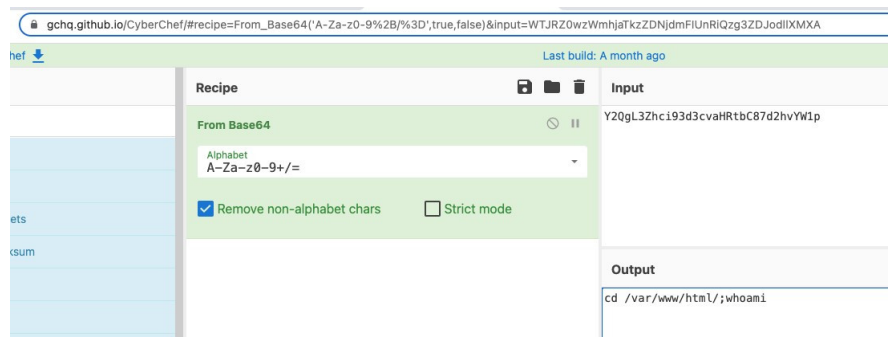

Figure 1.11: Decoded system command

2. **Manipulate Communication Protocols:** In addition to obfuscating payloads, webshells often manipulate communication protocols to masquerade as legitimate web traffic. Mimicking expected patterns allows webshells to bypass network monitoring and filtering defenses. Spoofing the characteristics of browser connections is a common tactic. To impersonate browsers, Webshell communications disguise their user-agent string. Accept headers, charset encoding, and other browser quirks are copied. As a result, webshell traffic blends in with normal user web activity. Webshells also mimic common webpage parameter names, values, and content types to avoid raising suspicion. For example, injecting commands into standard $_GET and $_POST variables bypasses filters looking for anomalies. Traffic shaping techniques are also common. Webshell communications are throttled to match expected human speeds and distributions, rather than robotic rapid connections. Randomised delays, burst timings, and jitter are used to avoid pattern detection. Protocol-level manipulations, such as HTTP request smuggling or header splitting, are also used to bypass firewall rules and evade WAF inspection. HTTPS encryption blinds sensors to malicious payloads.

3. **Environmental Exploitation:** Webshells often exploit weaknesses in the target environment to hide their presence and activities. Knowledge of security gaps and platform intricacies allows attackers to covertly blend in. A common tactic is to disguise webshells by using file names and extensions that are identical to legitimate scripts in the web root. During directory scans, for example, naming a PHP webshell as wp-load.php masks the payload as a WordPress bootstrap script. Webshell authors also exploit opaqueness in interpreted web languages like PHP that lack strong typing and runtime coupling. One can directly inject malicious commands into loosely structured code and seamlessly integrate them during dynamic execution. Platform architectural quirks are also leveraged. Webshells abuse normal operations, such as register variable writing in PHP, to force the execution of stealthy shell commands with no visible code hints. Attackers continue to hide webshells using compromised but trusted platforms and routes. Servers with prior malware implantation allow injected webshells to bypass suspicious-path filters. Client device pivoting and anonymous VPNs mask the origin. Slow traffic rates help to avoid flooding, which triggers detection rules. Distribution across multiple servers and accounts makes blacklisting difficult. Targeted execution timing during peak traffic periods increases stealth.

4. **Anti Analysis or Debugging:** Sophisticated webshells often incorporate antianalysis countermeasures designed to detect inspection environments and evade detection. These techniques target security researchers attempting to analyze webshell behavior through static and dynamic techniques. Webshells commonly employ the tactic of fingerprinting the runtime environment characteristics to detect sandboxes and debugging sessions. Webshells check properties like hardware configurations, installed software versions, process runtimes, and file metadata to detect simulated or manipulated conditions. Timing side channel checks are also common. Webshells measure code block durations and compare them against known benign thresholds. Low-level anti-debugging techniques are also employed to identify debuggers. Webshells check for invalid instruction exceptions, memory breakpoints, observation hooks, and other artifacts introduced by debuggers. Debugging and monitoring tools inherently impact system state. In order to detect manipulation, webshells test CPU register contents, stack contents, network sockets, and other attributes for expected values. Webshells can evade an inspection environment by disabling malicious functionality, deleting traces, refusing to execute, or taking other actions. Obfuscation, combined with anti-analysis, maximizes evasion.

These evasion techniques pose significant challenges for security professionals. They make it more difficult for security tools such as intrusion detection systems (IDS), web application firewalls (WAF), and antivirus software, as well as human analysts, to identify and mitigate webshells. Evasion webshells are more likely to remain undetected for longer periods, allowing attackers to maintain access to compromised servers. To defend against webshell evasion, organizations need a multi-layered security approach that includes proactive monitoring, behavioral analysis, security audits, and the use of security tools capable of identifying anomalies and suspicious behaviors in web server traffic and activities.

### 1.1.2 Webshell Feature

Webshells closely resemble benign webshells, which complicates their differentiation. Previous studies have employed three types of metadata and five sets of features to distinguish malicious webshells. Three distinct types of metadata are commonly associated with webshells, each providing unique insights into their characteristics and behaviors: source code, instruction sequence, and HTTP requests.

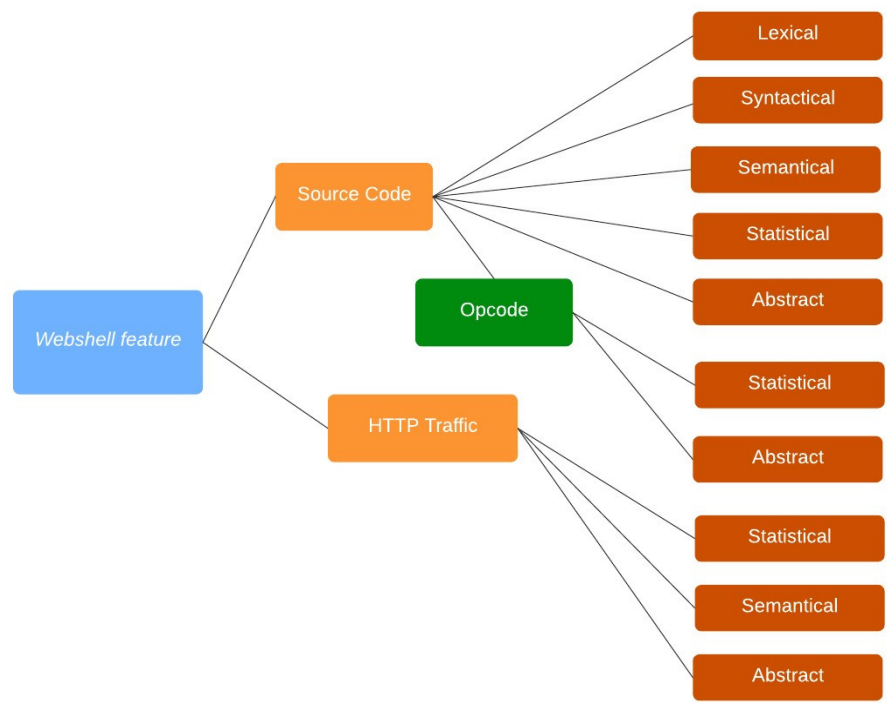

Figure 1.12: Classification of webshell features

- **Source code** represents the textual programming statements that can be compiled to achieve functional goals on web servers. It offers syntactical, statistical, and semantic features crucial for detecting malicious webshells. Various studies utilize source codes as metadata. Nevertheless, source code data may contain noise due to attackers adding meaningless or encrypted codes, and it lacks the reflection of dynamic behaviors during webshell executions.

- **Opcode** refers to the sequence of machine-level instructions or commands that the webshell executes when it executes on a web server. Structured instruction data is convenient for any analyzer. Analyzing such data helps to understand a webshell's dynamic behaviors and real goals. The authors in [30, 73] showed that instruction-based detections can achieve a higher accuracy than source code-based ones because sensitive function calls can be found more in the instruction sequence than in the source codes of a malicious webshell. Therefore, instruction data have recently attracted wide attention from researchers.

- **HTTP Traffic** are textual messages sent by clients to trigger actions on web servers, offering statistical and semantic features useful for identifying malicious requests associated with webshell activities. Analyzing HTTP traffic for webshell detection involves scrutinizing the communication between clients and web servers to identify patterns, anomalies, or signatures indicative of webshell activity. This analysis typically involves capturing and inspecting HTTP requests and responses exchanged between clients and servers, focusing on various attributes such as request methods, URIs, headers, parameters, payloads, and response codes. However, extensive HTTP traffic may encrypt and bury malicious HTTP requests, similar to textual source codes, resulting in a relatively low number of studies addressing this aspect.

We organize the features extracted from Webshell into five distinct classes based on their characteristics: lexical, syntactical, semantical, statistical, and abstract features.

- **Lexical feature:** These features pertain to the script's textual structure. Attackers often employ specific keywords and obscure instructions and parameters within the code to conceal their true intentions. As a result, such scripts exhibit a limited number of strings and specific patterns within tags or comments. Lexical features encompass factors like the count of strings and the presence of malicious text patterns in comments.

- **Syntactical feature:** Syntactical features concern the expressions, variables, and functions used within the scripts. Attackers may use system calls to gain elevated privileges or employ hazardous functions for uploading malware or retrieving critical files. Attackers can adapt the webshell to the target server platform by using conditional statements, and they can expose passwords by using loops. Common examples of syntactical features include the proportion of conditional statements and loops, the invocation of risky functions, and the usage of specific language components.

- **Semantic feature:** While lexical and syntactical features provide insights into how webshells are structured, semantical features derive the intentions behind the code from these lexical and syntactical elements. For instance, instead of measuring the abundance of arbitrary loops within scripts, semantical features might focus on the presence of loops associated with port scanning function calls or login access.

- **Statistical feature:** Malicious webshells often employ encrypted and obfuscated code to evade firewall detection. However, benign webshells may also use encryption for security purposes. As a result, statistical features are critical in distinguishing malicious webshells by comparing their statistical values to those of normal files. Examples of such statistical features include measures like information entropy and compression ratios.

- **Abstract feature:** Abstract features transcend the realms of lexical, syntactical, and semantical analysis. They are particularly useful in revealing concealed elements of webshells that may remain undetected through syntactical and semantic scrutiny. In the context of this study, abstract features encompass vectorized data, such as source code, opcodes, and web traffic. Deep learning-based approaches predominantly harness these features.

Based on the correlation between the metadata and the features as shown in the Fig. 1.12, two main webshell detection approaches are proposed: **Source Code Analysis Approach** involves analyzing the source code or opcode of a web application without executing it, and the **Network-based Analysis Approach** works with network traffic to scrutinize the communication between clients and web servers.

It is important to note that no single approach can guarantee 100% detection of webshells, and a combination of them may be necessary to effectively detect and prevent webshell attacks. Regular monitoring and auditing of web applications and servers are also crucial to detecting and preventing webshell attacks.

### 1.1.3     Operation Code (OpCode)

**Opcodes** [10], an abbreviation for Operation Code, represents the core instructions processed by a computer's central processing unit (CPU). It serves as a pivotal link between the software and hardware layers of a computing system, encapsulating the basic operations that the CPU can perform. Opcodes typically correspond to elementary tasks such as arithmetic, logical operations, data movement, and control flow alterations. These operations form the foundation upon which complex computational tasks are built, making the opcode an indispensable component of modern computing.

The structure of an opcode typically consists of two main components: the opcode itself, which specifies the operation to be performed, and additional operands or parameters that provide context or data for the operation. The length and format of opcodes can vary significantly depending on the CPU architecture and instruction set design. For example, some CPUs use fixed-length opcodes, where each instruction occupies a predetermined number of bits, while others employ variable-length opcodes to accommodate a wider range of instructions and address larger memory spaces. According to the statistical results of Bragen and Simen Rune [14], who disassembled 67 malware samples and 20 non-malicious ("goodware") samples, and extracted the frequency of different x86 opcodes in each sample, they found that the opcode distributions differed significantly between malware and benign samples. Specifically, around 1/3 of the common opcodes occurred more frequently in malware, 1/3 occurred less frequently, and 1/3 were similar to goodware. For rare opcodes (frequency < 0.2%), around 70% occurred with similar frequency in malware and goodware, 30% occurred more frequently in malware, and 10% occurred less frequently. Statistical tests revealed that the association between rare opcodes and malware class was stronger than that for common opcodes, explaining 12-63% of the frequency variation for rare opcodes versus 5-15% for common opcodes. The authors suggest that opcode frequency analysis could potentially provide a faster way to detect malware, complementing other malware detection techniques like signatures and heuristics. They discuss potential improvements, such as looking at opcode sequences rather than just individual opcodes, compiler-specific opcodes, and obfuscation/packing analysis. The Table 1.1 shows the list of the 15 most used opcodes by malicious files.

Within the context of web servers, opcodes, which are the fundamental building blocks of instruction execution within web servers, play a pivotal role in facilitating the

---

[10] https://en.wikipedia.org/wiki/Opcode

identification and mitigation of web shell activities. Web shells typically consist of obfuscated or disguised code fragments that blend into legitimate web application files, making them challenging to detect using traditional signature-based approaches. Security analysts, on the other hand, can develop heuristic algorithms and machine learning models to distinguish between benign and malicious code behaviors by analyzing the opcode patterns generated by web shell scripts during execution. Opcodes Table 1.1: Top 15 opcodes used exclusively used by malware

| Opcode | Description |
|--------|-------------|
| stosq | Store String |
| syscall | Fast System Call |
| setno | Set Byte on Condition - not overflow (OF=0) |
| cvtsd2si | Convert Scalar Double-FP Value to DW Integer |
| movmskpd | Extract Packed Double-FP Sign Mask |
| prefetcht1 | Prefetch Data Into Caches |
| fprem | Partial Remainder (for,compatibility with i8087 and i287) |
| cmpsq | Compare String Operands |
| lodsq | Load String |
| scasq | Scan String |
| cvtss2si | Convert Scalar Single-FP Value to DW Integer |
| fnsave | Store x87 FPU State |
| orpd | Bitwise Logical OR of Double-FP Values |
| fxsave | Save x87 FPU, MMX, XMM, and,MXCSR State |
| movmskps | Extract Packed Single-FP Sign Mask |

serve as an intermediary representation of the executed code, allowing security tools to monitor and analyze the runtime execution flow for anomalous or suspicious activities that indicate web shell presence. Security tools employ techniques like opcode sequence analysis, control flow graph traversal, and opcode frequency profiling to characterize the behavioral traits of web shells and develop effective detection strategies.

Opcode sequence analysis entails examining the sequential patterns of opcodes generated during web application code execution and identifying characteristic sequences commonly associated with web shell activities, such as file system manipulation, remote command execution, or data exfiltration. Control flow graph traversal techniques look at how the control flow depends on other opcode instructions in the web application code. They look for oddities or changes from normal control flow patterns that could mean someone is trying to inject or exploit the web shell. Opcode frequency profiling aims to identify outlier opcodes or opcode combinations that occur more frequently in web shell scripts than benign code, leveraging statistical analysis and machine learning algorithms to discern malicious intent. Through techniques such as n-gram modeling, Markov chains, and hidden Markov models (HMMs), security practitioners can capture the inherent structure and dynamics of opcode sequences, discerning recurring motifs and discernible patterns characteristic of web shell behavior. Moreover, machine learning algorithms, including recurrent neural networks (RNNs), long short-term memory (LSTM) networks, and convolutional neural networks (CNNs), offer advanced capabilities for learning and recognizing complex opcode sequence patterns, enabling the development of robust detection models.

*In conclusion, opcode sequence analysis represents a sophisticated approach to detecting web shells by leveraging the intrinsic behavioral characteristics encoded within opcode sequences. Advanced analytical techniques and machine learning algorithms are used by security professionals to find small problems that point to web shell activity. This makes web servers safer by lowering the risk of hacking, data breaches, and other cyber threats.*

### 1.1.4 Yara

Yara, a powerful and versatile pattern-matching tool, has emerged as a cornerstone in the fields of malware research, threat detection, and incident response. Developed by Victor Alvarez of VirusTotal in 2007, Yara provides security analysts with a flexible and expressive language for creating custom rules to identify and classify malicious software based on behavioral patterns, code structures, and other artifacts. At its core, Yara operates on the principle of signature-based detection, where analysts define rulesets consisting of strings, byte sequences, and logical conditions that characterize the unique features and behaviors of specific malware families or threat actors. These rulesets, written in a human-readable syntax, allow analysts to encapsulate their knowledge of malware characteristics and detection techniques into reusable and extensible patterns, enabling the automated identification of known threats across diverse environments and datasets.

One of the key strengths of Yara lies in its support for advanced pattern-matching capabilities, including regular expressions, wildcards, and Boolean operators, which empower analysts to craft precise and granular detection rules tailored to the nuances of targeted malware variants. By leveraging these expressive features, security practitioners can construct rules that encompass not only code signatures but also behavioral indicators, such as API call sequences, file system interactions, and network traffic patterns, thereby enhancing the resilience and efficacy of their detection strategies against polymorphic and obfuscated malware specimens. Furthermore, Yara's extensible architecture and integration capabilities enable seamless collaboration and interoperability with other security tools and platforms, facilitating the sharing of detection rules, threat intelligence, and analysis findings within the cybersecurity community.

Yara's versatile pattern-matching capabilities and flexible rule syntax empower security analysts to develop custom detection rulesets tailored to the distinct characteristics and behaviors exhibited by web shell instances. Leveraging Yara, analysts can define rules that encapsulate unique strings, byte sequences, and behavioral patterns associated with known web shell variants, enabling the automated detection of these malicious artifacts across diverse web server environments and configurations.

One of the primary strengths of Yara in web shell detection lies in its ability to incorporate both static and dynamic indicators of compromise (IOCs) into detection rules, thereby enhancing the resilience and accuracy of detection mechanisms against polymorphic and obfuscated web shell payloads. Yara rules can encompass not only code signatures and string patterns commonly found in web shell scripts but also behavioral indicators, such as file system interactions, network traffic patterns, and command execution sequences, which are indicative of malicious activity. By combining these diverse indicators within a single detection rule, analysts can construct comprehensive and adaptive detection strategies capable of identifying web shell instances across different stages of the cyber kill chain, from initial infection to post-exploitation activities. Furthermore, Yara's extensible architecture and integration capabilities enable seamless collaboration and information sharing among security teams, enabling the rapid dissemination of web shell detection rules, threat intelligence, and analysis findings within the cybersecurity community.

It's simple to write and understand Yara rules, and their structure is a lot like the C language. Each rule in Yara begins with a keyword rule and a rule number come at the beginning of each rule in Yara. Identifiers must follow the same rules for words as the C computer language. They can have any letter, number, or underscore, but the first character can't be a number. Case matters when it comes to rule names, and they can't be longer than 128 characters.

Rules usually have two parts: the string description and the condition. The condition section is always needed, but the string description section can be left out if the rule doesn't depend on any strings. It is in the string definition area that the strings that will be used in the rule are set up. Each string has an identifier that is made up of a $ character followed by a string of letters, numbers, and underscores. These identifiers can be used in the condition part to find the string that goes with them.

Like in the C programming language, text strings are wrapped in double quotes. Hex strings are made up of a series of hexadecimal numbers that can appear next to each other or spaced out. They are surrounded by curly brackets. Hex strings can't have decimal numbers in them.

The logic of the rule is in the part called "condition." This part needs to have a boolean statement that says when a file or process follows the rule and when it doesn't. Most of the time, the condition will use the strings' identifiers to link to strings that have already been defined. The string number works like a boolean variable in this case; it returns true if the string was found in the file or process memory and false otherwise. Below is an example of a Yara rule to detect one of the most used webshell families today, which is B374k.

```
rule webshell_B374kPHP_B374k {
    meta:
        description = "Web Shell - file B374k.php"
        author = " Florian Roth" date = "
        2014/01/28 " score = 70
        hash = " bed7388976f8f1d90422e8795dff1ea6 "
    strings:
        $s0 = " Http://code . google .com/p/b374k-shell " fullword
        $s1 = "$_=str_rot13 ( 'tm'. ' vas '. ' yngr ') ;$_ = str_rot13 ( strrev
            ( 'rqb '. ' prq '. '_'. '46 r '. '
            fno ' "
        $s3 = " Jayalah Indonesiaku & Lyke @ 2013 " fullword
        $s4 = "B374k Vip In Beautify Just For Self " fullword
```

```
        condition:
                1 of them
}
```

*In conclusion, Yara stands as a versatile and indispensable tool in the arsenal of cybersecurity professionals for detecting, analyzing, and responding to web shell threats. Its robust pattern-matching capabilities, flexible rule syntax, and broad applicability make it a cornerstone technology in safeguarding web server environments against unauthorized access, data breaches, and other malicious activities perpetrated through web shells. Through its integration into security workflows and collaboration platforms, Yara enables security teams to stay ahead of evolving web shell threats, protect critical web-based assets, and uphold the security and trustworthiness of online services and applications.*

### 1.1.5 Model Evaluation

Model evaluation is a crucial step in the process of building an ML/DL model, ensuring that the developed models perform effectively and reliably on unseen data. It involves assessing the performance of a trained model using various metrics and techniques to understand its strengths, weaknesses, and generalization capabilities. Webshell detection is a binary classification problem that involves distinguishing between two classes of webshell and benign code. For classification models, some commonly used metrics include: **The confusion matrix** is a fundamental tool in the evaluation of classification models, including those used in webshell detection. It provides a tabular representation of the performance of a classification model by summarizing the counts of true positive (TP), true negative (TN), false positive (FP), and false negative (FN) predictions. Each row of the matrix corresponds to the actual classes, while each column corresponds to the predicted classes. In the context of webshell detection: **True Positive (TP):** Instances are correctly classified as webshells. **True Negative (TN):** Instances correctly classified as benign. **False Positive (FP):** Instances are incorrectly classified as webshells when they are actually benign. Also known as a Type I error. **False Negative (FN):** Instances are incorrectly classified as benign when they are actually webshells. Also known as a Type II error. The confusion matrix provides a concise summary of the model's performance, allowing for the calculation of various evaluation metrics, including:

**Accuracy:** The proportion of correctly classified instances out of the total number of instances.

$$Accuracy = \frac{TP + TN}{TP + FP + FN + TN}$$

**Precision:** The proportion of true positive predictions out of all positive predictions made by the model. It measures the model's ability to avoid false alarms.

$$Precision = \frac{TP}{TP + FP}$$

**Recall (Sensitivity):** The proportion of true positive predictions out of all actual positive instances. It measures the model's ability to capture all positive instances.

$$Recall = \frac{TP}{TP + FN}$$

**F1-Score:** The harmonic mean of precision and recall provides a balanced measure of a model's performance.

$$F1 - score = \frac{2TP}{2TP + FP + FN}$$

**False Positive Rate (FPR):** The proportion of false positive predictions out of all actual negative instances. FPR is calculated as FP / (TN + FP). It is complementary to specificity and measures the model's tendency to generate false alarms.

$$FPR = \frac{FP}{FP + TN}$$

By analyzing the confusion matrix and calculating these evaluation metrics, we can gain insights into the strengths and weaknesses of the classification model, identify areas for improvement, and make informed decisions regarding model selection, hyperparameter tuning, and deployment strategies.

## 1.2 Webshell Detection Approaches

Webshell detection approaches can be broadly categorized into two types: static analysis and dynamic analysis. Each of them leveraging different methodologies to identify malicious webshell activity.

### 1.2.1 Static Analysis

For webshell detection, static analysis of source code involves examining the code without executing it, focusing on identifying patterns, structures, and anomalies indicative of webshell presence. Source code analysis entails scrutinizing the textual programming statements comprising the webshell and leveraging syntactic, semantic, and statistical features to distinguish between benign and malicious code. This analysis may involve identifying specific keywords, functions, or patterns commonly associated with webshells, such as system command executions, file manipulations, or network communications. Additionally, this analysis techniques extract structural information from the source code, including control flow graphs, data flow analysis, and syntax parsing, to identify potential vulnerabilities or suspicious behaviors. Control flow graphs depict the code's execution flow, allowing for the detection of branching or looping structures commonly used in webshells to execute malicious commands. Similarly, data flow analysis traces the flow of data within the code, identifying variables or inputs manipulated by malicious code to perform unauthorized actions.

As a more in-depth form of source code analysis, opcode analysis is involves examining the sequence of machine-level instructions or commands comprising the webshell, offering insights into its behaviors and execution flow. This analysis entails disassembling or decompiling the binary representation of the webshell to extract opcodes, function calls, and control flow structures. By analyzing opcode sequences, researchers can identify characteristic patterns or signatures indicative of webshell activity, such as system calls, file operations, or network interactions. Furthermore, opcode analysis enables the identification of obfuscated or encrypted code segments within the webshell, which may evade traditional source code analysis techniques. However, opcode analysis may be more complex and resource-intensive compared to source code analysis, requiring specialized tools and expertise to interpret and analyze low-level machine instructions effectively.

Beyond exact signature matching, machine learning and deep learning models can analyze code syntax and structure to detect statistical anomalies in code segments that are likely to contain obfuscated webshell logic. In the context of source code analysis, ML/DL algorithms can extract syntactic, semantic, and structural features from the code, such as function calls, variable assignments, and control flow structures. These features serve as inputs to the ML/DL model, which learns to differentiate between benign and malicious code based on learned patterns. For instance, we can train supervised learning algorithms such as support vector machines (SVMs) or neural networks to classify source code snippets as either benign or indicative of webshell activity. In the case of opcode analysis, ML/DL techniques can be applied to analyze the sequence of machine-level instructions comprising the webshell. ML/DL models can learn to recognize patterns in webshell behavior by taking out features like opcode frequencies, control flow patterns, and function calls from opcode sequences. It is possible to find webshells more accurately with deep learning architectures like convolutional neural networks (CNNs) or recurrent neural networks (RNNs). These can effectively capture complex dependencies and patterns within opcode sequences.

In conclusion, the application of ML/DL in source code analysis represents a promising approach to webshell detection, offering the potential for automated, scalable, and accurate detection of malicious code in web environments. Continued research and development in this area are essential to address the challenges and limitations of current techniques, further enhancing the security posture of web systems against webshell threats.

### 1.2.2 Dynamic Analysis

Dynamic analysis involves monitoring and analyzing the behavior of web applications during execution to detect webshells. Unlike static analysis, which inspects the code without running it, dynamic analysis focuses on identifying malicious activities based on the runtime behavior of the code. In dynamic analysis, there are two main objects to monitor and analyze to detect webshell attacks: internal behavior on the webserver and network traffic with the webserver.

Monitoring and analyzing the internal behaviour of a web server, also known as **host-based analysis**, is the practice of tracking system calls, unusual file operations, abnormal network communications, unauthorized data access, and other runtime actions that may indicate a webshell's presence. However, this approach is best suited for detecting malware in general, as it lacks many specific characteristics associated with detecting webshell attacks.

On the other hand, **HTTP traffics analysis** for webshell detection involves scrutinizing the network communication between clients and web servers to identify patterns, anomalies, or signatures indicative of webshell activity. HTTP traffic analysis is an important part of NetIDS [2, 35, 18] and WAFs [23] systems that try to find and stop bad things like webshell deployments, command and control communications, and attempts to steal data. This analysis typically involves capturing and inspecting HTTP requests and responses exchanged between clients and servers, focusing on various attributes such as request methods, URIs, headers, parameters, payloads, and response codes.

Signature-based detection, which matches predefined patterns or signatures representing known webshell characteristics against the observed HTTP traffic, is one technique for detecting webshells through HTTP traffic analysis. These signatures may include unique strings, keywords, or regular expressions commonly associated with webshells, such as commonly used webshell file names, function calls, or encoded payloads. Signature-based detection techniques are effective for identifying known webshells but may struggle to detect polymorphic or obfuscated variants.

Another technique is anomaly-based detection, which flags deviations from normal HTTP traffic patterns as potentially malicious. Anomaly detection algorithms analyze various attributes of HTTP requests and responses, such as request frequency, size, timing, user-agent strings, and HTTP methods, to identify suspicious behavior indicative of webshell activity. For example, sudden spikes in requests to non-standard URIs, abnormal payload sizes, or unusual patterns of HTTP headers may suggest the presence of a webshell.

Machine learning (ML) and deep learning (DL) techniques have emerged as powerful tools for analyzing HTTP traffic to detect webshells, offering the potential to automate the process and improve detection accuracy. These techniques leverage the rich and complex patterns present in HTTP traffic to discern between benign and malicious activities, enabling proactive detection and mitigation of webshell threats.

Typically, ML-based techniques train models on labeled datasets of HTTP traffic, explicitly annotating benign and malicious activities. These models learn to identify subtle patterns, anomalies, or signatures indicative of webshell activity, enabling them to classify incoming HTTP requests and responses in real-time. Features extracted from HTTP traffic, such as request methods, URIs, headers, parameters, payloads, and response codes, serve as inputs to the ML models, allowing them to learn complex relationships and make informed decisions about the presence of webshells. We can apply various ML algorithms, including supervised and unsupervised learning algorithms, to analyze HTTP traffic for webshell detection. While unsupervised learning models identify anomalies or deviations from normal traffic behavior without explicit labeling, supervised learning models train on labeled data to classify HTTP traffic as benign or malicious based on learned patterns.

Deep learning, a subset of ML that utilizes artificial neural networks with multiple layers, has shown promising results in analyzing HTTP traffic for webshell detection. Convolutional neural networks (CNNs), recurrent neural networks (RNNs), and their variants, such as long short-term memory (LSTM) networks and gated recurrent units (GRUs), are commonly employed for this purpose. These deep learning models are good at detecting complex and changing webshell behaviors because they can handle temporal dependencies, sequential patterns, and hierarchical representations found in HTTP traffic.

ML/DL techniques hold tremendous promise for enhancing webshell detection capabilities, offering proactive and scalable solutions for identifying and mitigating webshell threats in cybersecurity contexts. Continued research and development in this area are essential to further improve detection accuracy, reduce false positives, and address emerging challenges in webshell detection.

### 1.2.3 Webshell Dataset Collection

Datasets play a crucial role in the detection of webshells, providing the foundational resource for training and evaluating detection models, understanding the characteristics and behaviors of web shell malware, and developing effective detection algorithms and techniques. Webshell datasets encompass a diverse array of samples. These datasets serve as invaluable assets to gain insights into the tactics, techniques, and procedures (TTPs) employed by threat actors to infiltrate web servers, evade detection mechanisms, and execute malicious operations, thereby identifying common patterns and trends and extracting meaningful features for use in detection models.

Firstly, webshell datasets serve as a valuable resource for the development and evaluation of signature-based detection mechanisms, which rely on predefined patterns and heuristics to identify known web shell artifacts. By curating datasets comprising samples of web shell scripts, encoded payloads, command-and-control (C2) communications, and other related artifacts, researchers can extract characteristic features and patterns that distinguish web shell activities from legitimate web server operations. These features serve as the foundation for defining detection rulesets using signaturebased languages such as Yara, enabling the automated identification of web shell instances based on their unique behavioral signatures and code structures. Moreover, web shell datasets facilitate the validation and refinement of detection rules through real-world testing against live web server environments, enabling researchers to assess the robustness and efficacy of signature-based detection mechanisms in detecting and mitigating web shell threats.

Other primary functions of webshell datasets are to facilitate the training and evaluation of machine learning and statistical models for webshell detection. By providing labeled samples of benign and malicious webshell instances, datasets enable researchers to train ML/DL algorithms to discriminate between legitimate web server activities and malicious behaviors indicative of web shell presence. Moreover, datasets allow for the evaluation of detection models against a diverse range of web shell variants, ensuring their generalizability and effectiveness across different environments and threat landscapes. Additionally, webshell datasets enable researchers to assess the performance of detection algorithms in terms of key metrics such as accuracy, precision, recall, and false positive rate, providing insights into the strengths and limitations of various detection approaches and guiding the refinement of detection strategies.

Table 1.2: Some widely used Webshell datasets

| Dataset | Last Update Time | Number of Sample | Programming Language(s) | Noisy |
|---|---|---|---|---|
| Tennc | 2022 | About 2.000 | PHP, JSP, ASP, ASPX, PY, etc | Numerous |
| JohnTroony | 2020 | 119 | PHP | Few |
| WebSHArk 1.0 | 2015 | 809 | PHP, ASP, ASP.NET, JSP, CFM, PERL, etc | Few |
| Cycle183 PHP-Webshell-Dataset | 2020 | 2.917 | PHP | Few |
| MWF Dataset | 2023 | 1.359 | PHP | Few |

Some previous datasets are available for malicious webshell research, as shown in Table 1.2. Developed in 2013, Tennc [11] is a pioneering dataset of malicious webshells. The most significant aspect of this dataset is its continuous updates and current growth, made possible by the backing of an open-source webshell project. However, because it is challenging to control the quality of the samples uploaded by contributors from around the world, this dataset contains a large number of noisy samples, including mixed benign and duplicate webshell samples. A tiny dataset called JohnTroony [12] was produced in 2014 and at first included 132 malicious webshells. A few additional webshells have been added, duplicate samples have been eliminated, and sample names have been sharpened in the most recent updates to the dataset. Right now, the dataset has 119 malicious webshell examples and is of good quality. Another early malicious webshell dataset was released in 2015, called WebSHArk 1.0 [42]. Rare, noisy samples can be found in this collection. But WebSHArk 1.0 is no longer relevant. Studies on harmful webshells focus on the use of recently discovered samples. Cyclel83 PHP-Webshell-Dataset [13] is a sizable and recently released dataset. The Tennc and JohnTroony PHP webshell datasets are among the twelve prior datasets from which the malicious samples in this dataset were carefully chosen. The authors in [105] introduce a malicious webshell family dataset called MWF. It contains 1,359 PHP malicious webshell samples, grouped into 78 families and 22 outliers. Each sample has metadata about the dynamic function calls captured when executing the webshell in a sandbox. It also provides explicit family labels to enable multi-classification.

---

[11] Tennc, A webshell open source project, https://github.com/tennc/webshell

[12] JohnTroony PHP Webshell, https://github.com/JohnTroony/php-webshells

 The common weakness of these datasets is that they only include popular webshells and do not include new webshells that use obfuscation techniques, particularly those used in APT attacks. Therefore, there is a need for a richer data set to build a model to effectively detect new types of webshells. To solve this problem, we proactively built a webshell dataset collected from many reliable sources to serve as research. There are two main data sources used to build the dataset:

 • Collecting a wide range of webshells from reliable and most-starred sources on Github. However, testing shows that there are still benign files mixed in the webshell dataset, which leads to the noisy-dataset problem. Therefore, these files need to be removed because the model training process will be noisy and inaccurate if there are clean files in the webshell dataset. This data set cleaning process is carried out with support from cybersecurity experts and research institutes to ensure the highest quality.

 • The second source of data is collected from my actual work. Although the number is small, it is extremely important because these are mostly results collected from APT attacks. These samples are all unknown webshells equipped with many complex evasion techniques. Analyzing these samples allows for further understanding of tactics, techniques, and evade mechanisms in practice, thereby improving models

---

[13] Cyc1e183 PHP Webshell Dataset, https://github.com/Cyc1e183/PHP-Webshell-Dataset

to detect new types of webshells more effectively and accurately.

## 1.3 Related Works

Statistics of research related to **Webshell Detection** from reputable sources [14]show that from those 41 studies, 17 of them (42%) adopted machine learning, 12 studies (29%) used deep learning technology, and 12 studies (29%) proposed other kinds of solutions.

### 1.3.1 Non-AI Approaches

Besides the research applying AI algorithms to improve the ability to detect new types of webshells, there are still some studies using other approaches [104, 50, 85, 19], which also have notable points.

The authors in [58] introduced Cubismo as a means to enhance the detection capabilities of existing malware detection tools. Cubismo employs counterfactual execution to explore all possible execution paths, uncovering concealed code within PHP scripts. The process begins with the normalization of original scripts by removing extraneous lines, comments, and white spaces. These scripts are then subjected to counter-factual execution. During this analysis, exceptions, runtime errors, and nested predicates are disregarded to pinpoint new paths. Recursive deobfuscation is applied to tackle multi-layer encryption. Each new explored path and encountered dynamic construct resulted in the creation of new program files. These new programs undergo execution within a sandbox environment for potential detection. The resulting program files serve as inputs for existing malware detection tools, with the identification of a webshell contingent on any input file being flagged as malware.

Inherited from Cubismo, PHPMalScan [59] serves as a malware detection tool, encompassing webshells. PHPMalScan employs counterfactual execution in a sandbox and virtual environment to explore every possible code execution path. Sandboxes may collaborate by sharing the necessary artifacts for their respective executions. PHP functions are categorized as either safe or potentially malicious. Two metrics, the maliciousness score (MS) and the potentially malicious functions ratio (PMFR), are introduced based on the number and intensity of malicious functions. The maliciousness of scripts is determined using thresholds associated with MS and PMFR values.

---

[14] *IEEE Xplore, *ACM Digital Library, *SpringerLink, *Wiley Online Library, *ScienceDirect

A webshell classification tool is proposed by the authors in [88]. The proposed tool uses similarity analysis to detect derivatives of well-known webshells. Following the proposed method, PHP scripts need to be decoded first in order to reveal any deobfuscation layer of code. Second, all user-defined function names and bodies are extracted by analyzing PHP script tokenization. Third, scripts are fuzzy hashed and stored within the source files. Finally, similarity matrices of function names, function bodies, and file hashes are generated in request. Visualization tools such as heatmaps and dendograms are incorporated and used to discover the similarity among samples.

A search software for ASP webshells is proposed by the authors in [56]. The proposed tool has the ability to recognize several features of ASP webshells and report suspicious files to the administrator for further examination. Specifically, the tool recognizes calls to specific ASP components and functions, suspicious statements, and customized encryption functions. The tool is language-dependent and follows a semi-automatic approach for the detection of webshells.

A sandbox-based environment is described by the authors in [89] for the static and dynamic analysis of PHP scripts with the aim of semi-automatically detecting webshells. In the proposed environment, PHP shells are first deobfuscated and normalized. Deobfuscated scripts are statistically analyzed for specificness. Malicious scripts are indexed, saved in a database, and safely executed in a sandbox environment for behavioral and dynamic analysis. Sandbox execution enables reporting calls to exploitable functions (i.e.; command and code execution, information disclosure, and filesystem functions) and where they were called. The proposed environment is found to be failing in the execution of PHP files incorporating other scripts such as JavaScript and CSS. Moreover, the deobfuscation process is restricted to eval() and preg_replace() functions with explicit string arguments, which restricts its detection ability to specific kinds of webshells.

GuruWS, introduced by the authors in [45], is a hybrid platform designed to detect malicious webshells and vulnerabilities within web applications. It operates with two distinct modules: grMalwrScanner for webshell detection and grVulnScanner for identifying web application vulnerabilities. The grMalwrScanner module employs taint analysis for simple PHP scripts to identify risky function calls and their associated arguments. For more complex scripts, it relies on Yara rules to match malicious scripts. Additionally, a statistical-based analysis that ranks files based on five statistical features is optionally provided for users.

## 1.3.2 AI-Powered Source Code Analysis Approaches

Source code analysis methods provide a holistic view of webshell code [39, 109, 63, 84], allowing for the examination of its structure, patterns, and attributes. This comprehensive analysis enables high-accuracy detection of innovative webshells that may employ sophisticated obfuscation techniques or novel attack vectors to evade detection.

The author in [34] proposes a six-layer deep learner model for the detection of multi-language webshells. To prepare the input for the model, they remove all special characters from source codes and only consider tokens, including alphabetic characters or unicode strings. Then they use some vectorization methods, Term FrequencyInverse Document Frequency (TF-IDF), Hash Vectorization (HashVect), One-Hot encoding (One Hot) and Doc2Vec to vectorize the source code into vectors that range in size from 100 to 400. Their experiment results show that the deep learner model could archive 98.27% of accuracy. However, this approach will encounter two major problems. Firstly, the model is not efficient with an obfuscated webshell. Secondly, using too much redundant data in the source code will result in the model requiring a lot of computational resources.

The authors in [86] introduced two types of ensemble learners tailored for safeguarding IoT networks. One is a lightweight ensemble learner based on Random Forest (RF), designed for devices with moderate computing resources. The other is a heavyweight ensemble learner that combines the outputs of six classifiers (including five single learning classifiers: K-Nearest Neighbors, Naive Bayes, Decision Trees, K-Means, and Support Vector Machine, along with a deep learning classifier: MultiLayer Perceptron or MLP), ideal for devices with more robust computing capabilities. Both of these learners are trained using 100-dimensional feature vectors derived from TF-IDF vectorization of PHP opcode sequences. The model archived the best performance of F1-Score at 98.32%.

The authors in [110] propose a method for detecting web shell scripts based on multiview feature fusion. It presents a multiview feature fusion mechanism that extracts lexical features (e.g. global variables used), syntactic features (e.g. proportion of conditional/loop statements), and abstract features (e.g. sensitive function usage) from PHP scripts to effectively represent web shells. It uses the Fisher score to rank the importance of extracted features and determines the best set of 16 features for classification through experiments.The model trains an optimized support vector machine (SVM) classifier using the selected 16 features to distinguish web shells from benign scripts. The method achieves 92.18% overall accuracy and 95.26% detection rate for web shells on a large dataset of 1056 web shells and 1056 benign scripts. It outperforms well-known tools like VirusTotal,

ClamAV, LOKI, CloudWalker, and other state-of-the-art methods. The authors discuss the advantages of their multiview feature extraction and selection approach, as well as potential future work to improve coverage across languages and use deep learning for automatic feature engineering.

The authors in [25] propose a model called FRF-WD for detecting PHP webshells. The model combines two techniques: first, using the fastText algorithm to train a text classifier on the sequences of PHP opcodes (bytecode instructions) extracted from files. Secondly, using a random forest classifier trained on static features like longest string, entropy, and dangerous function signatures, along with the opcode sequence predictions from fastText, The PHP opcode sequences were found to be an effective feature for webshell detection, providing improved performance compared to just using the static features alone in a random forest model. The fastText model works well for classifying the variable-length opcode sequences, with 4-grams providing the best n-gram size. Experimental results on a dataset of 1587 webshells and 6934 benign PHP files showed that the FRF-WD model achieves 99.23% accuracy, 97.65% recall, and 97.92% precision in detecting webshells using 10-fold cross-validation.

The authors in [47] introduced an RNN-GRU model designed to identify malicious webshells. What sets this model apart is its specific focus on capturing associations between words within individual lines of source scripts. Consequently, the model is supplied with vectorized words extracted from each line in webshell sources. Their experiments revealed that this model surpasses existing detectors designed for multi-language webshells, including the matrix decomposition learner proposed in [69]. However, it's worth noting that the model exhibits slightly lower performance when dealing with PHP webshells compared to the RF-GBDT and FRF-WD ensemble learners proposed in [20] and [25], respectively.

The authors in [83] propose two detection algorithms for detecting PHP variable webshells based on information entropy. Firstly, the PHP Special String Information Entropy Detection Algorithm selects special characters like function names, variables, etc. in PHP files as test objects, then calculates the information entropy of these special strings in normal benign files to get a threshold. If the entropy is higher than the threshold, it flags the file as potentially containing a variable WebShell. Secondly, Quotes Information Entropy Detection Algorithm designed to detect nonASCII and digital variable WebShells. It calculates the information entropy of quotes (like single/double quotes) in normal files to get a threshold. If the entropy of the quotes in a file is higher than the threshold, it indicates that the file may contain a nonASCII or digital variable in WebShell. The tests showed that both algorithms were better at finding variable WebShells than traditional methods based on characteristic value matching. They did this with high accuracy and low false positive rates.

The authors of [74] also came up with a way to use thresholds. Their method uses a database that has malicious functions and signatures that are common in webshells, along with the danger scores (high, medium, and low) that go with them. All files on the web server undergo scanning, with a focus on matching malicious functions and signatures from the database. Additionally, the method determines the length of the longest word within header tags for each file and tallies blacklisted keywords in comment lines. An optimal threshold value for each feature is determined through the examination of a set of benign files. Files exceeding the established threshold values for the number of malicious functions, malicious signatures, or string word length are flagged as suspicious and reported to administrators for further scrutiny. Files that don't match their original counterparts are also considered suspicious.

Due to the shortcomings of traditional detection methods, researchers began to look for detection solutions from a new perspective. Applying ML/DL techniques to source code analysis to detect improved webshells brings many benefits. One of the primary benefits lies in the ability of ML and DL models to automatically learn and adapt to complex patterns and features present in source code, enabling the detection of novel and polymorphic webshell variants that may evade traditional signature-based methods. By leveraging sophisticated algorithms and architectures such as convolutional neural networks (CNNs) and recurrent neural networks (RNNs), ML and DL models can capture complex relationships and dependencies within source code, effectively distinguishing between benign and malicious instances with high accuracy and precision. However, the detection speed will not be as fast as the signature-based technique, which requires computational resources to perform.

The authors in [87] propose a method for detecting webshells based on a multiclassifier ensemble model using machine learning techniques. TF-IDF vectorization is applied to the features that are extracted from PHP files, including static characters, grammar tokens, and opcode sequences. Several base classifiers, like random forest, AdaBoost, Gradient Boosting are trained on the feature vectors. An improved classifier based on RF-AdaBoost is proposed that combines random forest and AdaBoost to enhance performance. Then a dynamic ensemble model is proposed using stacking that selects the top performing base classifiers for each feature and combines their predictions using a meta-model. Their experiments on a dataset collected from GitHub showed the proposed method achieved 98.447% accuracy and 99.227% precision. Although, the method can detect unknown webshells without needing source code, it overcomes the limitations of traditional detection approaches that rely heavily on rules and signatures.

The authors in [4] proposed an ensemble learning model consisting of Logistic Regression, Support Vector Machine, Multi-layer Perceptron and Random Forest classifiers to detect webshell files. It uses a weighted voting method to determine the final classification, where the weights are based on the accuracy of each base classifier. These base classifiers were trained and tested on TF-IDF vectorization of opcodes employing a well-defined feature selection algorithm. A feature selection method based on information entropy and Gini coefficient to construct an optimal feature subset that can detect both encrypted and unencrypted WebShell files. The experiments shows that WS-LSMR achieves 99.14% recall rate and 94.28% accuracy, outperforming traditional single classifiers on both encrypted and unencrypted samples.

The authors in [5] propose a deep super learner approach for detecting malicious WebShell files in PHP web applications. The deep super learner is an ensemble that combines logistic regression, multilayer perceptron, and random forest as base classifiers, with weights optimized using a cross-entropy loss function. It combines static features (string length variance, index of coincidence, information entropy, file compression ratio, eigencode matching) and dynamic features (opcode sequences) to construct a comprehensive feature set for WebShell detection. It uses Word2Vec for feature vectorization, a genetic algorithm for feature selection to reduce feature dimensionality, and SMOTE algorithm to oversample the malicious class to handle the data imbalance issue. Experimental results on a dataset of 571 WebShell samples and 5,379 benign PHP files show the proposed method achieves an accuracy of 98.90% in detecting WebShell attacks. The authors in [107] presented a two-step deep learning detection model. Initially, an SRNN with an attention mechanism was used to reduce input sizes. Subsequently, a 4-dimensional capsule neural network (CapsNet) made predictions using fused vectors generated from vectorizing both source and opcodes.

The authors in [61] propose a novel approach for detecting JSP webshells using a machine learning model that combines BERT for word vector extraction from the bytecode of JSP files and XGBoost for classification. The BERT model is applied to extract word vectors from the bytecode, which is proposed for the first time for JSP webshell detection. The BERT-based word embeddings are shown to outperform the traditional Word2Vec embeddings. The XGBoost algorithm is used as the classifier to detect whether the word vectors belong to a benign JSP file or a webshell. Experiments on a dataset of 2,903 JSP samples (2,073 benign and 830 webshells) show that the proposed BERT+XGBoost model achieves 99.14% accuracy, 98.68% precision, 98.03% recall, and 98.35% F1-score.

The authors in [106] propose The Webshell Detector (WSLD), which is a prototype system engineered for the real-time detection and prevention of webshell attacks. It employs a trio of heuristic methods, including fuzzy hash similarity, signature matching, statistical feature-based similarity, and deep learning. These methods rely on a synchronized webshell feature library. Input scripts are first classified into highrisk and suspicious categories using these three heuristic models. If an input script is detected as a webshell by any of these models, it is classified as high-risk; otherwise, it falls under the suspicious category. For the latter, a secondary round of detection using Long Short-Term Memory (LSTM) is conducted, focusing on extracting all system call sequences. Isolated webshells undergo preprocessing by a cloud analysis module to update the feature library. The authors use a dataset with 1050 malicious samples taken from a 638-star project on GitHub, and a total of 1050 normal samples were randomly selected from an open-source CMS. In experiments, the accuracy of WSLD can reach 98.86%, this is a relatively good result compared to other single approaches.

The Table 1.3 shows the comparison of some machine learning models related to webshell detection. Many recent webshell detection studies have used CNN models to enhance the ability to detect new webshells [90, 71]. One of the primary strengths of CNNs lies in their ability to perform automatic learn and extract hierarchical features from raw input data. This enables CNNs to identify complex patterns that making them more effective in detecting unknown webshells. CNNs also are highly scalable and can be trained on large datasets using modern hardware accelerators Table 1.3: Summary of related works

| Method | Approach | Dataset Benign(B) Webshell(W) | Best Performance |
|---|---|---|---|
| Weighted-vote (LR,SVM, MLP,RF) [4] | Weighted voting ensemble model that incorporates LR, SVM, RF and MLP | 5.379B 566W | Accuracy 94.28% |
| CNN, RF [40] | Mixing of Random Forest (RF) and Convolutional Neural Networks (CNN) | 11.161B 4.406W | F1-Score 94.77% |
| RF [37] | Using 21 high risk functions and variables with TF-IDF vectorization of opcode as input for RF model | 11.397B 912W | F1-Score 98.40% |
| Stack(2*RF, Ada-Cost) [41] | Combining two RFs and an AdaCost classiffier | 2.236B 2.162W | Accuracy 95.30% |
| MatDec [69] | A scoring matrix built from a set of source code statistical features | 1.002B 347W | F1-Score 98.95% |
| Stack(2*RF, GBDT) [20] | Combining two RFs and a GBDT classifier | 2.388B 2.232W | F1-Score 99.09% |
| Stack(X,Y,Z DT) [87] | Dynamic selection of top 3 best classifiers and a DT classifier. | Benign 3.478 Webshell 9.939 | F1-Score 99.10% |
| DNN-LSTM [63] | Employing the Deep Neural Network composed of LSTM and pooling layers. | 6.669B 6.669W | Accuracy 98.54 |
| DeepSuperLearner (LR,RF,MLP) [5] | A probabilistic stacking model built on 3 base classifiers: LR, RF and MLP | 5.379B 571W | Accuracy 98.90% |
| MLP [70] | Comparing CNN, LSTM, MLP using vectorized PHP source code | - | Accuracy 99.56% |
| CNN [54] | CNN models each is fed with a different vectorization of One-hot, bag-of-words, and Word2Vec | 8.361B 3.944W | F1-Score 99.30% |
| LSTM [54] | Heuristic methods, including fuzzy hash similarity, signature matching, statistical feature-based similarity, and LSTM | 1.050B 1.050W | Accuracy 98.86% |

like GPUs. This scalability is crucial for practical deployment in real-world scenarios, where web applications generate vast amounts of code that need to be analysed for potential webshells. *Inspired by this, the dissertation focuses on enhancing CNN model in analyzing web application source code to detect webshell.*

### 1.3.3 AI-Powered Network Analysis Approaches

Network Intrusion Detection/Prevention Systems (NetIDPS) [62, 44, 108, 13, 94, 53, 38] are pivotal in identifying and mitigating webshell attacks, which pose significant threats to web server security by providing unauthorized remote access to attackers. NetIDS functions by monitoring network traffic for patterns indicative of malicious activity, employing both signature-based and anomaly-based detection techniques to identify potential threats. In the context of webshell attacks, NetIDPS plays a vital role by inspecting HTTP and HTTPS traffic for anomalies or known malicious signatures associated with webshell operations. Signature-based detection leverages databases of known webshell patterns, enabling the NetIDPS to recognize and alert administrators to the presence of these malicious scripts or even automatically block the webshell connection. This approach is effective in identifying previously documented webshells but requires regular updates to the signature database to remain relevant against emerging threats. Anomaly-based detection methods enhance the capability of NetIDPS by establishing baselines of normal web server behavior and flagging deviations that may suggest a webshell attack. Such deviations could include unusual traffic patterns, unexpected file modifications, or atypical server responses. Anomaly detection is particularly valuable for identifying novel or customized webshells that do not match existing signatures. But the reality is not so simple, today's webshells are equipped with many advanced techniques to make the network traffic they generate difficult to detect. This means that NetIDPS systems [27] will no longer be effective. However, the NetIDPS solution has an outstanding advantage, which is the ability to accurately detect and block in real-time anomalous network data flow that matches its ruleset.

Machine learning (ML) and deep learning (DL) techniques offer significant advantages when applied to network traffic analysis [1] for detecting innovative webshells, addressing the challenges posed by the evolving landscape of cyber threats and the sophistication of modern attacks. One of the primary benefits of ML/DL in this context is their ability to effectively handle large volumes of network traffic data, encompassing diverse protocols, formats, and behaviors. Moreover, ML/DL techniques offer the flexibility to adapt and learn from new data in real-time, allowing for continuous improvement and refinement of detection models as threats evolve. ML/DL-based network traffic analysis enables the integration of contextual information and metadata, such as source IP addresses, geolocation data, and historical traffic patterns, enriching the detection process and improving the overall efficacy of webshell detection systems.

The authors in [17] propose three deep learning models: that are Artificial Neural Network (ANN), Convolutional Neural Network (CNN), and Recurrent Neural Network (RNN) to detect web attacks. The DATASET CSIC 2010 is preprocessed by removing missing values, duplicate values, encoding fields, and normalizing (Min-Max scaling) then fed into the deep learning classifiers to build a prediction model. The performance evaluation shows that the RNN model provides 94% accuracy and 6% error rate, outperforming other methods like ANN and CNN. RNN also has higher precision, recall, positive predictive value (PPV), and negative predictive value (NPV) compared to other models.

The authors in [99] propose a model that combines a convolutional neural network (CNN) for extracting local features, with a long short-term memory (LSTM) network for capturing sequence patterns in the traffic content. For each request, a character-level feature transformation method convert the URL and post body fields into a 300-character-based vector. The CNN+LSTM model with character-level features achieves high precision, recall, and F1-score in detecting Webshell traffic. It also shows good generalization ability in discovering unknown Webshell attacks. The character-level representation allows faster training and detection compared to other text embedding methods like word2vec.

The authors in [96] introduced a runtime webshell detection system for analyzing HTTP requests. It employs a Support Vector Machine (SVM) classifier trained on preprocessed and vectorized HTTP requests, with preprocessing encompassing GET and POST parameter decoding and SVM categorizing requests as suspicious, attack, or benign.

The authors in [48] proposed two distinct deep learning models, namely CNN and RNN-LSTM, to detect various network attacks, including webshell attacks. The CNN model is constructed based on preprocessed and encoded payloads using a word embedding technique, while the RNN-LSTM model is built upon the initial sequence of characters in payloads. Both models outperformed traditional ones like LR, SVM, and RF. Moreover, RNN-LSTM exhibited a slight improvement over CNN.

The authors in [91] proposed a model named SB-LSTM, consisting of four LSTM layers with 60 neurons each and a dense layer. They adopted a session-based detection approach, identifying sessions through statistical analysis and the calculation of time intervals between log file entries. The SB-LSTM model was fed with 6-length vectors encoding each session log entry.

The authors in [24] make a comparative study of using deep learning algorithms in NetIDS. Six deep learning models were selected including: DNN, CNN, RNN, LSTM, GRU, and a Hybrid CNN-LSTM architecture. While other approaches, including RNN, CNN, LSTM, GRU, and the Hybrid CNN-LSTM approach, archive commendable results. The DNN approach stands out with outstanding performance and proves to be the most effective in handling complex network patterns. DNNs are typically composed of multiple fully connected (dense) layers, where each neuronee in one layer is connected to every neuron in the subsequent layer. This architecture allows DNNs to learn complex representations of the HTTP traffic that often contains subtle and intricate patterns indicative of webshell activity. DNNs can automatically learn these patterns from the raw data without requiring extensive feature engineering, making them highly effective for detecting sophisticated or obfuscated webshells. DNNs also are well-suited for handling large volumes of information and extract relevant features automatically. This allows for more accurate and comprehensive analysis of HTTP requests in real-time. *From these advantages of DNN, the dissertation focuses on researching and improving the DNN model in analyzing HTTP traffics to detect webshell.*

## 1.4 Dissertation Research Direction

The comprehensive review of existing literature described above identifies current challenges, trends, and gaps in the field of webshell detection.

Traditional webshell detection methods often use pattern matching techniques, which have the advantage of quickly and accurately detecting known webshell patterns, but they are easily bypassed by new types. To solve this problem, the related researches now mainly focus on applying AI techiques to analyze web application source code and HTTP traffic to enhance the efficiency of webshell detection. Although these studies have achieved significant results, each method still has its own advantages and disadvantages and there is still room for further improvement.

For the source code analysis method, the advantage is the ability to detect webshell types accurately, but it is limited because it depends heavily on the webshell programming language and consumes a lot of time and resources. The application of ML/DL techniques in source code analysis enables them to detect previously unseen or polymorphic webshells that evade traditional signature-based detection methods. These techniques can learn from large-scale datasets of labeled code samples, enabling them to generalize to new and evolving threats. However, challenges remain that are the need for labeled training data, the interpretation of model decisions, and the potential for false positives or false negatives. Moreover, the computational complexity and resource requirements of ML/DL models may limit their applicability in certain environments. The combination of pattern matching techniques and ML/DL algorithms in source code analysis will improve the efficiency and performance of webshell detection.

For the HTTP traffic analysis method, the fast detection speed, programming language independence, and the ability to integrate seamlessly with NetIDPS systems are advantages, but the trade-off is that the accuracy will not be as high as the source code analysis method. One advantage of ML/DL approaches is their ability to adapt and learn from new data, allowing them to detect previously unseen or polymorphic webshells that evade traditional signature-based detection methods. Moreover, these techniques can scale to analyze large volumes of HTTP traffic in real-time, making them suitable for deployment in high-throughput web environments. There are still problems with using ML and DL to properly look at HTTP traffic for webshell detection. These include the need for labeled training data, figuring out what model decisions mean, dealing with encrypted traffic, the chance of getting false positives or negatives, and keeping up with new ways for webshells to hide their activity.

Therefore, the research direction of the dissertation is to enhance the webshell detection efficiency of both source code analysis and HTTP traffic analysis methods to cover the shortcomings of each, specifically as follows:

- Researching on the method of combining the advantages of signature-based techniques and ML/DL algorithm in source code analysis that are able to detect innovative webshells with very high accuracy. From there, propose a framework that provides a guideline for developing specific models tailored to various programming languages. This hybrid framework enables the rapid and accurate detection of both known and unknown webshell types. To demonstrate the effectiveness of the framework, the study build language-specific webshell detection models and compares the results with related studies. However, due to the diversity in the number of server-side programming languages, the dissertation focus on the two most popular server-side programming languages: PHP as interpreted language and ASP.NET as compiled language.

- Researching on the ML/DL model that perform in-depth analysis of HTTP traffic queries directed at web application systems, effectively identifying queries that indicate both known and unknown webshell attacks. The study experiment the model and compare the results with those of other studies using the public dataset to showcase its effectiveness. Furthermore, the model is capable of integrating into NetIDPS to automatically add the suspicious source addresses to the blacklist and block the URI of the webshell on the web server.

## 1.5     Summary of Chapter 1

In chapter 1, we presented a general overview of webshells, webshell detection techniques, and related research. So, the study looks at the pros and cons of the current methods for finding webshells, as well as the difficulties and emerging trends in using ML/DL models to make finding new types of webshells more effective. The clearly defined research goal of the dissertation is to build a solution based on source code analysis methods and a solution based on network analysis methods that allow comprehensive detection of webshell attacks. These solutions will combine the advantages of pattern matching-based detection techniques with machine learning and deep learning models to enhance the ability to detect new webshell attacks.

In the following chapters, the study presents research results, evaluates them, and proposes solutions based on deep learning to solve the webshell detection problem based on two approaches: source code analysis and network-based analysis.

# Chapter 2

# DL-POWERED WEBSHELL DETECTION BY SOURCE CODE ANALYSIS

In this chapter we will focus on analyzing web application source code approach to detect attacks webshell. Due to the limitations of the traditional technique relies on predefined signatures and patterns which can limit its effectiveness in detecting new or obfuscated webshells that do not match known patterns. The dissertation states the problem and clearly defined the goals that need to be completed. Then we will propose an framework for detecting webshell attacks based on the source code analysis approach. Pattern matching detection methods and a deep learning model will be used together in the framework to make it better at finding unknown webshell types. Based on the framework, the dissertation will build and experiment with two complete systems to detect webshells written in PHP and ASP.NET, which are the two most popular server-side languages today.

## 2.1 Problem Statement

Traditional detection techniques often rely on signature-based approaches, which involve matching known patterns or signatures of known webshells. As a result, it tends to very quickly and accurately distinguish between benign and malicious code based on specific signatures. However, this approach only detects previously identified webshells and may fail to detect new or modified variants that deviate from known signatures. As a result, these methods are vulnerable to evasion by attackers who continuously modify or obfuscate webshell code to avoid detection. High false-positive rates, which mistakenly flag benign code or legitimate activities as malicious, may plague signature-based approaches. This can lead to alert fatigue and unnecessary investigation efforts, reducing the efficiency and effectiveness of the detection system. Below is a list of notable studies that utilized traditional techniques.

Typical research works in analyzing application source code to detect webshells have shown that traditional methods and methods using ML/DL both have different advantages and disadvantages, but currently there are not many studies using a combined approach to take advantage of the advantages of these two methods. However, this is really a promising approach, the dissertation determines the research direction that will focus on proposing an architecture that combines signature-based detection techniques with detection techniques based on AI algorithms. The architecture will allow very fast detection of known webshells and the ability to accurately detect innovation-unknown webshells. And in this chapter, the study selects the two most popular server-side programming languages today, PHP and ASP.NET, to prove the correctness and feasibility of the architecture.

The problem in this chapter will be stated as follows:

**Given:** *x = web application source file* **Find:**

*P(α) : pattern matching algorithm, where α is the dataset of webshell signatures*

*D(β) : deep learning model, where β are the parameters of the model*

$$FW(P,D)(x) = \begin{cases} 1, & \text{if } x \text{ is a webshell} \\ 0, & \text{if } x \text{ is benign} \end{cases}$$

Let *x* be a web application source code file. Let *P* be a pattern matching algorithm with a set of *α* rules as patterns to recognize webshells. Let *D* be a deep learning model with the parameters that make up the model being *β*. We need to find a function *FW(P,D)* with the optimal *α* and *β* for P, D such that if *x* is a webshell then *FW(x)* = 1 else if *x* is a benign then *FW(x)* = 0.

**Three specific goals are as follows:**

- Proposing an DL-Powered Source Code Analysis Framework, namely ASAF, that mainly combines two techniques, signature-based and ML/DL algorithms, to allow fast and accurate detection of webshell types, including known and unknown types. The framework will be the orientation for building each specific model applicable to each different type of programming language.

- Proposing a complete interpreted language PHP webshell detection model built from ASAF. This model includes an algorithm that converts a PHP source file into a flat vector containing all the webshell features. The model also includes an ML/DL model with parameters tuned to best suit the PHP webshell detection problem, to ensure effective detection without requiring too much computational resources. Evaluating the effectiveness of the proposed model based on measurement criteria and comparing it to relevant studies.

- Proposing a complete compiled-language ASP.NET webshell detection model built from ASAF. This model includes an algorithm that converts an ASP.NET source file into a flat vector containing all the webshell features. The model also includes an ML/DL model with parameters tuned to best suit the ASP.NET webshell detection problem to ensure effective detection without requiring too much computational resources. Evaluating the effectiveness of the proposed model based on measurement criteria and comparing it to relevant studies.

## 2.2 Proposed DL-Powered Source Code Analysis Framework

As analyzed above, the increasing sophistication and prevalence of webshells lead to the need for a common source code analysis framework that can be applied to many different programming languages and is capable of fast detection with a low false positive rate for known webshell types. At the same time, it is the ability to detect with high accuracy new types of webshells. Based on previous research results, this study proposes an **A**I-powered **S**ource Code **A**nalysis **F**ramework, namely ASAF, that combines Yara rules for known webshell detection with a Convolutional Neural Network (CNN) model for detecting new, sophisticated webshell variants. By leveraging the strengths of both signature-based and deep learning-based methods, this framework aims to provide comprehensive and effective webshell detection. The structure of the framework include three modules: YARA-Based webshell detection, DL-powered webshell detection model learning and DL-Powered webshell detection.

All are linked together as shown in the Fig. 2.1.

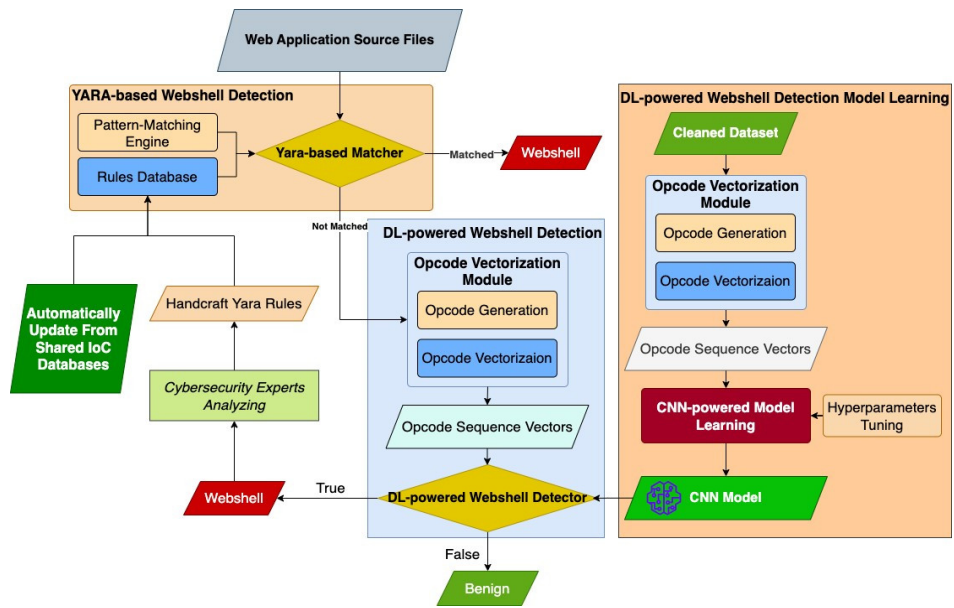

Figure 2.1: Correlational links between ASAF components

The process begins with web application source files undergoing a pattern-matching analysis using Yara components, which consist of a pattern-matching mechanism and a Yara-rules database. If a match is found, the file is immediately flagged as a webshell. If no match is detected, the file is deemed benign and proceeds to the next stage, which involves deeper analysis using opcode generation and vectorization modules. These modules convert the source code into opcode sequences, providing a low-level representation of the code's behavior.

The opcode sequences are then vectorized and fed into a Convolutional Neural Network (CNN) for further analysis. The cleaned dataset plays a critical role in training, validating, and testing the CNN model. The model has also been finely tuned through hyperparameter tuning, predicting whether the code is a webshell or benign. If CNN detects a webshell, this prediction is forwarded to cybersecurity experts for verification and rule updating, ensuring that new patterns are incorporated into the Yara-rules database. The framework also allows us to automatically update Yara rules from shared IoC databases. Conversely, if CNN predicts the code as benign, it is confirmed as safe. This dual-layered approach, leveraging both Yara rules for known threats and CNN models for unknown threats, ensures robust and dynamic detection of webshells, enhancing the security of web applications.

- **Yara Module:** The architecture of the Yara module in ASAF revolves around the Yara system. The main function of this system is to detect known webshells based on predefined patterns. Yara is made up of two components: the patternmatching mechanism and the Yara-rules database.

*The pattern-matching mechanism* is based on a combination of textual and binary pattern matching, allowing for the precise identification of specific sequences of characters or bytes within a file. At its core, Yara uses string matching, which can include plain text strings, hexadecimal patterns, and even wildcards for variable sections within strings. Additionally, Yara supports regular expressions, enabling complex pattern matching for strings that follow specific formats or structures. The tool also incorporates conditions and Boolean logic, allowing rules to specify that a file should be flagged if it contains multiple patterns in combination or certain patterns but not others. Furthermore, Yara can define where in the file the patterns should be found using offsets and ranges, ensuring patterns are matched at specific locations within the file. Modules extend Yara's functionality, such as the PE module for parsing Portable Executable files, and metadata provides additional context or information about the pattern being matched. By combining these features, Yara's pattern matching mechanism offers a flexible and powerful approach to defining and detecting known webshells, making it an effective tool for source code analysis. How the pattern matching technique works is described in Algorithm 2.1.

---

**Algorithm 2.1** Pattern-Matching mechanism

**Input:** $W$ - list of web application source files; $R$ **Output:** t of Yara-Rules Database $\omega$ - list of webshell

1: $\omega \leftarrow \emptyset$
2: **for** $f \in W$ **do**
3:     **for** $r \in R$ **do**     ▷ if file $f$ match with rule $r$
4:         **if** *RuleMatch(r,f)* = *true* **then**
5:             *ω.append(f)*     ▷ append $f$ as webshell to $\omega$
6:         **end if**
7:     **end for**
8: **end for**
9: **return** $\omega$

*The Yara-Rules database* consists of a comprehensive collection of meticulously crafted rules that define specific patterns, signatures, and characteristics associated with known threats. These rules leverage a combination of textual and binary pattern matching, utilizing plain text strings, hexadecimal patterns, wildcards, and regular expressions to capture the diverse forms of webshells. The database is continually updated with new rules as new threats are identified automatically from the shared Indicators of Compromise (IOC) database or manually by security experts. By maintaining an up-to-date Yara rules database, the framework ensures a reliable first line of defense against known webshells, allowing for efficient and accurate analysis of web source code.

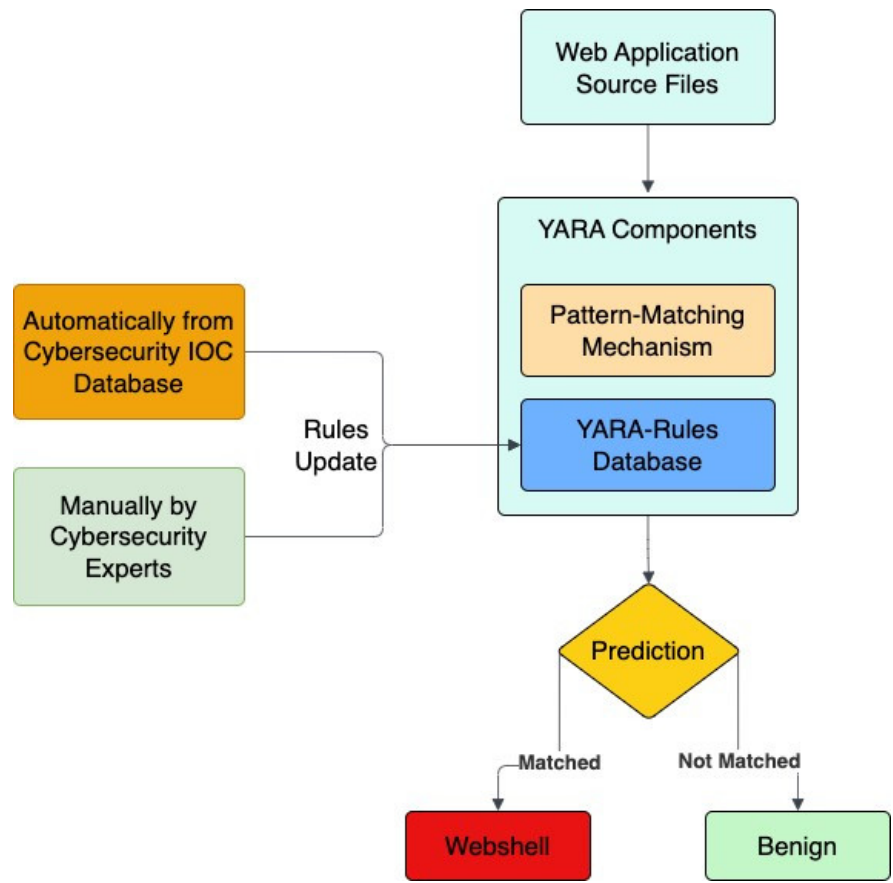

Figure 2.2: Yara architecture

The Fig. 2.2 shows the architecture of the Yara module in ASAF. To detect the webshell, the source code files will be scanned by the Yara system. The system applies all the rules in the database to the web source code to identify matches. All the files that match the patterns of the Yara rules are marked as potential webshells. The others are classified as benign.

- **Opcode Vectorization Module:** The purpose of the module within the webshell detection framework is to enhance the accuracy and depth of source code analysis by converting web source code into its corresponding opcode sequences. This lower-level representation of the code exposes the fundamental operations performed by the code, which can reveal hidden malicious patterns that may not be apparent through higher-level code analysis. By translating the source code into opcode sequences in the form of vectors, the framework can apply ML/DL models to detect advanced evasion and obfuscated webshells [55]. The module is made up of two main components: Opcode Generation and Opcode Vectorization as shown in Fig. 2.3.

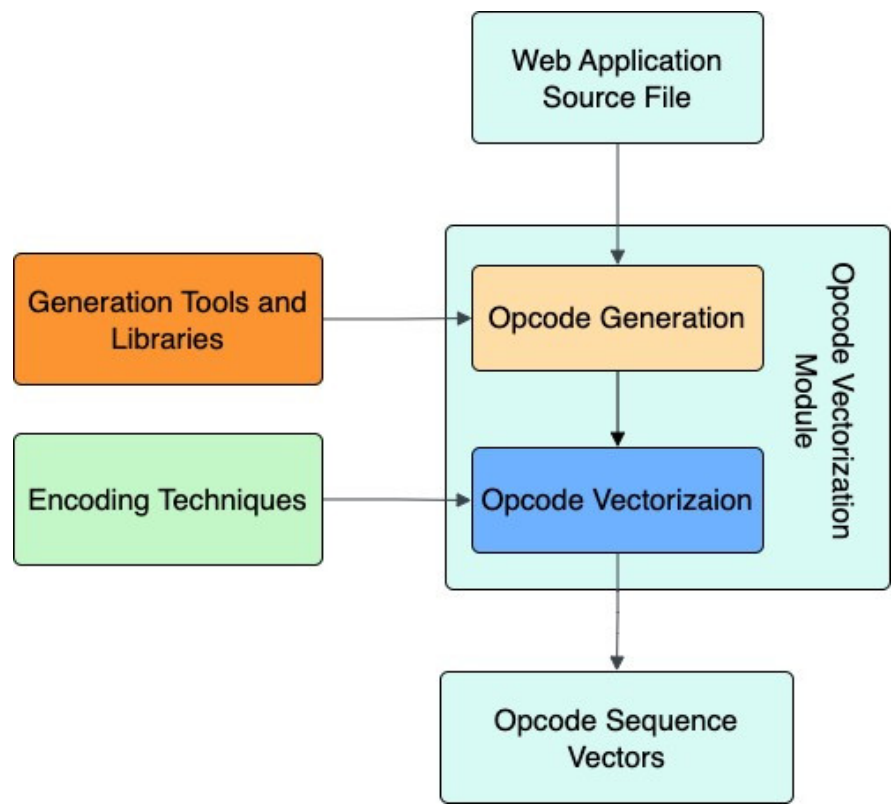

Figure 2.3: Opcode vectorization module

*The Opcode Generation* is responsible for translating source code into a sequence of opcodes, which are low-level representations of the instructions executed by the CPU. The web source code is first parsed to generate an Abstract Syntax Tree (AST) or Intermediate Representation (IR). This parsing step helps in breaking down the code into its basic components. Using the AST/IR, the code is then translated into opcodes. This step involves mapping high-level code constructs to their corresponding low-level operations. There are tools and libraries that provide the means to generate opcodes from source code across various programming languages. Each of these tools offers unique capabilities tailored to the specific requirements and nuances of the respective programming languages, thereby enabling comprehensive and effective opcode analysis, such as:

- PHP: VLD (Vulcan Logic Disassembler)
- .NET (C#): ILDasm, Mono.Cecil
- Python: dis module
- JavaScript: Esprima or Acorn (with custom scripts)
- Java: javap
- Ruby: RubyVM::InstructionSequence

*The Opcode Vectorization* is responsible for converting the sequence of opcodes into a numerical format that can be used as input for machine learning models. This step is crucial for leveraging the power of Convolutional Neural Networks (CNNs) and other models to analyze opcode sequences. The sequence of opcodes is tokenized into individual numerical vectors. Common encoding techniques include one-hot encoding, frequency encoding, and more advanced methods like Word2Vec or embeddings specifically trained on opcode sequences. To ensure uniform input size for the CNN, the sequences are padded or truncated to a fixed length. In ASAF, we propose the *Opcode Index Vectorization Algorithm*, or **OIVA** for short, to represent the sequence in which the opcodes are called in vector form. The opcode sequence vectors can fully represent the features of a web application source code, including data that can be considered characteristic of webshell, allowing CNN models to take advantage of their greatest advantage, the ability to automatically learn featured.

The algorithm is stated as follows: assuming the programming language (*L*) has (*n*) opcode functions, let (*p*) be the set containing opcode functions (*l*).

$$p = l_1, l_2, ... l_n \tag{2.1}$$

Therefore, the *Opcode$_i$* can be defined as a vector *OCI$_i$* that is a set of index *o* of the instructions in *p*, where *m* is the number of called instructions.

$$OCI_i = o_{i1}, o_{i2}, ... o_{im} \quad (2.2)$$

Since the $n_i$ of each $OCI_i$ is different, the lengths of these vectors need to be identical. Let *max_length* is the maximum of $(n_1, n_2, ...)$, then we padding all the $OCI_i$ to the same *max_length* length by value 0. With this representation, the OCI vectors will be able to represent most properties of the source code. The Algorithm 2.2 illustrates our strategy to compute an OCI vector as output, from the opcode file in text format as input.

---

**Algorithm 2.2** OIVA

  **Input:** $f$ - Opcode file; *IS* - List of opcode **Output:** OCI vector

1: $OCI \leftarrow \emptyset$
2: **for** *line* $\in f$ **do**
3:     **for** $i \in IS$ **do**
4:         **if** $i \in line$ **then**
5:             *OCI.append(index)*
6:         **end if**
7:     **end for**
8: **end for**
9: *OCI.padding(max_length, 0)*
10: **return** *OCI*                                                                                                   ▷ index of $i$ in *IS*

- **Dataset Collecting and Cleaning**

  In the ASAF, the dataset plays a critical role in training, validating, and testing the Convolutional Neural Network (CNN) model. The quality, diversity, and size of the dataset directly influence the effectiveness and accuracy of the webshell detection system. The dataset should include both benign and malicious web application source files to train the CNN effectively. For collecting benign datasets, the data source is relatively abundant and easy to access. There are some methods to gather source files from the internet, such as Open-Source Repositories (GitHub, GitLab, Bitbucket, ...) or Open-Source Frameworks and CMS, ... The collected benign source files will be added to the intermediate dataset, ready for preprocessing. For webshell datasets, the number of files will be more limited, especially for new webshell types. Therefore, collecting a webshell dataset requires gathering data from many different sources.

Public Malware Repositories (VirusTotal, MalShare, TheZoo, Hybrid Analysis, ...) are popular services that aggregate malware samples. We can search for webshells by using relevant keywords. Honeypots are a proactive way to capture webshells in the wild. Honeypots are decoy systems designed to attract attackers, allowing you to collect and analyze their payloads. Collaborating with incident response teams can provide access to webshell samples discovered during security incidents. We can also access webshell data sources through security forums and open-source repositories, such as: Exploit Database, Hack Forums, GitHub Repositories, etc. Finally, personal and professional networks provide access to new types of webshells that are not yet widely shared.

Before the webshell collection is added to the dataset, each file needs to go through the webshell confirmation process, which includes two stages. At the first stage, files that match Yara rules are added to the preprocessed dataset. Those that do not match are flagged for manual review by cybersecurity experts at the second stage. In the event that the file is confirmed to be a webshell, it is added to the dataset, and new Yara rules are updated in the database to detect this type of webshell. After collecting the webshell and benign datasets from multiple sources, the collected dataset needs to go through preprocessing and cleaning. Preprocessing the collected data to remove duplicates and irrelevant files. The final result is a high-quality, cleaned dataset, as shown in Fig. 2.4.

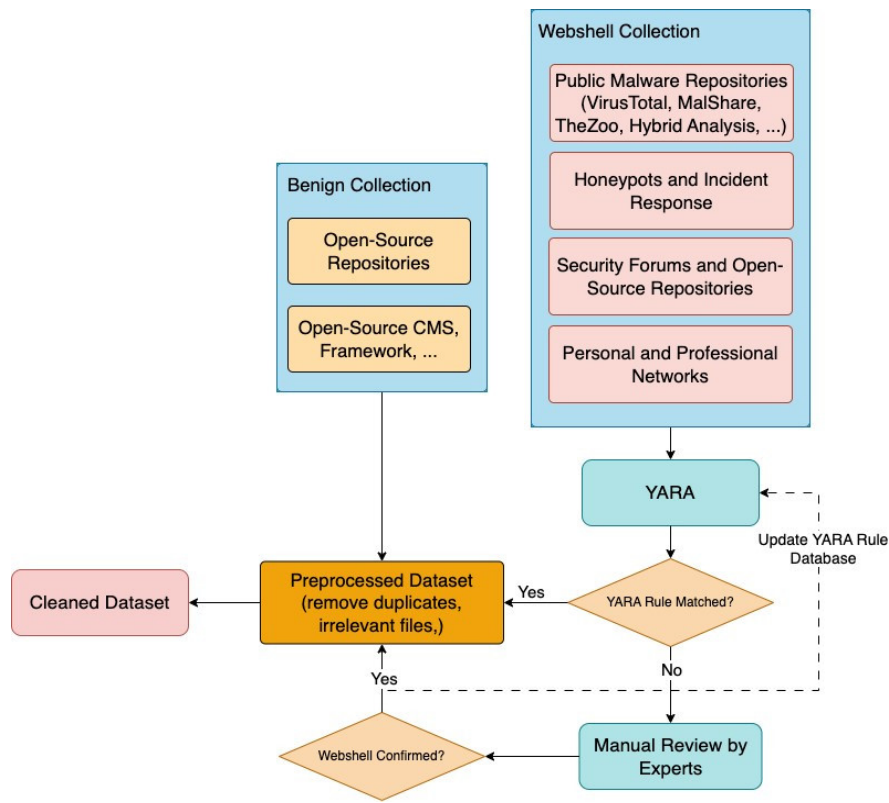

Figure 2.4: Dataset collecting and cleaning

- **CNN Model Architecture**

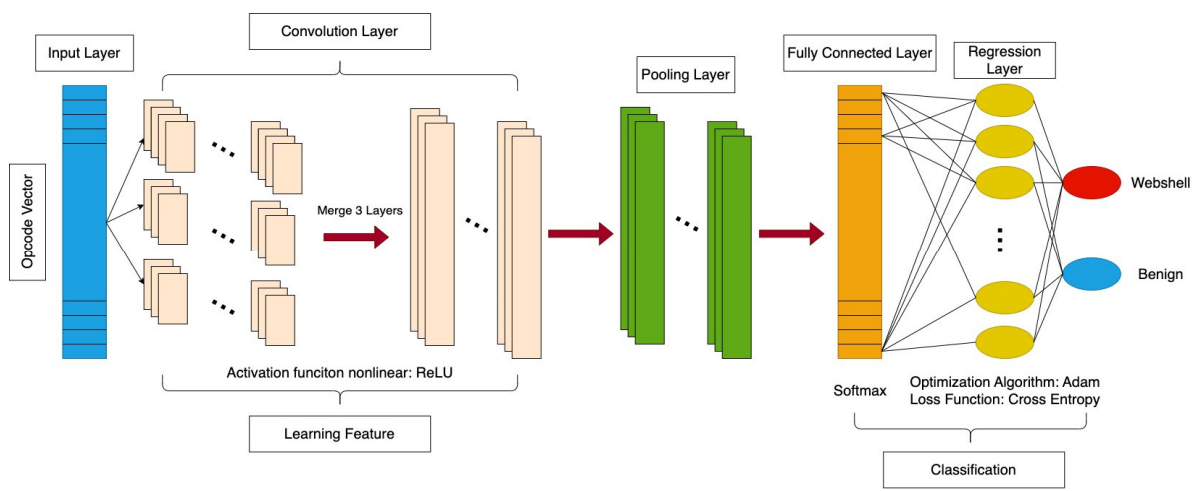

Figure 2.5: CNN model architecture

In an ASAF, CNN model architectural design plays an important role. The architecture of a proposed CNN model is composed of layers, relationships between layers, and also hyperparameters whose value is set before the learning process begins. Usually, for each specific problem, there will be certain architectures that shows outstanding advantages. However, it needs to go through a process called hyperparameter tuning to achieve the best efficiency, performance, and speed. Hyperparameter tuning consumes quite a lot of time and resources, so not all hyperparameters will be refined when we know they are optimal for the problem. Therefore, at this step, we draft the CNN model architecture using the same structure and optimal hyperparameters as shown in Fig. 2.5. The other hyperparameters will be selected after we make the tun at the next step. The draft architecture of the CNN model is composed of five layers.

1. The network starts with an **input layer**, which takes the vectorized opcode sequences. This layer simply feeds the data into the network without any transformations.

2. Following the input layer are a series of **convolutional layers**. The model uses three parallel one-dimensional convolutional layers, each with different kernel sizes, that slide over the input data, performing convolutional operations to capture features of varying lengths from the input sequences. All these convolution layers will then be merged into a concatenate unified convolution layer to combine the features extracted by different kernels, providing a richer representation of the input data. The nonlinear activation function *ReLU f(x) = max(0,x)* where *x* is the input value to create abstract information at a higher level for the following layer. By providing a huge amount of training data, we can expect the neural network to learn specific patterns of the malware family as well as powerful invariant features over time to distinguish the malware from benign files.

3. The third layer is the **pooling layer** which reduces the number of parameters that we need to calculate, thereby reducing computation time. The proposed model uses the global_max_pooling layer. The main purpose of this layer is to reduce the dimension of the previous layer, remove features that are no longer needed, and preserve the features that are sufficient to classify the object.

4. The fourth layer is **fully connected layers**. After several convolutional and pooling layers, the output is flattened and passed through one or more fully connected layers. These layers function similarly to traditional neural networks, where each neuron is connected to every neuron in the previous layer. Fully connected layers integrate the high-level features learned by the convolutional layers and map them to the final output.

5. The last layer is the **classification layer** that converts the feature matrix in the previous layer into a vector containing the probabilities of the source files with the following parameters:

    – *Softmax activation function:* In the classification problem, the output needs to predict which class the input data belongs to; the result of 100% is divided equally among all classes; the class with the greatest probability is the output. Another problem encountered is that signals from the neural network layers can have negative values. So the requirement is to find a function that returns a positive probability of what class the signal belongs to, and the sum of the probabilities equals 100%. The Softmax activation function solves this problem and is often used at the last layers of the classification network to evaluate the classification probability of the input data. The formula for the Softmax function is as follows:

    $$\text{Softmax } X_{ij} = \frac{\exp X_{ij}}{\sum_{k}^{PC} \exp X_{ij}}$$

    – *Loss function cross-entropy:* Calculates the difference between the probability distributions of the prediction and the actual distribution of each class. The model predicts a probability distribution $(p, 1 - p)$ for each class and compares it with the real distribution $(y, 1 - y)$ of each class. In the webshell detection problem, cross-entropy is used to compare the distance between softmax outputs and the probability that the file is benign or webshell in one-hot encoding. One-hot encoding is the process of converting each value into binary features containing only the value 1 or 0, or, in other words, one-hot encoded labels tell us what kind of source file it is with 100% certainty. The lower the cross-entropy, the more accurate the prediction results. If the prediction is perfect, its value will be 0. As such, the cross-entropy is used as a loss function to help a neural network evaluate the probability (the certainty) of predicting a data sample corresponding to a class. The cross-entropy function is defined by the following formula:

    $$-(y \log((p)) + (1 - y) \log(1 - p)))$$

- *Optimizers:* for the purpose of smoothing the steps of the descending gradient so that it can converge faster, the model using the Adaptive Moment Estimation (adam) algorithm is common in CNN architectures; The parameters set for this optimizer include *learning_rate* = 0.05; $\beta_1$ = 0.9; $\beta_2$ = 0.999 and $\varepsilon = 10^{-8}$.

• **Hyperparameter Tuning**

The CNN model architecture above is just a basic architectural framework designed to best suit the webshell detection problem; however, the programming language has a great influence on the characteristics of each type of webshell. Therefore, it is essential to perform hyperparameter tuning to build a CNN model for each type of webshell written in different programming languages.

Hyperparameter tuning is a critical process in optimizing Convolutional Neural Network (CNN) models to achieve the best possible performance. Hyperparameters are configuration settings used to control the learning process of the model, and they are not learned from the data. Key hyperparameters for CNNs include the number of layers, the number of filters in each convolutional layer, filter sizes, stride, padding, activation functions, dropout rates, batch size, learning rate, and the number of epochs for training. Tuning these parameters involves systematically experimenting with different values to find the optimal configuration that minimizes the loss function and enhances the model's accuracy. Common techniques for hyperparameter tuning include grid search, random search, and more advanced methods like Bayesian optimization and genetic algorithms. Grid search exhaustively searches through a manually specified subset of the hyperparameter space, while random search samples hyperparameters from a specified distribution. Bayesian optimization builds a probabilistic model of the objective function and uses it to select the most promising hyperparameters to evaluate next. Proper hyperparameter tuning can significantly impact the CNN's ability to generalize from the training data to unseen data, thereby improving its effectiveness in tasks such as detecting unknown webshells. This process is often computationally intensive, requiring robust infrastructure and efficient use of resources to balance exploration of the hyperparameter space with the computational cost.

*The problem is stated as follows*: let $f : \chi \to \mathbb{R}$ be the objective function needed to optimize, where $\chi$ is the set of hyperparameters we want to search over. There are many techniques for selecting a set of hyperparameter values for optimization, from simple ones like Grid Search, Random Search to complex ones such as Evolutionary Optimization or Bayesian Optimization. We use *k*-fold cross validation to calculate the score for a given set of hyperparameter values and MSE (Mean Square Error) as the score function that will be minimized. Since the cost of computation per *k*-fold cross validation is expensive, Bayesian Optimization is suitable for not calculating all values of hyperparameters. The main idea is to consider model *f* as a probability distribution, We compute *f* at parameter $x_1, x_2, ..., x_D$. Then, $f(x_1), f(x_2), ..., f(x_D)$ are observed variables; any *f(x)* that cannot be computed by high cost is considered a hidden variable. Then we use the Gaussian process to select the probabilistic model $P(f(x)|f(x_1), f(x_2), ... f(x_D))$ to compute *f(x)*.

There are some popular libraries, such as Keras Tuner, Optuna, and ScikitOptimize provide robust frameworks for this purpose. Keras Tuner is specifically designed for optimizing Keras models, offering various search algorithms like Random Search, Hyperband, and Bayesian Optimization to explore the hyperparameter space. Users define a HyperModel or use a prebuilt one, then set up a tuner to execute the search. Optuna is another powerful library that employs an adaptive approach, using efficient sampling and pruning strategies to optimize hyperparameters. It integrates seamlessly with many machine learning frameworks and allows users to define an objective function that Optuna optimizes. Scikit-Optimize (or skopt) provides simple yet flexible tools for hyperparameter optimization, including Bayesian Optimization, which builds a surrogate model to make intelligent decisions about which hyperparameters to try next. To use these libraries, users typically define the search space for each hyperparameter, specify the objective function to minimize, and configure the search strategy. These libraries help automate the tuning process, significantly enhancing the performance and efficiency of CNN models by systematically identifying the best hyperparameter configurations. They also support visualization tools to analyze the search process and outcomes, making them invaluable for developing highperforming CNN models in Python.

*In conclusion, the proposed framework leverages the strengths of Yara's rule-based detection for known webshells and the adaptive learning capabilities of CNNs for detecting unknown webshells. The framework provides a general guideline, thereby allowing us to build solutions to effectively detect webshell attacks for each type of programming language using source code analysis.*

## 2.3  PHP Webshell Detection

### 2.3.1  Approach Direction

As of the latest data, W3Techs (World Wide Web Technology Surveys) reports that PHP is used by approximately 78.9% [15] of all the websites. This is a substantial majority, indicating PHP's dominance in the server-side language market, making it a common target for webshells. Traditional detection methods often fall short due to the evolving nature of webshells, necessitating a more sophisticated approach. Using ASAF, in this section, the study proposes to build a model to effectively detect webshells programmed in PHP.

### 2.3.2  Yara-Based Analysis

The Yara-Rules dataset used in this study was collected from many sources and over a very long period of work in the field of information security. The first is from GitHub and GitLab, which are platforms that host numerous valuable and updated Yara-Rule repositories shared by cybersecurity researchers and organizations. The second is from my professional networks with specialized agencies in the field of information security, professional associations, research institutes, conferences, and webinars. This kind of source can provide access to custom-developed Yara rules that allow for the detection of several new webshells that are not yet widely shared. The data set contains a total of *699 rules*, allowing detection of many popular PHP, JSP, ASP, ASP.NET, Python webshells today. This set of rules will continue to be updated regularly to enhance the ability to detect new webshell patterns.

### 2.3.3  Opcode Vectorization

VLD, short for Vulcan Logic Disassembler, is a powerful PHP extension designed to disassemble compiled PHP code, providing a detailed representation of its internal opcode. VLD reveals the underlying operation codes (opcodes) that the PHP engine generates from the source code during execution. This tool is particularly useful for developers and security analysts who need to understand the low-level operations of PHP scripts, optimize performance, or detect anomalies and vulnerabilities such as webshells. Security experts can use VLD to analyze obfuscated or malicious PHP code by examining the opcodes for suspicious operations.

---

[15] https://w3techs.com/technologies/details/pl-php

By executing a PHP script using the "*-d*" flag to generate and view the opcodes, as follows:

```
php −d vld . active=1 −d vld . execute=0 your_script .php
```

Here, *"vld.active=1"* activates the VLD extension, and *"vld.execute=0"* prevents the script from running after disassembly, focusing on opcode generation. The output will display the opcodes for each line of the PHP script, including detailed information about each operation performed by the PHP engine.

### 2.3.4 Dataset Collecting and Cleaning

For the benign dataset, different PHP frameworks, forums and content management systems were collected from their official sites. They includes Laravel, Wordpress, Joomla, phpMyAdmin, phpPgAdmin, phpbb [16]. After removing non-PHP files, the benign set contains totally **7,400** files.

To build the webshell dataset, we collected a wide range of webshells from reliable and most stars sources on Github [17]. In addition, some webshell samples we collected during working were also used. There are totally **4,171** PHP webshell files.

Following ASAF, in the step of reviewing the collected webshell dataset with Yara and reviewed by experts, we eliminated 27 false positives. After removing irrelevant and duplicate files, the total number of benign files is **7,275** and the total number of webshell files is **4,087**.

In order to train and validate our proposed method of detecting PHP webshells, we divided the benign and webshell datasets into two parts with a ratio of 8:2 as the rule of thumb [75]. Based on the distribution of files in the dataset sources, the split of training and testing sets is chosen by the whole source. Thus, the following table shows our final datasets for training and testing.

---

[16] Benign repos on Github: Laravel; WordPress; Joomla; phpmyadmin; phppgadmin; phpbb.

[17] Webshell repos on Github: Tennc, PHP-backdoors, B374k, php-webshells, xl7dev/WebShell, BlackArch/webshells, fuzzdb, webshell-collector, webshell-sample, awsome-webshell, WebShellBypass-WAF, indoxploit-shell.

Table 2.1: Non-duplicate benign and webshell datasets

|                  | Training Set | Testing Set |
|------------------|:------------:|:-----------:|
| Benign Dataset   | 5,820        | 1,455       |
| Webshell Dataset | 3,270        | 817         |

### 2.3.5 Hyperparameter Tuning CNN Model

The CNN model for detecting PHP webshells is built on the basis of ASAF's CNN model; the detailed parameters of the model are selected through the hyperparameter tuning process. We choose to use the *grid search* technique that exhaustively searches over a manually specified subset of the hyperparameter space. It systematically works through multiple combinations of parameter values, cross-validating as it goes to determine which set gives the best performance. The advantages of this technique are that it is simple, easy to understand, and guarantees finding the optimal combination within the specified grid. However, it is computationally expensive, especially with a large number of hyperparameters or values, and may not be feasible for large-scale problems. There are six hyperparameters that can be tuned, and they take two types of values: range and choice. Especially for *filter size*, since the model uses three layers merged together in a convolution layer, each layer receives a different filter size, so its value will be a 3-dimensional vector of the form $[x, x+1, x+2]$.

Table 2.2: PHP-ASAF hyperparameters tuning value

| Hyperparameter   | Value                  | Type   | Optimal Value |
|------------------|------------------------|--------|---------------|
| learning rate    | [0.001, 1.0]           | range  | 0.001         |
| dropout rate     | [0.01, 0.8]            | range  | 0.5           |
| batch size       | [8, 16, 32, 64, 96, 128] | choice | 96          |
| epoch            | [8, 16, 32, 64, 96, 128] | choice | 64          |
| filter size      | [[2,3,4], [14,15,16]]  | range  | [3,4,5]       |
| number of filter | [1, 300]               | range  | 128           |

After performing hyperparameters tuning, the optimal CNN values are shown in Table 2.2.

### 2.3.6 Experimental Results and Evaluation

#### 2.3.6.1 Implementation Details

Based on the proposed method, we built and implemented our solution, namely **PHP-ASAF**, in python language. The experiments were performed in a computer having 2 x Intel(R) Xeon(R) CPU E5-2697 v4 @ 2.30GHz (45MB Cache, 18-cores per CPU), 128GB for the main memory, CentOS Linux release 7.4.1708, python release 2.7. For the deep learning platform, we use tensorflow v.1.14.0, scikit-learn v.0.20.4, scipy v.1.2.2, numpy v.1.16.5 and Yara-python v.3.10.0.

The experimental part is performed with 3 scenarios along with the test dataset built in Section 2.3.4.

- S1: Evaluate the PHP webshells detection efficiency of Yara component in PHPASAF.

- S2: Evaluate the PHP webshells detection efficiency of CNN model in PHPASAF.

- S3: Evaluate the PHP webshells detection efficiency of PHP-ASAF.

#### 2.3.6.2 Results and Evaluation

*a*. Yara Detection

Using the Yara ruleset in Section 2.3.2, we evaluate the ability to detect PHP webshell types on a test dataset of 1455 benign files and 817 webshell files collected in Section 2.3.4. Table 2.3 shows the results we got in the matrix confusion.

Table 2.3: Confusion matrix of PHP webshell detection by using Yara

|  | **Real Webshell** | **Real Benign** |
|---|---|---|
| Predicted Webshell | 709 | 8 |
| Predicted Benign | 108 | 1447 |

To evaluate the performance of your Yara-based detection system, we can calculate several key metrics based on the confusion matrix provided:

Table 2.4: Key metrics of of PHP webshell detection by using Yara (%)

| **Measure** | **Value (%)** |
|---|---|
| Accuracy | 94.89 |
| Precision | 98.88 |
| Recall | 86.76 |
| Specificity | 99.45 |
| F1-Score | 92.43 |
| False Positive Rate | 0.55 |
| False Negative Rate | 13.24 |

The prediction results underscore a significant limitation of the current Yara-based webshell detection system: its reliance on known webshell patterns. While the system demonstrates high accuracy (94.89%), precision (98.88%), recall (86.76%), and specificity (99.45%) in detecting webshells, these metrics primarily reflect its capability to identify webshells that match predefined patterns in the Yara rules. This inherent dependence on known signatures means that the system is adept at recognizing previously identified webshells but falls short when encountering new or obfuscated variants that do not align with the established rules.

The high precision rate signifies that the system effectively avoids false positives, ensuring that flagged files are almost always webshells. However, the recall rate, while commendable, reveals that approximately 13.24% of actual webshells go undetected. This gap highlights the system's vulnerability to novel webshells that deviate from known patterns. Such undetected threats can pose significant security risks, as attackers continually evolve their techniques to bypass signature-based detection methods.

The specificity of 99.45% indicates that benign files are rarely misclassified as webshells, a crucial attribute for minimizing unnecessary alerts and maintaining operational efficiency. However, this specificity does not compensate for the system's inability to adapt to new threat landscapes dynamically. As webshells evolve, so must the detection methodologies, moving beyond pattern matching to incorporate more adaptive and heuristic approaches.

*In conclusion, while the Yara-based system excels at identifying known webshell patterns, its effectiveness against unknown or modified webshells remains limited. To enhance its robustness, the system should integrate advanced machine learning techniques capable of learning from data and identifying anomalies beyond predefined patterns.*

**b. CNN Detection**

Same as the previous experiment, we used also the datasets described in Section 2.3.4 to evaluate the CNN model by using the tensorflow engine. The results illustrated by the matrix confusion in Table 2.5 and the key metrics in Table 2.6.

Table 2.5: Confusion matrix of PHP webshell detection by using Yara

|  | **Real Webshell** | **Real Benign** |
|---|---|---|
| Predicted Webshell | 807 | 17 |
| Predicted Benign | 10 | 1438 |

The recent evaluation results of the CNN-based detection system highlight its impressive capability in identifying known webshell patterns. With an accuracy rate of 98.81%, the model demonstrates strong overall performance in distinguishing between webshells and benign files. The high precision rate of 97.94% indicates that the model is highly effective in minimizing false positives, ensuring that most of the files flagged as webshells are indeed malicious. This is crucial in reducing the burden of false alarms on security analysts, allowing them to focus on genuine threats.

Furthermore, the recall rate of 98.78% reflects the model's robustness in detecting almost all actual webshells, showcasing its reliability in capturing known malicious patterns. The specificity, also at 98.83%, underscores the system's efficiency in cor rectly identifying benign files, thereby minimizing the risk of misclassification and potential disruption of legitimate activities.

Table 2.6: Key metrics of of PHP webshell detection by using CNN (%)

| Measure | Value (%) |
|---|---|
| Accuracy | 98.81 |
| Precision | 97.94 |
| Recall | 98.78 |
| Specificity | 98.83 |
| F1-Score | 98.35 |
| False Positive Rate | 1.17 |
| False Negative Rate | 1.22 |

The F1 score of 98.35% balances precision and recall, reinforcing the system's effectiveness in handling both false positives and false negatives. The low false positive rate of 1.17% and false negative rate of 1.22% further validate the model's accuracy and dependability.

***c*. PHP-ASAF Detection**

From the above two experiments, we have shown the advantages and disadvantages of Yara and CNN techniques in detecting PHP webshells. In this section, we will continue to experiment with the PHP-ASAF model on the same data set to prove its effectiveness over the above two techniques in detecting PHP webshell attacks. Experimental results are shown in the confusion matrix in Table 2.7 and the key metrics in Table 2.8.

Table 2.7: Confusion matrix of PHP webshell detection by using PHP-ASAF

| | Real Webshell | Real Benign |
|---|---|---|
| | | |

| Predicted Webshell | 809 | 17 |
| --- | --- | --- |
| Predicted Benign | 8 | 1438 |

The integration of Yara's pattern-matching capabilities with CNN's deep learning prowess allows PHP-ASAF to detect both known and unknown webshell patterns effectively. It can be seen that the detection results of PHP-ASAF are better in all measures when compared with PHP and CNN detection. Specifically, when looking at the number of 817 PHP webshell files, CNN has 10 cases of False Negatives (FN), Table 2.8: Key metrics of of PHP webshell detection by using CNN (%)

| **Measure** | **Value (%)** |
| --- | --- |
| Accuracy | 98.9 |
| Precision | 97.94 |
| Recall | 99.02 |
| Specificity | 98.83 |
| F1-Score | 98.48 |
| False Positive Rate | 1.17 |
| False Negative Rate | 0.98 |

while PHP-ASAF has 8 FNs. This is because there are 2 PHP webshell files out of 10 FN cases. This has been correctly identified by the Yara module in PHP-ASAF as a webshell.

In addition to the accuracy evaluation results, we also measured the detection time for the PHP-ASAF model. Accordingly, the total time for the PHP-ASAF model to detect webshell on the test dataset of 2272 PHP source files is 2945 seconds. That means it takes an average of 1.29 seconds for each PHP source file. For the source code-based webshell detection method, the time factor is not the most important factor, and it depends a lot on the size of the source file. Therefore, the average time we measured is only relative.

### 2.3.6.3 Comparisons

To justify our ASAF's performance, we compare our results to those of other approaches. For comparison, we chose two non-AI approaches [82, 76] and two ML/DL approaches [20, 72]. Due to the limitations of sharing source code, we have simulated the RF-GBDT and Word2Vec+CNN model as described by the authors, which are currently one of the best archivements for evaluation on our dataset aforementioned in Section 2.3.4. The actual results of the simulated RF-GBDT and Word2Vec+CNN are not as high as announced. The comparison results in Table 2.9 shows that the ASAF model achieves the best results in both accuracy of 98.9% and F1-Score of 98.48% when compared with other methods on our dataset.

Table 2.9: Comparison of different webshell detection approaches on our dataset (%)

| Method | Accuracy | F1-Score |
|---|---|---|
| Simulated Word2Vec+CNN[72] | 98.42 | 97.80 |
| Simulated RF-GBDT[20] | 98.59 | 98.05 |
| GuruWS[76] | 85.56 | 92.00 |
| Php-malware-finder [82] | 94.23 | 96.46 |
| **ASAF** (our) | **98.9** | **98.48** |

## 2.4 ASP.NET Webshell Detection

### 2.4.1 Approach Direction

According to W3Techs, ASP.NET is the second most used server-side programming language in the world after PHP. This means that the number of webshell attacks on web systems using ASP.NET is also very large. Unlike PHP, which is an interpreted language, ASP.NET is a compiled language, so the mechanism for creating opcodes from the source code of the web application is totally different. Building an ASP.NET webshell attack detection solution based on ASAF, namely ASP.NET-ASAF, will further demonstrate the ability to deploy a webshell attack detection solution that supports all server-site programming languages.

### 2.4.2 Yara-based Analysis

The Yara module in ASP.NET webshell attack detection solution is used in conjunction with the ruleset in Section 2.3.2. The rule set contains a total of 699 webshell patterns. We will regularly update this set of rules to improve our ability to detect new webshell.

### 2.4.3 Opcode Vectorizaion

Firstly, the source code of an ASP.NET application can be represented as MSIL, which is able to solve the problems related to code obfuscation. Unlike PHP, which interprets every time a web page is requested, ASP.NET compiles dynamic web pages into DLL files that the server can execute quickly and efficiently. Then, the DLL files will continue to be converted to Opcode. There are two commonly used tools: ILDasm (IL Disassembler) and Mono.Cecil. We choose to use Mono.Cecil, which is a powerful library for reading, manipulating, and writing .NET assemblies. It allows developers to inspect and modify the Intermediate Language (IL) code of .NET assemblies. Below is how to use Mono.Cecil in Python.

```
using Mono.Cecil; using
Mono.Cecil.Cil;

var assembly = AssemblyDefinition.ReadAssembly("
    MyAssembly.dll"); foreach (var module in assembly
.Modules)
{
    foreach (var type in module.Types)
    {
        foreach (var method in type.Methods)
        { if (method.HasBody)
            {
                foreach (var instruction in method.Body.Instructions)
                {                                  Console.
                    WriteLine(instruction); }
            }
        }
    }
}
```

The code below is a part of the Opcode corresponding to the functions *get_Request* in the source code of *insomnia_shell.aspx*.

```
[System.Web] System.Web.HttpRequest
        get_Request()    cil   managed
{
    .custom instance    void    [mscorlib] System.Diagnostics.
    DebuggerHiddenAttribute::.ctor() = (01 00 00 00)
```

```
        . maxstack     1
    . locals init ( class [ System .Web] System .Web. HttpContext V_0, class [ System
    .Web] System .Web. HttpRequest V_1) IL_0000 : call class [ System .Web] System
    .Web.
        HttpContext  [ System .Web] System .Web. HttpContext : : get_Current ()
        IL_0005 :       stloc .0
        IL_0006 :       ldloc .0
        IL_0007 :        brfalse . s     IL_0012
        IL_0009 :       ldloc .0
        IL_000a :        callvirt           instance     class    [ System .Web] System .
        Web. HttpRequest   [ System .Web] System .Web. HttpContext : : get_Request ()
        IL_000f :       stloc .1
        IL_0010 :       br . s          IL_0016
        IL_0012 :       ldnull
        IL_0013 :       stloc .1
        IL_0014 :       br . s          IL_0016
        IL_0016 :       ldloc .1
        IL_0017 :       ret
}
```

Currently, the list of MSIL has 229 instructions for .NET framework 5.0. By applying the OIVA algorithm, we obtain a set of Opcode vectorizations that fully represent the feature of ASP.NET webshell.

### 2.4.4    CNN Model Hyperparameter Tuning

Similar to the Hyperparameter Tuning process to build parameter selection for the CNN model of PHP-ASAF, we also perform this process to select optimal parameters for the ASP.NET-ASAF model. There are six hyperparameters that can be tuned, and they take two types of values: range and choice. The set of hyperparameters with their range to be tuned and optimal value is shown in Table 2.10.

Table 2.10: ASP.NET-ASAF hyperparameters tuning value

| Hyperparameter | Value | Type | Optimal Value |
|---|---|---|---|
| learning rate | [0.001, 1.0] | range | 0.001 |
| dropout rate | [0.01, 0.8] | range | 0.5 |
| batch size | [8, 16, 32, 64, 96, 128] | choice | 64 |
| epoch | [8, 16, 32, 64, 96, 128] | choice | 32 |
| filter size | [[2,3,4], [14,15,16]] | range | [4,5,6] |
| number of filter | [1, 300] | range | 128 |

### 2.4.5 Dataset Collecting and Cleaning

To build the dataset, we collect it from popular source code sharing platforms around the world, such as Github, Gitlab or Sourceforge. Collected source code files must ensure factors such as trustworthiness, reputable sharers, and good community appreciation.

For benign source files, the collection is relatively simple because there are quite a few open source CMS frameworks that use ASP.NET shared by the community; for example, DotNetNuke, Umbraco, Kentico, Sitefinity CMS, N2 CMS, and Orchard CMS. Obviously, collecting the webshells will be difficult, even though the reason for our use is research. When training the model, the dataset is as diverse as possible to achieve efficiency. Most research webshells can only be found on Github[18], but can also be obtained from other unorthodox sources; however, the reliability will not be high, especially with the ASP.NET webshell. After collecting **3.347** benign source files and **2.113** webshells from multiple sources, we went through the cleaning and preprocessing steps of the dataset. We use Yara, the same as the one in Section 2.3.2 to remove 49 benign files from webshell datasets.

Finally, the total number of ASP.NET source files we gathered was 2.064 webshells and 3.347 benign files. To train the model and test the effectiveness of the proposed method, the dataset will be divided into two parts, with a ratio of 8:2 corresponding to the training dataset and the test dataset. The following table shows our final datasets for training and testing.

Table 2.11: ASP.NET webshell and benign datasets

| | Training Set | Testing Set |
|---|---|---|
| **Webshell Dataset** | 1651 | 413 |
| **Benign Dataset** | 2678 | 669 |

---

[18] Webshell repositories: Tennc, xl7dev/WebShell, BlackArch/webshells, webshell-collector, webshell-sample, awsome-webshell.

## 2.4.6 Experimental Results and Evaluations

### 2.4.6.1 Implementation Details

Our ASP.NET-ASAF is conducted simulations in Python using TFLearn, a popular deep learning library to help programmers build and train deep learning models quickly and efficiently. The experiments were performed in a workstation with 2 x Intel(R) Xeon(R) CPU E5-2697 v4 @ 2.30GHz (45MB Cache, 18-cores per CPU), 128GB memory, Windows 10 Professional, Visual Studio 2019 Enterprise, Python release 2.7. For the deep learning platform, we use tensorflow v.1.14.0, scikit-learn v.0.20.4, scipy v.1.2.2, numpy v.1.16.5, and Yara-python v.3.10.0.

To evaluate the effectiveness of the ASP.NET-ASAF, we conducted experiments with the following three scenarios:

- S1: Evaluate the ASP.NET webshells detection efficiency of Yara component in ASP.NET-ASAF.

- S2: Evaluate the ASP.NET webshells detection efficiency of CNN model in ASP.NETASAF.

- S3: Evaluate the ASP.NET webshells detection efficiency of ASP.NET-ASAF.

### 2.4.6.2 Results and Evaluation

**S1: Evaluation of Yara component in ASP.NET-ASAF**: We use the test data at Section 2.4.5 to evaluate Yara component with 669 rules. As shown in Table 2.12 and Table 2.13, the prediction result indicates that Yara was not effective against unknown webshells with 67 files that were misclassified and the very high FNR up to 16.22%.

**S2: Evaluation of CNN model in ASP.NET-ASAF**:

After selecting the most suitable parameters and training the CNN model mentioned in the hyperparameter tuning step, we use the test data at Section 2.4.5. The Table 2.12: Confusion matrix of ASP.NET webshell detection by using Yara prediction result is shown in Tables 2.14 and Table 2.15. The model achieves a high accuracy rate, reflecting the proportion of total correct predictions out of all predictions made. An accuracy of 98.43% and an F1-score of 97.75% indicate that the CNN model is reliable and performs well in distinguishing between webshell and benign files. The false negative rate (FNR) is the proportion of webshells incorrectly classified as benign. With an FNR of 1.69%, the model shows a strong capability to detect webshells, ensuring that very few actual webshells are missed.

|  | Real Webshell | Real Benign |
|---|---|---|
| Predicted Webshell | 346 | 8 |
| Predicted Benign | 67 | 661 |

Table 2.13: Key metrics of ASP.NET webshell detection by using Yara (%)

| Measure | Value (%) |
|---|---|
| Accuracy | 93.07 |
| Precision | 97.74 |
| Recall | 83.78 |
| Specificity | 98.80 |
| F1-Score | 90.22 |
| False Positive Rate | 1.2 |
| False Negative Rate | 16.22 |

Table 2.14: Confusion matrix of ASP.NET webshell detection by using CNN

|  | Real Webshell | Real Benign |
|---|---|---|
| Predicted Webshell | 406 | 10 |
| Predicted Benign | 7 | 659 |

**S3: Evaluation of ASP.NET-ASAF**:

Combining the advantages of the two Yara methods and the CNN deep learning model, ASP.NET-ASAF will solve the problem of effectively detecting webshell attacks, including unknown patterns. Experiments on the data set in Section 2.4.5 give the results of the confusion matrix and key measures in Table 2.16 and Table 2.17. The result shows that the F1-score increased from 97.95% to 98.07%, accuracy in creased from 98.43% to 98.52% when compared to the CNN prediction result. With the advantage of being able to detect known webshells with very high accuracy, Yara will minimize the number of webshells that CNN misclassifies as benign. When combining Yara with the CNN model, one misclassified webshell will be corrected. Then, the FNR decreased from 1.69% to 1.49%.

Table 2.15: Key metrics of of ASP.NET webshell detection by using CNN (%)

| Measure | Value (%) |
|---|---|
| Accuracy | 98.43 |
| Precision | 97.60 |
| Recall | 98.31 |
| Specificity | 98.51 |
| F1-Score | 97.95 |
| False Positive Rate | 1.49 |

| False Negative Rate | 1.69 |
|---|---|

Table 2.16: Confusion matrix of webshell detection using ASP.NET-ASAF

|  | **Real Webshell** | **Real Benign** |
|---|---|---|
| Predicted Webshell | 407 | 10 |
| Predicted Benign | 6 | 659 |

Table 2.17: Key metrics of webshell detection by using ASP.NET-ASAF (%)

| **Measure** | **Value (%)** |
|---|---|
| Accuracy | 98.52 |
| Precision | 97.60 |
| Recall | 98.55 |
| Specificity | 98.51 |
| F1-Score | 98.07 |
| False Positive Rate | 1.49 |
| False Negative Rate | 1.45 |

In addition to the accuracy evaluation results, we also measured the detection time for the PHP-ASAF model. Accordingly, the total time for the PHP-ASAF model to detect webshell on the test dataset of 2272 PHP source files is 2945 seconds. That means it takes an average of 1.29 seconds for each PHP source file. For the source code-based webshell detection method, the time factor is not the most important factor, and it depends a lot on the size of the source file. Therefore, the average time we measured is only relative.

We measured the webshell detection time for the ASP.NET-ASAF model on a test dataset of 1082 files. It takes an average of 1.35 seconds for each ASP.NET source file, higher than the average time of PHP-ASAF. This is because ASP.NET is a compiled language, the opcode vectorization process will take more time than interpreted languages.

Currently, there are not many studies capable of detecting ASP.NET webshell using source code analysis techniques, and the number of studies willing to share source code is even more limited. Therefore, comparing results with many studies to ensure objectivity is relatively difficult.

## 2.5    Summary of Chapter 2

Chapter 2 of the dissertation provides an overview of the foundational knowledge of the approach to webshell detection using source code analysis. It then proposes an DL-Powered Source Code Analysis Framework at Section 2.2, namely ASAF, to effectively detect malicious code injection attacks into web application source code using known and unknown webshells.

The dissertation is directed toward combining the advantages of two popular detection techniques today, the pattern matching technique using Yara to effectively detect known webshells and the CNN deep learning model to detect new webshells. The framework includes a total of five components linked together through the ASAF workflow. This framework allows us to build each specific system to effectively detect webshell attacks developed in different languages.

To prove the feasibility of this framework, for each type of interpreted and compiled language, the study has experimented with building systems to detect the two most popular server-side programming languages today, PHP at Section 2.3 and ASP.NET at Section 2.4. Experimental results are compared with a number of other research results to demonstrate effectiveness.

Thus, the contributions in this Chapter have basically solved the five challenges mentioned in Section . The ASAF framework allows the construction of specific systems to effectively detect each type of webshell programmed in different languages. By combining the pattern matching-based detection method with the CNN deep learning model, ASAF allows the accurate detection of webshells including advanced webshells, thereby solving challenges number 2 and 4. In the experimental part, we also collected and cleaned a number of webshell samples to serve the training and testing process of the proposed model. PHP and ASP.NET webshell detection models have been applied in the national research project, code number KC01.19/16-20 to demonstrate the applicability in practice.

*The research results in this chapter have been presented in four publications, including one article in the SCI-E/Scopus journal [LVH-J1], one article in an international journal [LVH-J3] (indexed by E-SCI until 2023), one article in the national journal of science and technology on information security [LVH-J4], and two papers at the WoS/Scopus conference [LVH-C1][LVH-C3]. Methods for detecting malicious code in web application source code using PHP and ASP.NET have also been registered for patents at the Department of Intellectual Property, Ministry of Science and Technology [LVH-P1, LVH-P2].*

*In particular, the method for detecting ASP.NET webshells was granted a patent on May 19, 2023.*

Experimental results have shown that the framework has practical applicability in supporting cybersecurity experts in periodically checking the source code of web applications to accurately detect webshells. We can ignore the method's time and resource usage disadvantages. However, this also highlights the ongoing challenge of developing a system that can detect webshell attacks in near real-time and integrates with proactive defense systems to automatically block and filter these attack sources. The next chapter of the dissertation will propose a solution based on network traffic analysis to solve the above problem.

# Chapter 3

# DL-POWERED PROACTIVE WEBSHELL DETECTION AND PREVENTION BY HTTP TRAFFIC ANALYSIS

In previous chapter, we proposed a framework that allows for high-accuracy detection of webshell attacks using source code analysis of web application source code. Although this method can greatly support network security experts, it is only suitable for periodic testing sessions due to the disadvantage of consuming a lot of time and resources. Therefore, there is still a need for another solution that allows simple deployment, is capable of real-time detection of abnormal signs for web servers, and automatically blocks attack sources to minimize damage to the system.

First, in this chapter, we address the problem of detecting webshell attacks using a HTTP traffic analysis of the network traffic with the web server. We clearly define the challenges and objectives of the chapter that require achievement. From there, we propose a model to detect and actively prevent webshell attacks. Finally, we experiment with the model and compare the results with related works.

## 3.1 Problem Statement

The innovation of web development technology has made web applications more and more popular and is gradually replacing traditional native applications because of the advantage of not depending on the operating system. Therefore, protecting information security for the web system becomes more and more important but also challenging. Hackers constantly look for zero-day security vulnerabilities [52] to exploit the system, and security experts constantly create patches or security solutions to fix these vulnerabilities. Usually, security experts are the following, so the systems always have zero-day vulnerabilities. We can deploy a lot of security solutions for the system, such as intrusion prevention/detection systems, firewalls, web application firewalls, etc., but these are not effective against zero-day vulnerabilities. For a web server system, hackers will exploit zero-day vulnerabilities to inject webshells that allow them to take full control of the server system remotely. At that time, installing security patches for the system no longer makes sense because the system is already under the control of the hacker. So what is an effective solution to the problem of system security? The answer is security monitoring to detect abnormalities in the system as soon as possible. Security experts immediately fix the system vulnerabilities before things get worse.

Thanks to the development of ML/DL algorithms [93] that allow deep analysis of network traffic to quickly detect anomalies, it has motivated many studies. The authors in [97] propose a black-box method of webshell detection by analyzing the HTML feature of Webshell pages using a support vector machine (SVM) classification algorithm. This idea allows the authors to not directly perform source code scanning on the web server and can implement integration with IDS systems. By monitoring the request and response traffic to find abnormal behaviors, the authors in [99] propose a feature extraction technique from character-level traffic content. They then use a combination of convolutional neural networks (CNN) and long-short term memory network (LSTM) to detect webshells. The authors in [96] propose a supervised machine learning model to detect webshells based on HTTP traffic analysis. The authors in [49] propose the Difficult Set Sampling Technique (DSSTE) algorithm, which is a combination of Edited Nearest Neighbor and KMeans for intrusion detection. The authors also experimented to compare DSSTE with other machine learning and deep learning techniques using two popular IDPS network stream datasets, NSL-KDD and CSE-CIC-IDS2018.

The above research results have shown that there are three major challenges that need to be solved.

- Network intrusion detection systems based on signatures/rules are not effective in detecting and preventing new generation webshells equipped with encryption and obfuscation techniques.

- In intrusion detection deep learning models, especially for attacks like malware, webshells, etc., the training dataset often has a much smaller number of intrusion attack samples than benign samples. This requires technical mechanisms to minimize the effect of class imbalance when training deep learning models.

- Using MLA/DLAs to deeply analyze network flows enables accurate detection of webshell attacks. However, it must be able to integrate with the NetIDPS system for automatic blocking of suspicious webshell attack source addresses in real-time.

However, this approach holds significant promise. Inspired by the above research results, the dissertation determines the research direction that will focus on proposing a comprehensive solution that combines signature-based detection techniques with the deep neural network model to effectively real-time detect and prevent various types of webshell attacks.

The problem in this chapter will be stated as follows:

**Given:** *x = HTTP traffic* **Find:**

$R(α)$ : *Rule-based detector, where α is the ruleset of webshell*

$D(β)$ : *deep learning model, where β are the parameters of the model*

$$F(P,D)(x) = \begin{cases} 1, & \text{if } x \text{ is a webshell query} \\ 0, & \text{if } x \text{ is benign query} \end{cases}$$

Let $x$ be a HTTP traffic that exchanged with web application server. Let $P$ be a rule-based webshell detector with a set of $α$ rules as patterns to recognize webshells. Let $D$ be a deep learning model with the parameters that make up the model being $β$. We need to find a function $F(P,D)$ with the optimal $α$ and $β$ for P, D such that if $x$ is a webshell query to the web server, then $F(x) = 1$ else if $x$ is a benign query then $FW(x) = 0$.

**Three specific goals are as follows:**

- Proposing a deep neural network (DNN) model allows in-depth analysis of network traffics exchanged with web servers to detect real-time signs of webshell attacks. The DNN model will undergo hyperparameter tuning to optimize performance and accuracy.

- Proposing an improved loss-function algorithm to solve the problem of imbalance in the data set.

- Integrating the DNN model with the NetIDPS solution to enable automatically adding attack source IPs to a blacklist and proactively blocking URI queries to webshell on the web server.

## 3.2 Proactive Webshell Detection and Prevention 3.2.1

**Approach Direction**

The intrusion detection method combining rule matching and the deep learning model is illustrated in Fig. 3.1a and Fig. 3.1b. In this method, when the NetIDPS system device is deployed in inline mode, the network traffic, in both receiving and transmitting directions, is captured by the NetIDPS system. After performing the decoding step, it will proceed with the intrusion detection process based on two detectors. The first detector uses a rule set for intrusion detection. This ruleset is stored in files that have the same rule format as the rules of the Suricata. Each rule will have a pattern or a unique signature that identifies network traffic as under attack. Based on the ruleset, in the inline mode, each network packet will be inspected by the NetIDPS and granted four actions: (i) drop the packet; (ii) reject the packet (discard the packet and notify the source that sent the packet); (iii) pass and alert; (iv) pass without warning.

For network flows whose actions are determined by the rule-based detector to be Drop or Reject, the NetIDPS system will proceed to intercept the incoming data flows. Then, the network traffic flows will not be allowed on the outgoing side of NetIDPS. In the event that the matching result is an alert action, the system will acknowledge the alarm and still let the traffic flow to the outgoing side. Similarly, if the incoming network traffic matches the pass action rule, NetIDPS does not inspect the flows.

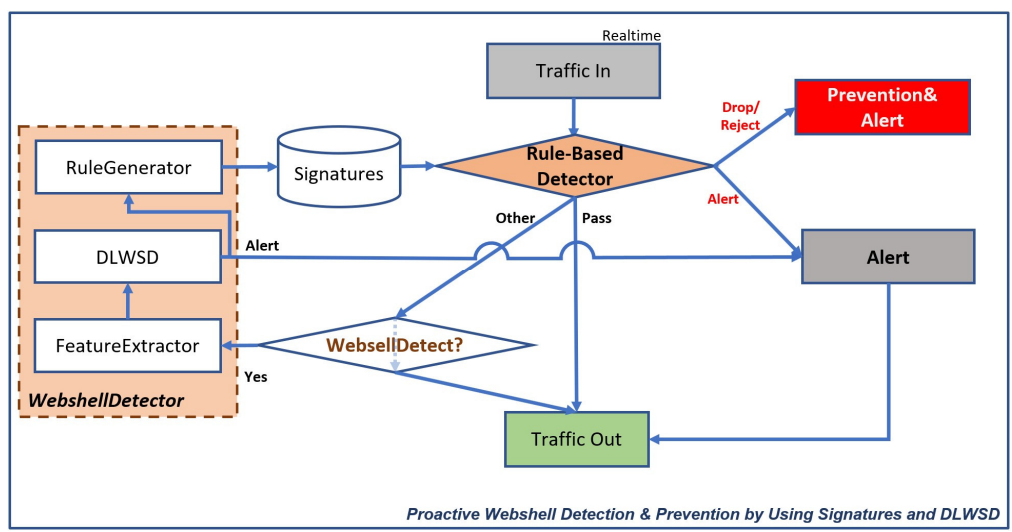

(a) IPS mode

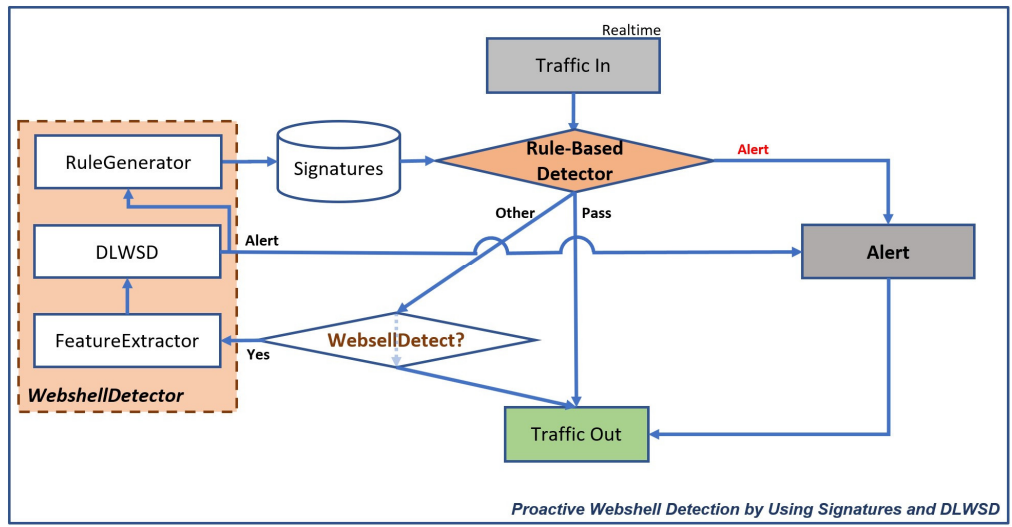

(b) IDS mode

Figure 3.1: Proactive webshell detection method based on signatures and DNN

The remaining cases, noted by **"Other"**, correspond to network traffic flows that do not match any rules of NetIDPS.

For the other cases, the traffic flows can be further examined by a detector, namely **DeepInspector**. This detector is based on a deep learning model to detect whether the traffic flow is possibly an intrusion by using a webshell. Activation of this deep detector is also set by us through the *deep_inspecting* selection parameter in the NetIDPS system configuration: *deep_inspecting* has the value "yes", then NetIDPS will investigate further with data streams that are out of the set. the law; value "no", the system will not integrate the deep verification process.

For deep inspection with a DNN machine learning model, we use six key parameters to control the deep inspection process: *inspection_frequency*, *frequency_min*, *frequency_max*, and *inspection_interval*, *interval_min*, *interval_max*. These parameters are all natural numbers with units of milliseconds and have the following meanings:

- *inspection_frequency*: sampling frequency for deep analysis when this value is determined to be greater than 0; when this parameter is 0, the NetIDPS system will sample deep analysis with random frequency in the range determined from [*frequency_min,frequency_max*]. The default value of the sampling frequency is 2 minutes, and the default interval is from 1 to 5 minutes.

- *inspection_interval*: sampling interval for deep analysis when this value is greater than 0. In case this parameter has the value 0, the system will sample for deep analysis in a random interval from [*interval_min,interval_max*]. The default value of the sampling time is 20 seconds, and the default interval is between 10 and 30 seconds.

These parameters will be selected and configured in the NetIDPS system. With a large number of network data streams (throughput of 10Gbps or more), the sampling frequency and time will determine the performance of the NetIDPS system. Too fast sampling frequency will result in the system having to execute many intrusion detections based on the deep learning model. Therefore, these parameters must also be selected according to the capacity of the NetIDPS computing system.

The DeepInspector is built from a deep learning model using a deep neural network (DNN). In this detector, the network traffic flows sampled in PCAP format will undergo feature extraction. Each traffic flow will be extracted into 83 features using the CICFlowMeter tool [31]. In our model, the four features *"Flow ID, Src IP, Src Port, Label"* are completely ignored. From the remaining 79 features, 2 features *"Dst Port, Protocol"* are used for the categorical variables, and the remaining 77 features are considered continuous variables in the DNN model.

The feature set of the sampled network flows will be passed to the DNN-based detector. Based on the DNN model trained from the specialized data set we collected and built, DeepInspector will classify each network data stream into one of two classes, of which one is a clean data flow (called Benign) and the other corresponds to webshell attacks (namely Webshell). In the case of the data flows classified as Webshell, DeepInspector first sends the alerts to the alarm generator. Then, it will generate a new rule or update an existing rule in the ruleset of NetIDPS in order to alert or even drop/reject future similar traffic flows.

In case the NetIDPS system is deployed passively for network intrusion detection, as illustrated in Fig. 3.1b, the network traffic flows will be delivered to the NetIDPS through a SPAN/Mirror port of the switch. In this deployment scenario, all Drop/Reject actions cannot affect the network traffic flows. Therefore, the intrusion prevention section is not involved in the operation of NetIDPS. And then, all rules with Drop/Reject action are converted into alert actions.

### 3.2.2 Deep Learning Intrusion Detection Model

In the combined method that we propose in this present invention, we choose to use a deep learning model using DNN. Based on our experimental results, the network intrusion detection with the DNN deep learning model using the *tabular_learner* technique of the FastAI development framework allows for the best accuracy, minimizing the false detection rate when compared to other deep learning models as well as other development frameworks like Keras, TensorFlow, Theano, etc. [64].

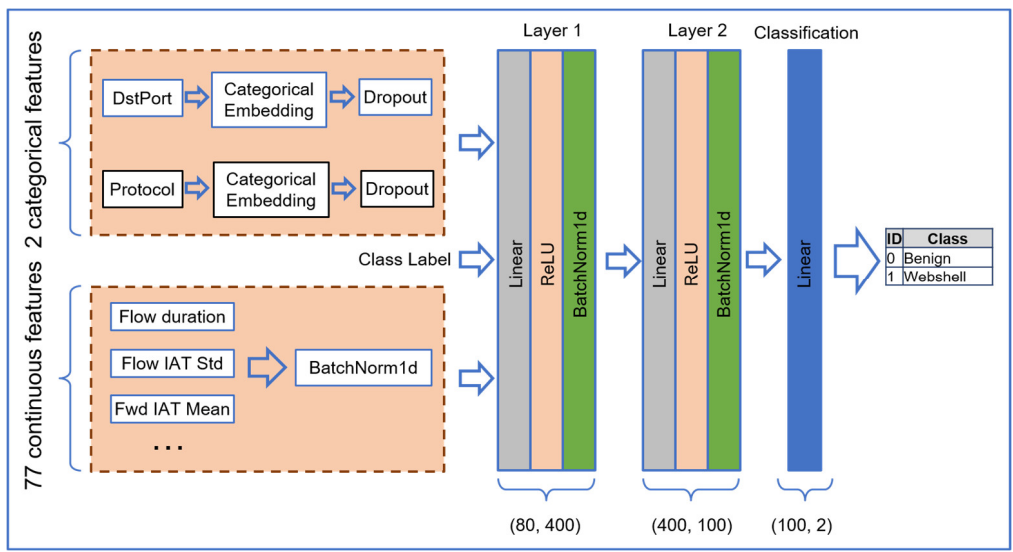

Figure 3.2: DNN architecture for webshell detection

Fig. 3.2 illustrates our DNN model. There are four important components, including: categorical variables, continuous variables, hidden layers, and output layers. Type variables are those that have a finite non-numeric value, such as an IP address, protocol, etc. Whereas continuous variables have any value within a range of values, preset value. Besides the above components, the architectural model has the following parameters:

- Batch Normalization: Batch normalization is one of the most popular normalization methods in deep learning modeling. It enables faster and more stable training of deep neural networks by stabilizing the distribution of input layers during training. Batch normalization also serves as a routine to help reduce overcrowding. Using batch normalization, the model training process does not need to use dropout, so there is no loss of information.

- ReLU: The ReLU function is often used when training neural networks. ReLU filters values less than 0 according to the trivial formula $f(x) = max(0,x)$. The latter has many advantages over *Sigmoid* and *Tanh* such as the convergence speed and calculation being much faster.

The architecture of the DNN model applied in our proposed method is established as follows:

- The two data flow characteristics, destination port (DstPort) and protocol (Protocol), are treated as two categorical variables. Each attribute will go through the categorical embedding and dropout.

- The 77 characteristics related to network data flows and traffic features (specifically 'Timestamp', 'Flow Duration', 'Tot Fwd Pkts', 'Tot Bwd Pkts', 'TotLen Fwd Pkts', 'TotLen Bwd Pkts', ' Fwd Pkt Len Max', 'Fwd Pkt Len Min', 'Fwd Pkt Len Mean', 'Fwd Pkt Len Std', 'Bwd Pkt Len Max', 'Bwd Pkt Len Min', 'Bwd Pkt Len Mean', 'Bwd Pkt Len Std', 'Flow Byts/s', 'Flow Pkts/s', 'Flow IAT Mean', 'Flow IAT Std', 'Flow IAT Max', 'Flow IAT Min', 'Fwd IAT Tot', ' Fwd IAT Mean', 'Fwd IAT Std', 'Fwd IAT Max', 'Fwd IAT Min', 'Bwd IAT Tot', 'Bwd IAT Mean', 'Bwd IAT Std', 'Bwd IAT Max', 'Bwd IAT Min', 'Fwd PSH Flags', 'Bwd PSH Flags', 'Fwd URG Flags', 'Bwd URG Flags', 'Fwd Header Len', 'Bwd Header Len', 'Fwd Pkts/s', 'Bwd Pkts/ s', 'Pkt Len Min', 'Pkt Len Max', 'Pkt Len Mean', 'Pkt Len Std', 'Pkt Len Var', 'FIN Flag Cnt', 'SYN Flag Cnt', 'RST Flag Cnt' , 'PSH Flag Cnt', 'ACK Flag Cnt', 'URG Flag Cnt', 'CWE Flag Count', 'ECE Flag Cnt', 'Down/Up Ratio', 'Pkt Size Avg', 'Fwd Seg Size Avg' , 'Bwd Seg Size Avg', 'Fwd Byts/b Avg', 'Fwd P kts/b Avg', 'Fwd Blk Rate Avg', 'Bwd Byts/b Avg', 'Bwd Pkts/b Avg', 'Bwd Blk Rate Avg', 'Subflow Fwd Pkts', 'Subflow Fwd Byts', 'Subflow Bwd Pkts', 'Subflow Bwd Byts', 'Init Fwd Win Byts', 'Init Bwd Win Byts', 'Fwd Act Data Pkts', 'Fwd Seg Size Min', 'Active Mean', 'Active Std', 'Active Max', 'Active Min', 'Idle Mean', 'Idle Std', 'Idle Max', 'Idle Min') were used as continuous variables.

The categorical class label of the network data stream is also included with the attributes in the first hidden class for training. This layer is composed of three standard blocks: "Linear", "ReLU" and "BatchNorm1D". From 80 input attributes (including layer label attributes), the output of the first hidden layer is set by us to 400 features. For the second hidden layer, this layer structure is the same as the first hidden layer. However, from the 400 input attributes, the output will be normalized to 100.The final output layer assumes the role of classification through the linear filter. This set maps from 100 input attributes to 1 unique value representing 1 class out of binary classes:
1 clean class (Benign) and 1 webshell attack class (Webshell).

### 3.2.3 Webshell Detection and Prevention

From the trained and stored DLWSD deep learning model, we built the *DeepInspector*, deploying a service running in kernel mode (daemon), to receive network traffic flows saved as PCAP files by NetIDPS. DeepInspector takes on the role of monitoring the sampling flows generated frequently *inspection_frequency* during *inspection_duration* by NetIDPS to analyze and detect webshell attacks. The process of performing webshell analysis and detection is illustrated in Algorithm 3.1. Accordingly, when the NetIDPS system is configured to enable intrusion detection mode that combines both law and machine learning, the network data streams will be received and sent to DeepInspector system software according to a multi-process communication mechanism using a Unix socket.

The PCAP data received by DeepInspector will be analyzed and converted into a set of features through CICFlowmeter [31]. Each traffic flow is modeled through 83 features. Next, we normalize this feature set and keep only exactly 79 features used to analyze and predict whether network flows are webshell attacks or not. Based on the prediction result, we use the function *argmax* to find the Benign/Webshell classes with the largest predicted probability. Then, for each webshell detected flow, DeepInspector generates an EVE-log-styled alert in the specific class "Webshell Attacking" of NetIDPS with a severity of 1. Depending on the current signature set of NetIDPS, DeepInspector also generates a new rule with the Drop/Alert action to quickly detect similar webshell-attacking intrusions. New rules generated from DeepInspector will be saved in the directory /etc/NetIDPS/rules/ which contains the NetIDPS system's rule sets. From there, DeepInspector will send information to the NetIDPS system **Algorithm 3.1** Proactive webshell detection and prevention

**Input:** *f* - PCAP traffic flows file.

**Output:** *Alerts* - EVE-log containing alerts for all webshell attack flows;

*Rules* - rulelist containing new rules or updating existing rules to alert/drop/reject traffic flows

1: *Alerts* ← ∅

2: *Rules* ← ∅

3: *F* ← *CICFlowmeter*(*f*)         ▷ extract 83 features of each flow from PCAP file *f*

4: *Fin* ← *F*\[*FlowID,SrcIP,SrcPort,Label*] ▷ remove 4 unused features 5: *preds* ← *DLWSD.Predict*(*Fin*) ▷ perform the prediction

```
 6: FC = preds.argmax(axis = 1)    ▷ get the flow classes: 0 - Benign; 1 - Webshell
 7: for i ← 0 to length(FC) do
 8:     if FC[i] == 1 then                           ▷ classified as Webshel flow
 9:         Alerts ← EveWSAlert(F[i]) ▷ Constitute an EVE alert by using metadata from the flow
            F[i]; set alert category being "Webshell" with severity of 1
10:         Rules ← WSRuleGenerator(F[i])    ▷ generate a new rule or update an existing rule
            to handle the next similar flows
11:     end if
12: end for
13: return Alerts;Rules
```

via the Unix socket mechanism so that NetIDPS updates and uses new rules, serving to prevent intrusion when there are similar attacks on network traffic flows later.

### 3.2.4   Handling Imbalanced Datasets

In fact, when implementing IDPS systems, compromised traffic flows are often tiny compared to "benign" network flows. It is also shown in statistics that most of the standard datasets related to network intrusion have a much lower number of attacked flows than benign flows. The imbalance [29] samples between classes severely affects both the training process and the classification process in machine learning models, especially for deep learning models. For the binary classification problem, the class having the most samples is called a "majority class"; the other is called a "minority class".

From there, it is necessary to have methods to handle data imbalances between classes. One of the methods of handling data imbalances between classes often used in machine learning models is to assign weights to classes. The sample imbalances are used to create the weights for each class in the training process. These weights are used by the cross-entropy loss function in order to ensure that the majority class is down-weighted accordingly. Thus, the loss function for the DNN training is built by Algorithm 3.2.

---

**Algorithm 3.2** Loss function to balance dataset

**Input:** *df* - Dataframe containing all samples, classified by the 'Label' column.

**Output:** *lossFunc* - cross entropy loss function in order to ensure that the majority class is down-weighted accordingly

1: *samples* ← *df.groupby*(´Label´).*count*()        ▷ count the samples for each class specified by 'Label'
2: *nB,nW* = *samples.iloc*[0,0],*samples.iloc*[1,0]
3: *wB* = (*nB* + *nW*)/(2.0 ∗ *nB*) ▷ weight for Benign class 4: *wW* = (*nB* + *nW*)/(2.0 ∗ *nW*) ▷ weight for Webshell class
5: *lossFunc* = *CrossEntropyLossFlat*(*weight* = [*wB,wW*])
6: **return** *lossFunc*

---

## 3.3    Experiments and Evaluation

### 3.3.1    Environment

To validate the proposed method, we deployed a NetIDPS in an appliance having a configuration of 2 x Intel Xeon-Platinum 8160 (2.1GHz/24-core); 384 GB DDR42666 RAM; NVIDIA Tesla T4 16GB Computational Accelerator;SmartNIC Napatech NT40E3-4-PTP (10Gb 4-port SFP+, 4 GB DDR3 RAM buffer). Suricata v6.0.3 is used as the core of NetIDPS. However, many important components have also been added in Suricata to be able to control the traffic flows and implement the deep inspection strategy.

We use Python version 3.8 as a programming language with the following libraries and frameworks: Fastai V2.3.0, Scikit-learn V0.24.1, Matplotlib V3.4.1, Pandas V1.2.3, Numpy V1.20.2. For selecting and tuning hyperparameters in the DNN model, we use the

Adaptive Experimentation Platform, termed Ax [19], which is a machine learning system to help automate this process. Using Bayesian optimization makes Ax suitable for a wide range of applications.

### 3.3.2   Dataset Preparation

To experiment, one of the biggest difficulties we encountered was data collection. There are many published network flow datasets since 1998 such as DARPA (Lincoln Laboratory 1998-99), CAIDA (Center of Applied Internet Data Analysis 20022016), ADFA (University of New SouthWales 2013), and CSE-CIC-IDS2017, CSECIC-IDS2018 (Canadian Institute for Cybersecurity). However, these datasets cover many types of network attacks, with very little data available about webshell attacks. So to collect data for training and testing the model, we built a testbed system as shown in Fig. 3.3 which is divided into two completely separated networks, namely DMZ-Network and Attack-Networks. On the former, we deploy all common and necessary equipment, including routers, firewalls, switches, and three servers, which are a web server, a web application server, and a web database server. On the AttackNetworks, we use the Kali Linux operating system as the attacker server, which uses more than 400 types of PHP, ASP, ASPX, JS webshells. We use several craws website tools to create the normally HTTP traffic as legal clients. Besides that, we also simulate webshell attacks by using Kali Linux to upload and execute webshell to create intrusion traffic. Suricata is used as a packet capture and HTTP filtering tool and saves network traffic into PCAP files.

Thus, in order to validate and evaluate the effectiveness of the DLWSD method, we use two datasets:

- Dataset 1 (DS1): This is the data set that we directly build through the testbed system described above. There are two labeled data types in the data set: *Webshell* representing a flow containing a Webshell attack embedded in packets and sent directly to the webserver using the HTTP method, and *Benign* for normal HTTP flow. Labeling data is also an important step that takes a considerable amount of time and effort. In this step, we create an automatic tool in Python for automatically labeling the data. There is a total of 180,089 Benign flows and 7,310 Webshell flows. The goal of using this dataset is to verify DLWSD's ability to correctly detect webshell attacks by analyzing the network traffic.

---

[19] https://ax.dev/

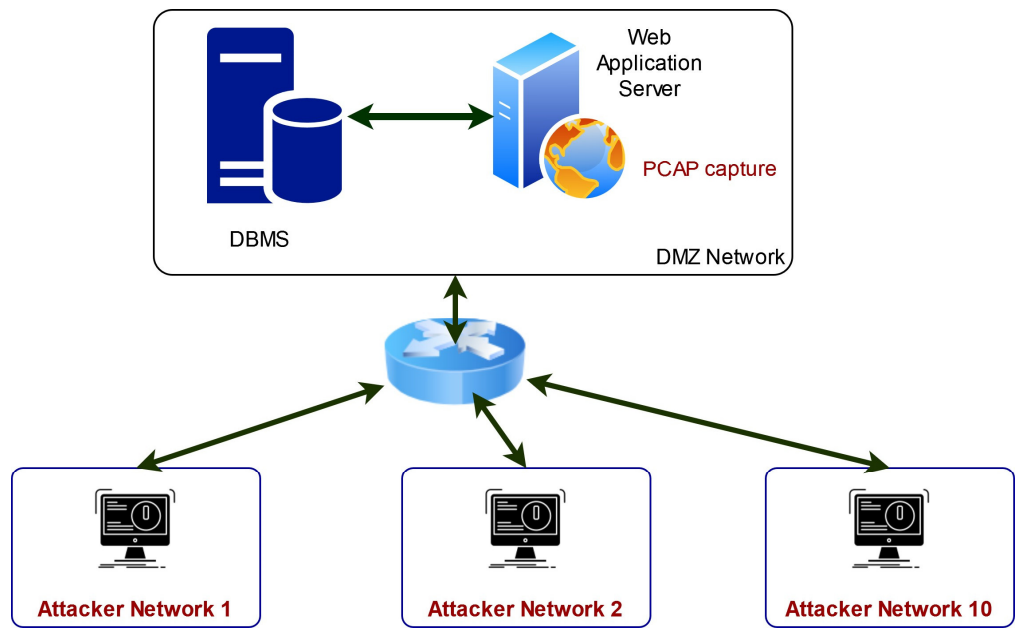

Figure 3.3: Architecture of testbed system

- Dataset 2 (DS2): We use a well-known and reliable dataset on CSE-CIC-IDS2018. This dataset is published by the Canadian Institute for Cybersecurity [46] and is used by many research projects. For our experiment, we chose the dataset *03-022018* due to this dataset only containing bot attacked flows. This kind of attack is considered a webshell-based attack. Thus, our goal when using this data set is to objectively compare the efficiency of our DLWSD method with that of other studies using the same dataset.

After obtaining datasets, they are generated from CICFlowMeter for each dataset. We convert timestamps to Unix epoch numeric values and just keep 79 features as described in Section 3.2.2. We then shuffle the data before saving it into one file containing all the labels. Finally, we obtained the two cleaned datasets presented in Table 3.1.

Table 3.1: Total flows in cleaned datasets

| Dataset | Benign Flows | Webshell Flows |
|---|---|---|
| DS1 | 180,079 | 7,210 |
| DS2 | 758,334 | 286,191 |

From these two datasets, we constitute a new dataset DS3 for global testing by combining DS1 and DS2. DS3 contains two parts: training and testing at a ratio of 7:3. Each part of DS3 is composed of DS1 and DS2 with the corresponding ratio.
Dataset split information is presented in Table 3.2.

Table 3.2: Number of training and testing samples

|  | Begin Samples | Webshell Samples |
|---|---|---|
| **Training** | 656,710[126,054 DS1 - 530,656 DS2] | 205,559[5,048 DS1 - 200,511 DS2] |
| **Testing** | 281,703[54,025 DS1 - 227,678 DS2] | 87,845 [2,162 DS1 - 85,680 DS2] |

### 3.3.3 Hyperparameter Optimization

We select model parameters based on a technique called Hyperparameter Optimization [57]. We use Ax for optimal parameters. Ax is a platform for optimizing any kind of experiment, including machine learning experiments, A/B tests, and simulations. We use a technique called Bayesian Optimization. Bayesian optimization starts by building a smooth surrogate model of the outcomes using Gaussian processes based on the observations available from previous rounds of experimentation.

Table 3.3: Hyperparameter optimization value

| Hyperparameter | Value | Type | Optimal value |
|---|---|---|---|
| Learning rate | [0.001, 1.0] | Range | 0.003 |
| Batch size | [16, 32, 48, 64, 96, 128] | Choice | 64 |
| Epochs | [1, 2, ..., 15, 16] | Choice | 2 |
| Layers | [[200, 100], [400, 100] [1,000, 500]] | Range | [400, 100] |

There are four hyperparameters that can be tuned, and they take two types of values: range and choice, as shown in Table 3.3. From that, we obtained the learning rate as 0.003; batch size as 96; number of epochs as 2; and [400, 100] for the features of layers.

To ensure the hyperparameters are optimized, we perform k-fold cross-validation with the DS1 and also measure the prediction time. We chose the number of folds as 5, and Table 3.4 shows our cross-validation results. The results of 5-fold cross validation are almost perfect with an average accuracy of 99.99%, F1-score of 99.99%, FPR ratio 0.216%, and execution time are also very short, under two seconds.

Table 3.4: Result of hyperparameter optimization with 5-fold cross validation for DS1

| Epoch | Accuracy | Recall | F1-score | FPR | Time (s) |
|---|---|---|---|---|---|
| 1 | 99.99 | 100 | 100 | 0.2 | 0.90 |
| 2 | 100 | 100 | 100 | 0.0 | 1.35 |
| 3 | 99.98 | 99.99 | 99.99 | 0.34 | 1.96 |
| 4 | 99.99 | 100 | 99.99 | 0.2 | 0.98 |
| 5 | 99.98 | 99.99 | 99.99 | 0.34 | 1.00 |
| **Average** | 99.99 | 100 | 99.99 | 0.216 | 1.238 |

### 3.3.4 Results and Evaluation

From the proposed method, we built a tool to evaluate our method with the cleaned datasets mentioned above. We implement DLWSD based on the FastAI framework. To validate the efficiency of our method, DLWSD, three scenarios were built and described as follows:

- S1: Use consecutively DS1 and DS2 to perform the training and testing the DLWSD method without applying the adjustment of the imbalanced dataset.

- S2: Use consecutively DS1 and DS2 to perform the training and testing the DLWSD method within handing the imbalanced dataset.

- S3: Use DS3 to train and test the DLWSD while balancing the classes.

We use the following evaluation metrics: accuracy, precision, F1-score, recall, AUC, and FPR, and we also measure the prediction time to show experimental results.

#### 3.3.4.1 S1 Results

In this scenario, we perform 5-fold cross-validations with the DS1 and DS2 without adjusting the imbalanced dataset by assigning the weights for classes. Cross-validation is a statistical method used to estimate the performance of machine learning models. The dataset will be randomly divided in an 80:20 ratio, corresponding to the training and testing datasets at each fold. The goal of this scenario is to evaluate the efficiency and performance of the DNN model that has been tuned for hyperparameter optimization.

Table 3.5: DLWSD 5-fold cross-validation with DS1

| Fold | Accuracy | Precision | F1-score | Recall | AUC | FPR | Time(s) |
|---|---|---|---|---|---|---|---|
| 1 | 99.90 | 99.11 | 99.55 | 100.00 | 99.99 | 0.89 | 3.74 |
| 2 | 99.98 | 99.52 | 99.76 | 100.00 | 99.99 | 0.48 | 3.59 |
| 3 | 99.98 | 99.38 | 99.69 | 100.00 | 99.99 | 0.62 | 3.59 |
| 4 | 99.98 | 99.58 | 99.69 | 99.79 | 99.79 | 0.42 | 3.65 |
| 5 | 99.98 | 99.45 | 99.72 | 100.00 | 99.99 | 0.55 | 3.62 |
| AVG | 99.98 | 99.41 | 99.68 | 99.96 | 99.95 | 0.59 | 3.63 |

Table 3.6: DLWSD 5-fold cross-validation with DS2

| Fold | Accuracy | Precision | F1-score | Recall | AUC | FPR | Time(s) |
|---|---|---|---|---|---|---|---|
| 1 | 99.97 | 99.92 | 99.94 | 99.96 | 99.99 | 0.03 | 20.03 |
| 2 | 99.92 | 99.70 | 99.85 | 99.99 | 100.00 | 0.11 | 20.12 |
| 3 | 99.99 | 99.98 | 99.97 | 99.97 | 100.00 | 0.01 | 19.96 |
| 4 | 99.93 | 99.84 | 99.88 | 99.92 | 99.98 | 0.06 | 20.52 |
| 5 | 99.98 | 99.93 | 99.96 | 99.99 | 99.99 | 0.03 | 20.38 |
| AVG | 99.96 | 99.87 | 99.92 | 99.97 | 99.99 | 0.05 | 20.20 |

The evaluation results are shown in Table 3.5 and Table 3.6. For the DS1, the results of 5-fold are almost perfect, with an average accuracy of 99.98%, F1-score of 99.96%, FPR ratio 0.59%, and an execution time are also very fast at 3,63 seconds for 20% of 56,187 samples. For the DS2, the results are even better when the number of samples increases to 313,358 with the average values of accuracy, F1-score and FPR being 99.96%, 99.92%, 0.05% respectively. The average time to classify a sample in both datasets is about 32 milliseconds.

### 3.3.4.2 S2 Results

As mentioned above, the amount of data related to a webshell attack is relatively low, which can lead to a misperception of model quality. The object of this scenario is to help us directly compare the DLWSD with and without applying the handling imbalance technique. Table 3.7 and Table 3.8 show the results of 5-fold cross validation of DLWSD on the DS1 and DS2 datasets that assigned the weights for classes.

When directly comparing the results of DLWSD in scenarios S1 and S2, it is easy to see that applying the class weighting technique gives better prediction results in almost all evaluation metrics, except that the execution time is almost unchanged. Specifically, our method gives a very high F1-score of 99.81% with DS1 and 99.98% with DS2. Moreover, the false positive rates are also tiny: only 0.14% and 0.01% respectively, with DS1 and DS2. This means our method allows us to minimize the rate of webshell mistake detection.

Table 3.7: Weighted-DLWSD 5-fold cross-validation with DS1

| Fold | Accuracy | Precision | F1-score | Recall | AUC | FPR | Time(s) |
|---|---|---|---|---|---|---|---|
| 1 | 99.98 | 99.59 | 99.76 | 99.93 | 99.93 | 0.02 | 3.44 |
| 2 | 99.98 | 99.72 | 99.79 | 99.86 | 99.86 | 0.01 | 3.44 |
| 3 | 99.99 | 99.65 | 99.83 | 100.00 | 100.00 | 0.01 | 3.45 |
| 4 | 99.98 | 99.59 | 99.72 | 99.86 | 99.86 | 0.02 | 3.45 |
| 5 | 99.99 | 99.86 | 99.93 | 100.00 | 100.00 | 0.01 | 3.70 |
| AVG | 99.98 | 99.68 | 99.81 | 99.93 | 99.93 | 0.14 | 3.50 |

Table 3.8: Weighted-DLWSD 5-fold cross-validation with DS2

| Fold | Accuracy | Precision | F1-score | Recall | AUC | FPR | Time(s) |
|---|---|---|---|---|---|---|---|
| 1 | 99.98 | 99.99 | 99.97 | 99.94 | 99.99 | 0.00 | 20.02 |
| 2 | 99.98 | 99.99 | 99.96 | 99.93 | 100.00 | 0.00 | 19.89 |
| 3 | 99.99 | 100.00 | 99.99 | 99.97 | 100.00 | 0.00 | 20.52 |
| 4 | 99.99 | 99.99 | 99.98 | 99.97 | 100.00 | 0.01 | 20.43 |
| 5 | 99.99 | 99.97 | 99.98 | 99.99 | 100.00 | 0.01 | 20.48 |
| AVG | 99.99 | 99.99 | 99.98 | 99.96 | 100.00 | 0.01 | 20.27 |

### 3.3.4.3 S3 Results

In this scenario, the dataset we use is DS3, which is a mixed dataset of DS1 and DS2. Through the comparison results between scenarios 1 and 2, it shows that DLWSD when applying the class weighting technique gives better results, so in this scenario, we will use DLWSD enhanced by balancing classes. The goal of this scenario is to provide the most objective evaluation results with a dataset that includes the CSE-CIC-IDS2018 dataset widely used in many studies and our built dataset in the testbed environment. The evaluation results in this scenario shows the practical Table 3.9: Experiment results with DS3 enhanced by balancing classes

(a) Confusion matrix

| | True Webshell | True Benign |
|---|---|---|
| **Predicted Webshell** | 87,794 | 48 |
| **Predicted Benign** | 58 | 281,645 |

(b) Performance indicators

| Metric | Value(%) |
|---|---|
| Accuracy | 99.97 |
| Precision | 99.94 |
| F1-score | 99.94 |
| Recall | 99.93 |
| FNR | 0.07 |
| FPR | 0.02 |
| ROC-AUC | 99.90 |

Table 3.10: Comparison of DLWSD with other methods with DS2

| Method | Accuracy(%) | Precision(%) | F1-score(%) | Recall(%) |
|---|---|---|---|---|
| **DLWSD** | **99.99** | **99.99** | **99.98** | **99.96** |
| DNN fast.ai [64] | 99.92 | 99.85 | 99.85 | 99.85 |

| DSSTE+ miniVGGNet [49] | 96.97 | 97.94 | 97.04 | 96.97 |

applicability of DLWSD.

From the results, we can see that DLWSD only mistakenly identified 58 out of 87,794 true webshell attacks, resulting in a relatively low FPR value of 0.02%. The high-value accuracy of 99.97% and low False Negative Rate of 0.07% mean that the rate of benign flow misclassified as a webshell flow is low.

### 3.3.5 Comparisons and Discussions

To objectively evaluate the effectiveness of the DLWSD method, we compared the results with the DNN model using fast.ai in [64] and the DSSTE+miniVGGNet model in [49] on the same CSE-CIC-IDS2018 dataset on 03-02-2018 (DS2). The results in Table 3.10 shows that all performance metrics of the DLWSD method are higher than those of the other methods.

From that, we have built a specialized DeepInspector suite for detecting Webshell exploit-type intrusion attacks. This detector is integrated into the NetIDSP system to enable automatically adding attack source IPs to a blacklist and proactively blocking URI queries to webshell on the web server. Real-time deep inspection of only traffic flows that do not satisfy any signature with periodic sampling at a defined frequency and interval in DLWSD allows for avoiding bottlenecks when dealing with large-scale network traffic. Based on the above results, our research team conducted experimental measurements on the time and performance of the NetIDPS device when deployed in practice and published at [79, 80, 81]. Actual test results allow us to control, detect, and prevent intrusions, especially with tons of Webshell exploits with 4x10Gbps network flows.

The source code and the dataset used in our experiment can be freely accessed from the GitHub link: https://github.com/levietha0311/DLWSD/.

## 3.4 Summary of Chapter 3

In this chapter, we have surveyed, analyzed, evaluated, and identified three challenges to the problem of detecting and preventing the intrusion of malicious Webshell code. In light of these challenges, we have proposed the DLWSD method based on the DNN deep learning network model combined with the traditional rule-based detection model. With the DNN deep learning model, we have modeled each network flow by 79 features, of which 02 features are used for classification. With the loss function correction to handle the training dataset imbalance problem, our proposed DLWSD method gives excellent results with both our generated dataset (DS1) and the prestigious dataset from the Canadian Institute for Cybersecurity (DS2). In-depth experimental results shows that DLWSD gives Webshell detection results with performance metrics (Accuracy, Precision, F1-Score, FPR) of (99.98, 99.68, 99.81, 0.14) and (99.99, 99.99, 99.98, 0.01) respectively. We also combined the two datasets above to form DS3, using 70% for DLWSD training and the remaining 30% for testing. The results obtained with the DS3 are also very good with the corresponding performance metrics (99.97, 99.94, 99.93, 0.02). Compared to the baseline results from previous studies, DLWSD also showed superior results on the same experimental dataset.

The contributions in this Chapter have basically solved the 5 challenges mentioned in Section . The webshell detection approach based on analysis of the HTTP traffic exchanged with the web server allows detecting webshells regardless of the program-

## 3.4. SUMMARY OF CHAPTER 3

ming language. The DNN model enables deep analysis of network traffic to detect webshell queries including new webshells at high speed. For experimentation, in addition to the CSE-CIC-IDS2018 dataset, we proactively deployed a testbed system to simulate webshell attacks to build our dataset. We also integrated the DNN model with the NetIDPS solution to enable proactive detection and filtering of webshell attacks with a bandwidth of up to 4x10Gbps.

*The research results in this chapter have been presented in two publications: one article in the SCI-E/Scopus journal [LVH-J2], one paper at the WoS/Scopus conference [LVH-C2].*

# CONCLUSION AND FUTURE WORKS

## Contribution Highlights

Webshell attacks pose significant dangers to organizations and individuals, primarily due to their potential to grant attackers unauthorized remote access to a server. Webshells are often difficult to detect because they can be hidden within legitimate web applications and can mimic normal traffic. Research aims to improve the detection accuracy of webshell attacks, but optimizing time and system resource usage is critical to maintaining the security and integrity of web applications. To generally solve this problem, the dissertation approaches both directions based on network traffic analysis and source code analysis techniques. Thanks to advances in deep learning models' ability to automatically learn and extract complex patterns from large datasets, that is particularly important for detecting unknown webshells, which may use sophisticated techniques to avoid detection. Our dissertation delves deeply into webshell attacks, their encryption, evasion, and obfuscation methods, as well as a comprehensive review of related works. From there, we clearly define the research challenges, direction and objectives. By the end of the study, the following contributions have clearly demonstrated the achievement of four research objectives and the way to address five research challenges.

- Proposing an DL-Powered Source Code Analysis Framework (ASAF) that integrates signature-based techniques with deep learning (DL) algorithms. This hybrid approach facilitates the rapid and precise detection of both known and unknown webshell types. The proposed architectural framework serves as a guideline for developing specific models tailored to different programming languages.

    Based on ASAF, proposing two comprehensive webshell detection solutions with

CNN models tailored for PHP as interpreted and ASP.NET as compiled languages. Each model includes an algorithm that transforms the respective source files into flat vectors, encompassing all webshell features. Furthermore, the models incorporate ML/DL algorithms optimized for their specific webshell detection problems to ensure effective detection with minimal computational resources. The effectiveness of these models will be evaluated based on defined measurement criteria and compared to relevant studies. *This contribution have been presented in four publications, including* *one article in the SCI-E/Scopus journal [LVH-J1], one article in an international journal [LVH-J3] (indexed by E-SCI until 2023), one article in the national journal of science and technology on information security [LVH-J4], and one paper at the WoS/Scopus conference [LVH-C1]. Methods for detecting malicious code in web application source code using PHP and ASP.NET have also been registered for patents at the Department of Intellectual Property, Ministry of Science and Technology [LVH-P1, LVH-P2]. In particular, the method for detecting ASP.NET webshells was granted a patent on May 19, 2023.*

- Proposing the DLWSD method to detection and proactively prevent the webshell attacks, which combines a DNN deep learning network model with a traditional rule-based detection model. By correcting the loss function to address the training dataset imbalance, our proposed DLWSD method achieves excellent results on both our generated dataset and the reputable dataset from the Canadian Institute for Cybersecurity. From that result, we developed a specialized DeepInspector suite for detecting webshell exploit-type intrusion attacks. This detector is integrated into the NetIDPS system using the Unix socket IPS communication mechanism. *This contribution have been presented in two publications: one article in the SCI-E/Scopus journal [LVH-J2], one paper at the WoS/Scopus conference [LVH-C2].*

## Dissertation Limitations

Although the dissertation has achieved good research results and made practical contributions as mentioned above, it still has limitations, specifically as follows:

- Most of the current research related to detecting webshell attacks uses autogenerated data sets. This shows that there is actually no webshell data set that is considered standard and widely used in the research community. Out of that general trend, the dissertation is also using self-collected data sets, thus causing many difficulties in objectively comparing the results of other studies.

- The diversity of server-side programming languages leads to a diversity of webshell types. Besides, each type of webshell, according to programming languages, has different characteristics, so different feature extraction methods need to be built. Because of time and resource limitations, the dissertation only chose the two most popular languages today, PHP and ASP.NET, as research and experimental subjects.

- The field of artificial intelligence is currently exploding and constantly making new advances. The introduction of advanced deep learning and machine learning models is continuously being made. Due to time limitations, the dissertation has not been able to research and test the latest current models to apply to the webshell detection problem.

## Future Works

Although the dissertation achieved significant results, this field still has a lot of room for expanding research. In the future, we would like to explore the following research directions:

- Conducting a general survey of webshell datasets used in current research, thereby building a good data set that can be used as a standard for later research related to webshells.

- Expanding research on webshells written in other languages, such as JSP, Ruby, Python, etc., towards building a general model that can effectively detect all types of webshells without depending on the programming language.

- Continue research and experimentation with the latest single DL/ML models and ensemble models to improve the ability to accurately detect advanced webshells.

# PUBLICATIONS

BIBLIOGRAPHY 112

## Conferences